\documentclass[aps,prb,onecolumn,superscriptaddress,amsmath,amsfonts]{revtex4}

\usepackage[dvipdfmx]{graphicx}
\usepackage{bm} 
\usepackage{braket}
\usepackage[usenames]{color}
\usepackage{comment}
\usepackage{amssymb}
\usepackage{mathrsfs}
\usepackage{multirow}
\usepackage{accents}
\usepackage[all]{xy}
\usepackage{amscd}

\newcommand{\C}{\mathbb{C}}
\newcommand{\R}{\mathbb{R}}
\newcommand{\Z}{\mathbb{Z}}

\newcommand{\bk}{\bm{k}}

\makeatletter
\def\widebar{\accentset{{\cc@style\underline{\mskip10mu}}}}
\makeatother
\makeatletter
\def\wideubar{\underaccent{{\cc@style\underline{\mskip10mu}}}}
\makeatother
\newcommand{\maru}[1]{\ooalign{
\hfil\resizebox{.8\width}{\height}{#1}\hfil
\crcr
\raise.1ex\hbox{\large$\bigcirc$}}}

\begin{document}


\title{Topology of nonsymmorphic crystalline insulators and superconductors
}

\author{Ken Shiozaki}
\email{shiozaki@illinois.edu}
\affiliation{Department of Physics, University of Illinois at Urbana
Champaign, Urbana, IL 61801, USA}

\author{Masatoshi Sato}
\email{msato@yukawa.kyoto-u.ac.jp}
\affiliation{Yukawa Institute for Theoretical Physics, Kyoto University,
Kyoto 606-8502, Japan}

\author{Kiyonori Gomi}
\email{kgomi@math.shinshu-u.ac.jp}
\affiliation{Department of Mathematical Sciences, Shinshu University,
Nagano, 390-8621, Japan}




\date{\today}

\begin{abstract}
Topological classification in our previous paper [K.\ Shiozaki and M.\ Sato,
Phys.\ Rev.\ B $\bm{90}$, 165114 (2014)] is extended to nonsymmorphic crystalline
 insulators and superconductors. 
Using the twisted equivariant $K$-theory, we complete the classification
of topological crystalline insulators and superconductors in the
 presence of additional order-two nonsymmorphic space group symmetries. 
The order-two nonsymmorphic space groups include half lattice translation with
 ${\bm Z}_2$ flip, glide, two-fold screw, and their magnetic space groups. 
We find that the topological periodic table shows modulo-2 periodicity in
 the number of flipped coordinates under the order-two nonsymmorphic
 space group.
It is pointed out that the nonsymmorphic space groups allow $\Z_2$
 topological phases even in the absence of time-reversal and/or
 particle-hole symmetries.
Furthermore, the coexistence of the nonsymmorphic
 space group with time-reversal and/or particle-hole
 symmetries provides novel
 $\Z_4$ topological phases, which have not been realized in ordinary topological
 insulators and superconductors.
We present model Hamiltonians of these new topological phases and
analytic expressions of the $\Z_2$ and $\Z_4$ topological invariants.
The half lattice translation with ${\bm Z}_2$ spin
 flip and glide symmetry are compatible with the existence of boundaries, 
leading to topological surface gapless modes protected by the order-two
 nonsymmorphic symmetries.
We also discuss unique features of these gapless surface modes. 
\end{abstract}



\maketitle

\tableofcontents

\section{Introduction}
Since the discovery of the topological crystalline insulator (TCI),
\cite{Fu2010, Hsieh2012, Tanaka2012a, Dziawa2012, Xu2012} it has
been widely recognized that material dependent crystal symmetry
stabilizes topological phases,\cite{Teo2008, Fang2012, Chiu2013,
Morimoto2013, Benalcazar2013, Alexandradinata2014, Shiozaki2014,
Ozawa2014, Chiu2015} in a
manner similar to general symmetries of
time-reversal symmetry (TRS) and particle-hole symmetry (PHS).\cite{Tanaka2011, Hasan2010, Qi2011, Volovik2003} 
The idea of the topological protection by additional symmetry
is also applicable to
superconductors\cite{Ueno2013, Teo2013, Zhang2013a, Tsutsumi2013,
Kobayashi2015, Hashimoto2015}(and
superfluids\cite{Mizushima2012, Wu2013, Tsutsumi2015, Mizushima2015, Mizushima2015b}), being dubbed as topological crystalline superconductor (TCSC).
Like topological insulators and superconductors, TCIs and TCSCs
support stable gapless boundary states associated with bulk topological
numbers, 
when the additional symmetry is compatible with the boundary.
Moreover, crystal symmetry also may stabilize bulk topological gapless modes in
semimetals\cite{Wang2012, Wang2013, Shiozaki2014, Yang2014, Morimoto2014,
Kobayashi2015, Yang2015, Koshino2014, Chiu2014, Chan2015, Youngkuk2015, Young2015} and nodal
superconductors.\cite{Kobayashi2014, Kobayashi2015}
Indeed, symmetry protected Dirac semimetals have been
demonstrated experimentally.\cite{Liu2014,
Xu2014,Borisenko2014,Neupane2014,Jeon2014,Liu2014b} 

In our previous paper,\cite{Shiozaki2014} based on the $K$-theory, we
have presented a 
complete classification of TCIs and TCSCs that are protected by an
additional order-two point group symmetry. The order-two point groups
include spin-flip, reflection, two-fold rotation, inversion, and their
magnetic versions,  
which are characterized by the number of flipped coordinates
under the symmetry. 
Like topological insulators and superconductors,\cite{Schnyder2008, Kitaev2009, Ryu2010} 
our topological table shows the Bott periodicity in the space dimensions, but 
it also exhibits the modulo-4 periodicity in the flipped coordinates.
As well as bulk topological phases, symmetry protected (defect) gapless modes
have been classified systematically.


Until very recently, in the study of TCIs and TCSCs, much attention had been
paid for point group symmetries. 
However, point groups are not only allowed symmetries in crystals. 
Space groups contain combined transformations of point
group operations and nonprimitive lattice translations.  
This class of symmetry is called nonsymmorphic space group (NSG).  
Whereas NSGs also
provide nontrivial topological phases, \cite{Mong2010, Liu2013a,
Fang2015, Shiozaki2015, Dong2015, Wang2015, Varjas2015, Sahoo2015, Lu2015} a systematic
classification has been still lacking.

The purpose of the present paper is to extend our previous
classification to
that with order-two NSG symmetries.
The order-two NSGs include half lattice
translation with ${\bm Z}_2$ flip,  glide, two-fold screw rotation, and their
magnetic symmetries.
We employ the twisted equivariant
$K$-theory\cite{Freed2013, Thiang2014}  to classify TCIs and TCSCs with
order-two NSG. 
In general, NSGs give rise to
momentum-dependent ``twist'' in algebraic relations between the space
group operators. 
The twisted equivariant $K$-theory provides an unbiased and
computable framework to treat any twist in TCIs and TCSCs.  



Using the twisted equivariant $K$-theory, we complete the classification of TCIs and
TCSCs in the presence of additional order-two NSG
symmetries.
We identify $K$-groups that provide topological
numbers for the TCIs and TCSCs. 
From isomorphisms connecting different dimensions of the base space, the
$K$-groups are evaluated by those in one-dimension, which is the
lowest dimension consistent with the twist of order-two NSGs.
Computing all the building block $K$-groups in one-dimension, we 
present the topological table of the TCIs and TCSCs.

The resultant topological table shows several interesting features, which
have not been observed in others. 
First, the NSGs allow $\Z_2$ topological phases even in the absence of
time-reversal and/or particle-hole symmetries. 
The $\Z_2$ phase of glide symmetric insulators\cite{Fang2015,Shiozaki2015}
belongs to this phase.
Second, we find that $\Z_4$ phases are realized by the coexistence of
NSGs with
time-reversal and/or particle-hole symmetries.
Such $\Z_4$ phases have not been discussed before,
for non-interacting
fermions.
In the presence of glide and time-reversal symmetries, 
insulators in three dimensions and superconductors in two dimensions,
respectively, may support the $\Z_4$ phases. 
Finally, like the
order-two point group case,\cite{Shiozaki2014}
the new topological table
hosts the periodicity in 
flipped dimensions under order-two NSGs, although 
the period in this case reduces to half.

For half lattice translation with ${\bm Z}_2$ flip and glide reflection,
there exist one- and two-dimensional boundaries preserving these
symmetries.
From the bulk-boundary correspondence, such boundaries may host 
topological gapless modes protected by these symmetries.
We also discuss the
characteristic features of these gapless boundary modes.
We point out unique spectra of surface modes in the nonsymmorphic $\Z_2$
and $\Z_4$ phases and present analytic expressions of the relevant bulk
topological invariants. 


The organization of this paper is as follows. 
In Sec.\ref{Sec:NSG}, we explain symmetries discussed in this
paper. Our main results are summarized in Sec.\ref{sec:bulk}.
We present relations between $K$-groups with different order-two
NSG symmetries and space dimensions.
From the relations, $K$-groups in one-dimension can be regarded as
building blocks of $K$-groups in higher dimensions.
Complete lists of $K$-groups in one-dimension are also presented.
Topological periodic tables in the presence of order-two
NSG are shown in Sec.\ref{sec:topological_table}. 
The periodicity of topological table is also discussed.
In Sec.\ref{sec:topological_surface_state}, we discuss topological
surface states protected by order-two NSGs.
In general, not all elements of bulk $K$-groups for TCIs and TCSCs
ensure the existence of topological surface states.
Boundary $K$-groups, in which all elements are relevant to the existence
of topological surface states,
are identified in Sec.\ref{Sec:BulkBoundary}. 
Surface states protected by half lattice translation and those protected
by glide symmetry are examined in Secs.\ref{Sec:half-translation} and \ref{Sec:3D}, respectively. 
We conclude this paper in Sec.\ref{Sec:Conc}. 
All technical details on the $K$-theory are presented in the Appendices.




\section{Order-two nonsymmorphic space groups}
\label{Sec:NSG}
A NSG is a combined symmetry of a point group
and a
nonprimitive lattice transformation.
In this paper, we consider order-two NSGs where the symmetry operation
gives a trivial $U(1)$
factor when one applies it twice.

For an order-two NSG, the constituent point group
is also order-two, and the constituent nonprimitive
lattice translation should be a half translation.
Therefore, without loss of generality, an order-two NSG acts on the $d$-dimensional coordinate space as
\begin{align}
(x_1, x_2, \dots, x_d) \mapsto (x_1+\frac{1}{2}, x_2, \dots, x_{d-d_{\parallel}}, -x_{d-d_{\parallel}+1}, \dots, -x_{d}), 
\label{eq:nonsymmorphic_coordinate}
\end{align}
where the lattice constant is set to be $1$. 
Here the constituent order-two point group flips $d_{\parallel}$ coordinates 
$(x_{d-d_{\parallel}+1}, \dots, x_d)$, and 
the direction of the half translation is chosen as the
$x_1$-axis.
It should be noted that the flipped coordinates do
not include $x_1$: Actually, if this
happens, the half translation reduces to a primitive lattice translation
by shifting the origin of the $x_1$-axis, so the corresponding space
group is not nonsymmorphic.    
The dimension $d$ of the system and the number $d_{\parallel}$ of the
flipped coordinates characterize order-two NSGs.
For example, glide in the three dimensions and two-fold screw in three
dimensions are given by
$(d,d_{\parallel})=(3,1)$ and $(d,d_{\parallel})=(3,2)$, respectively.
Also, a combination of a half translation with a ${\bm Z}_2$ global
transformation, which we call as
nonsymmorphic ${\bm Z}_2$ symmetry, 
is specified by $(d,d_{\parallel})=(d,0)$.


Below, we consider band insulators or superconductors which are
described by the Bloch or Bogoliubov-de Gennes Hamiltonian $H({\bm k})$.
The Hamiltonian is supposed to be
invariant under an order-two NSG. 
Because of the periodic structure of the Brillouin zone, 
the base space of $H({\bm k})$ is a $d$-dimensional
torus $T^d$ in the momentum space ${\bm k}$.
The order-two NSG in
Eq.(\ref{eq:nonsymmorphic_coordinate}) acts on $T^d$ as
\begin{align}
\bk \mapsto \sigma \bk \equiv (k_1, k_2, \dots, k_{d-d_{\parallel}},
 -k_{d-d_{\parallel}+1}, \dots, -k_{d}).
\end{align}
Whereas the half translation of $x_1$ in Eq.(\ref{eq:nonsymmorphic_coordinate})
does not change the
momentum ${\bm k}$, 
it provides a non-trivial $k_1$-dependence for the symmetry
operator, as we shall show below. 
To distinguish $k_1$ and the flipped components
${\bm k}_{\parallel}=(k_{d-d_{\parallel}+1},\dots,k_{d})$ from others,
we denote the base space $T^d$ as a direct product $\widetilde{S}^1\times
T^{d-d_{\parallel}-1}\times \widebar{T}^{d_{\parallel}}$, where $k_1\in
\widetilde{S}^1$
and ${\bm k}_{\parallel}\in \widebar{T}^{d_{\parallel}}$. 
Without loss of generality, we assume that the Brillouin zone has the
2$\pi$-periodicity in the $k_i$-direction and $k_i\in [-\pi, \pi]$.

The order-two NSG symmetry implies 
\begin{align}
&U(\bk) H(\bk) U(\bk)^{-1} = c_{\sigma} H(\sigma \bk), && U(\sigma \bk) U(\bk) = \epsilon_{\sigma} e^{-i k_1}, && c_{\sigma}, \epsilon_{\sigma} \in \{1,-1\}, 
\label{eq:nonsymmorphic_unitary}
\end{align}
where $U({\bm k})$ is a unitary operator.
The half translation of $x_1$ in Eq.(\ref{eq:nonsymmorphic_coordinate})
provides
the Bloch factor $e^{-ik_1}$ in the second equation of
(\ref{eq:nonsymmorphic_unitary}).
For faithful representations of order-two NSGs, the sign
$\epsilon_{\sigma}$ should be $1$, but the spinor representation of
rotation makes it possible to have $\epsilon_{\sigma}=-1$.
Here, in addition to ordinary symmetry with $c_{\sigma}=1$, we consider
``antisymmetry'' with $c_{\sigma} = -1$. 
For magnetic order-two NSGs, 
we have
\begin{align}
&A(\bk) H(\bk) A(\bk)^{-1} = c_{\sigma} H(-\sigma \bk), && A(-\sigma
 \bk) A(\bk) = \epsilon_{\sigma} e^{i k_1}.  && c_{\sigma},
 \epsilon_{\sigma} \in \{1,-1\}, 
\label{eq:nonsymmorphic_magnetic}
\end{align}
instead of Eq.(\ref{eq:nonsymmorphic_unitary}), where $A(\bk)$ is an
anti-unitary operator.
In a suitable basis, 
$H({\bm k})$, $U({\bm k})$ and $A({\bm k})$ are
periodic in the Brillouin zone, as shown in Appendix \ref{sec:Periodic_Bloch}.

\begin{table*}[!]
\begin{center}
\caption{AZ symmetry classes and their classifying spaces. 
The top two rows ($s=0,1$ (mod 2)) are complex AZ classes, and the bottom eight
 rows ($s=0,1,\dots,7$ (mod 8)) are
 real AZ classes. The second column represents the names of the AZ classes. 
The third to fifth columns indicate the absence (0) or the presence
 $(\pm 1)$ of TRS, PHS and CS, respectively, where $\pm 1$ means the
 sign of $T^2 = \epsilon_T$ and $C^2 = \epsilon_C$.
The sixth column shows the symbols of the classifying space.
}
\begin{tabular}[t]{ccccccccccc}
\hline \hline
s & AZ class & TRS & PHS & CS & $\mathcal{C}_s$ or $\mathcal{R}_s$ & classifying space & $\pi_0(\mathcal{C}_s)$ or $\pi_0(\mathcal{R}_s)$ \\
\hline
0 & A & $0$ & $0$ & $0$ & $\mathcal{C}_0$ & $\cup_{m \in \Z} (U(n)/U(n+m)\times U(n-m))$ & $\mathbb{Z}$ \\
1 & AIII & $0$ & $0$ & $1$ & $\mathcal{C}_1$ & $U(n)$ & $0$ \\
\hline
0 & AI & $+1$ & $0$ & $0$ & $\mathcal{R}_0$ & $\cup_{m \in \Z}(O(n)/O(n+m)\times O(n-m))$ & $\mathbb{Z}$ \\
1 & BDI & $+1$ & $+1$ & $1$ & $\mathcal{R}_1$ & $O(n)$ & $\mathbb{Z}_2$ \\
2 & D & $0$ & $+1$ & $0$ & $\mathcal{R}_2$ & $O(2 n)/U(n)$ & $\mathbb{Z}_2$ \\
3 & DIII & $-1$ & $+1$ & $1$ & $\mathcal{R}_3$ & $U(2 n)/Sp(n)$ & $0$ \\
4 & AII & $-1$ & $0$ & $0$ & $\mathcal{R}_4$ & $\cup_{m \in \Z}(Sp(n)/Sp(n+m)\times Sp(n-m))$ & $2\mathbb{Z}$ \\
5 & CII & $-1$ & $-1$ & $1$ & $\mathcal{R}_5$ & $Sp(n)$ & $0$ \\
6 & C & $0$ & $-1$ & $0$ & $\mathcal{R}_6$ & $Sp(n)/U(n)$ & $0$ \\
7 & CI & $+1$ & $-1$ & $1$ & $\mathcal{R}_7$ & $U(n)/O(n)$ & $0$ \\
\hline \hline
\end{tabular}
\label{Classifying_space}
\end{center}
\end{table*}

In addition to the order-two NSGs, we also consider
the Altland-Zirnbauer (AZ) symmetries,\cite{Altland1997, Kitaev2009} i.e. TRS $T$, 
PHS $C$, and chiral symmetry (CS) $\Gamma$: 
\begin{align}
&T H(\bk) T^{-1} = H(-\bk), && T^2 = \epsilon_T,&& \epsilon_T \in
 \{1,-1\}, 
\nonumber\\ 
&C H(\bk) C^{-1} = -H(-\bk), && C^2 = \epsilon_C&& \epsilon_C \in
 \{1,-1\}, 
\nonumber\\
&\Gamma H(\bk) \Gamma^{-1} = -H(\bk), && \Gamma^2=1. 
\label{eq:AZsymmetry}
\end{align}
Here $T$ and $C$ act on the single particle Hilbert space as anti-unitary
symmetry, while $\Gamma$ is unitary.
When $T$ and $C$ coexist, we choose the phase of $C$ so that $[T,C]=0$.
The absence of any AZ symmetry or the presence of CS defines the
two-fold complex AZ classes, and 
the presence of TRS and/or PHS defines the 8-fold real AZ classes.
(See Table \ref{Classifying_space}. )
The order-two NSGs subdivide these AZ classes:  
A different commutation or anti-commutation relation between $U(\bk)$
($A(\bk)$) and the AZ symmetries $T,C,\Gamma$ provide a different subclass.
We specify these relations by $\eta_T, \eta_C, \eta_{\Gamma} \in
\{1,-1\}$: 
\begin{align}
&T U(\bk) = \eta_{T} U(-\bk) T, && C U(\bk) = \eta_{C} U(-\bk) C, &&
 \Gamma U(\bk) = \eta_{\Gamma} U(\bk) \Gamma, 
\nonumber\\
&T A(\bk) = \eta_{T} A(-\bk) T, && C A(\bk) = \eta_{C} A(- \bk) C, && \Gamma A(\bk) = \eta_{\Gamma} A(\bk) \Gamma. 
\label{eq:AZNSG}
\end{align}
The data $\{c_{\sigma}, \epsilon_{\sigma}, \eta_T, \eta_C,
\eta_{\Gamma}\}$ defines refined symmetry classes for each AZ class. 

\section{Classification of order-two nonsymmorphic crystalline insulators and
 superconductors}
\label{sec:bulk}

As was explained in the previous paper,\cite{Shiozaki2014} 
$K$-groups for Hamiltonians define topological phases.
In this section, we summarize useful relations of $K$-groups, which are
relevant to the classification of TCIs and TCSCs in the presence of
order-two NSGs.

\subsection{Nonsymmorphic crystalline insulators in complex AZ classes}

We first consider crystalline insulators with an order-two unitary
NSG in complex AZ classes (namely, class A ($s=0$) and class AIII
($s=1$) in Table \ref{Classifying_space}). 
No AZ symmetry exists in class A, and CS exists in class AIII.
We denote the unitary operator $U$ as follows,
\begin{align}
\mbox{in class A} : \left\{ \begin{array}{ll}
U, &\mbox{for $c_{\sigma} = 1$}  \\
\widebar U, & \mbox{for $c_{\sigma} = -1$}  \\
\end{array} \right., && 
\mbox{in class AIII} : \left\{ \begin{array}{ll}
U_{\eta_{\Gamma}},& \mbox{for $c_{\sigma} = 1$} \\
\widebar U_{\eta_{\Gamma}},& \mbox{for $c_{\sigma} = -1$} \\
\end{array} \right.,
\end{align}
where the subscript in $U$ specifies $\eta_{\Gamma}$ in Eq.(\ref{eq:AZNSG}).
Each AZ class has  two nonequivalent subclasses, $t=0, 1$ (mod $2$)
shown in
Table \ref{Symmetry_type_UC}.
Note that $\bar{U}_{\eta_{\Gamma}}$ and
$(-1)^{\eta_{\Gamma}}\Gamma 
U_{\eta_{\Gamma}}$ give the same constraint on
Hamiltonians. 
Thus $U_{\eta_\Gamma}$ and $\widebar{U}_{\eta_\Gamma}$ gives the same
subclasses in class AIII.
%
%
By multiplying $U({\bm
k})$ by $i$,  
we can also change $\epsilon_\sigma$ in
Eq.(\ref{eq:nonsymmorphic_unitary}) 
without changing other symmetry
relations. Therefore, the classification in the present case does not depend on
$\epsilon_\sigma$.
The data $(s,t; d, d_{\parallel})$ specifies the imposed NSG and the
dimension of the system.   

\begin{table*}[!]
\begin{center}
\caption{(Color online)
Possible types [$t=0,1$ (mod 2)] of order-two nonsymmorphic unitary symmetries in complex AZ class [$s=0,1$ (mod 2)].  
$U$ and $\bar{U}$ represent symmetry and antisymmetry, respectively. 
The subscript of $U_{\eta_{\Gamma}}$ and $\bar{U}_{\eta_{\Gamma}}$ specifies the relation $\Gamma U = \eta_{\Gamma} U \Gamma$. 
Symmetries in the same parenthesis are equivalent. 
``$2\Z$'' is a weak topological index from zero-dimension. 
$\textcolor{blue}{\Z_2}$ (blue) and $\Z$ represent an emergent
 topological phase by the additional NSG symmetry and 
a unchanged topological phase under the additional NSG symmetry, respectively (see Appendix \ref{Sec:Sym_Forget} for details). }
\begin{tabular}[t]{ccccccccccc}
\hline \hline
$s$ & $t$ & AZ class & Coexisting symm. & Example of physical realization & $K_{\C}^{(s,t)}(\widetilde S^1)$ \\
\hline
$0$ & $0$ & A & $U$ & TCI with order-two NSG & ``$2\Z$'' 
\\
$0$ & $1$ & A & $\widebar U$ & TCI with order-two NSG antisymmetry&
		     $\textcolor{blue}{\Z_2}$ 
\\
\hline
$1$ & $0$ & AIII & ($U_+$, $\widebar{U}_+$) & TCI with order-two NSG
		 preserving sublattice& $\Z$ 
\\
$1$ & $1$ & AIII& ($U_-$, $\widebar{U}_-$) & TCI with order-two NSG
		 exchanging sublattice& $0$ 
\\
\hline \hline
\end{tabular}
\label{Symmetry_type_UC}
\end{center}
\end{table*}

We use the twisted equivariant $K$-theory to classify the crystalline
insulators.
Each nonsymmorphic crystalline insulator with the
data $(s,t;d,d_{\parallel})$ has its own $K$-group, which we denote 
$K_{\C}^{(s,t)}(\widetilde S^1 \times T^{d-d_{\parallel}-1} \times
\widebar T^{d_{\parallel}})$. 
The $K$-groups provide possible topological numbers for the
nonsymmorphic TCIs. 
%

To evaluate the $K$-groups, we can use the isomorphisms
shifting the dimension of the system:\cite{Shiozaki2014}(see also
Appendices \ref{sec:Dimensional reduction} and \ref{Sec:Dim_Raise_Map})
\begin{align}
&K_{\C}^{(s,t)}(\widetilde S^1 \times T^{d-d_{\parallel}-1} \times \widebar T^{d_{\parallel}}) \cong 
K_{\C}^{(s,t)}(\widetilde S^1 \times T^{d-d_{\parallel}-2} \times \widebar T^{d_{\parallel}}) \oplus 
K_{\C}^{(s-1,t)}(\widetilde S^1 \times T^{d-d_{\parallel}-2} \times
 \widebar T^{d_{\parallel}}), 
\label{eq:KUcomplex1}
\\
&K_{\C}^{(s,t)}(\widetilde S^1 \times T^{d-d_{\parallel}-1} \times \widebar T^{d_{\parallel}}) \cong 
K_{\C}^{(s,t)}(\widetilde S^1 \times T^{d-d_{\parallel}-1} \times \widebar T^{d_{\parallel}-1}) \oplus 
K_{\C}^{(s-1,t-1)}(\widetilde S^1 \times T^{d-d_{\parallel}-1} \times \widebar T^{d_{\parallel}-1}), 
\label{eq:KUcomplex2}
\end{align}
where $s$ and $t$ in the superscript $(s,t)$ are defined modulo 2. 
In the above isomorphisms, the first $K$-groups of the right hand side
gives so-called weak topological
indices, which are obtained as topological indices of stacked lower
dimensional systems. So 
only the second $K$-groups
provide strong topological indices. 
%
By using the isomorphisms, the $K$-group in $d$-dimensions can reduce to
those in lower dimensions: 
Any $K$-group $K_{\C}^{(s,t)}(\widetilde S^1 \times
T^{d-d_{\parallel}-1} \times \widebar T^{d_{\parallel}})$  
can be constructed from a set of the $K$-groups $\{
K_{\C}^{(s,t)}(\widetilde S^1)\}_{s=0,1, t=0,1}$ in one-dimension.
In other words, $K_{\C}^{(s,t)}(\widetilde S^1)$ is the
building block of the $K$-groups.
For instance, the strong topological index in $d$-dimensions is obtained as  
\begin{eqnarray}
K_{\C}^{(s,t)}(\widetilde{S}^1\times T^{d-d_{\parallel}-1}\times
 \widebar{T}^{d_{\parallel}})|_{\rm
 strong}=K_{\C}^{(s-d+1, t-d_{\parallel})}(\widetilde{S}^1)|_{\rm strong}. 
\label{eq:strongKUcomplex}
\end{eqnarray}
We compute $K_{\C}^{(s,t)}(\widetilde S^1)$ 
in Appendix
\ref{Appendix:D}, and summarize the results in the Table
\ref{Symmetry_type_UC}. 
In general, $K_{\C}^{s,t}(\widetilde S^1)$ also contains
weak topological indices originate from topological numbers in zero-dimension. 
For example, 
$K_{\C}^{s=0,t=0}(\widetilde S^1) = 2\Z$ corresponds to the number of occupied
states, where the even number is required by the nonsymmorphic symmetry.
In Tables \ref{Symmetry_type_UC}-\ref{Symmetry_type}, we write such weak
indices as
``$2\Z$'' or ``$4\Z$'' to distinguish those from strong indices which occur
only in one-dimension. 



\subsection{Nonsymmorphic magnetic crystalline insulators in complex AZ classes}
Second, we consider crystalline insulators with
an order-two magnetic NSG symmetry in complex AZ classes. 
We denote the anti-unitary operator $A$ as
\begin{align} 
\mbox{in class A} : \left\{ \begin{array}{ll}
A, & \mbox{for $c_{\sigma} = 1$}  \\
\widebar A, & \mbox{for $c_{\sigma} = -1$}  \\
\end{array} \right., && 
\mbox{in class AIII} : \left\{ \begin{array}{ll}
A_{\eta_{\Gamma}},& \mbox{for $c_{\sigma} = 1$} \\
\widebar A_{\eta_{\Gamma}},& \mbox{for $c_{\sigma} = -1$} \\
\end{array} \right., 
\end{align}
where the subscript in $A$ and $\widebar{A}$
specifies 
$\eta_{\Gamma}$ in
Eq.(\ref{eq:AZNSG}).
It should be noted here that $\epsilon_{\sigma}$ in
Eq.(\ref{eq:nonsymmorphic_magnetic}) change the sign by shifting
$k_1$ by $\pi$, i.e. $k_1\to k_1+\pi$.
Therefore, the topological classification does not depend on
$\epsilon_{\sigma}$. 
There are four nonequivalent symmetry classes labeled by $t =
0,\dots, 3$ (mod $4$) in Table \ref{Symmetry_type_AC}. 
In the present case, the data $(t; d, d_{\parallel})$ specifies the
imposed NSG and the dimension of the system.

\begin{table*}[!]
\begin{center}
\caption{(Color online)
Possible types [$t=0,1,2, 3$ (mod 4)] of order-two additional
 anti-unitary symmetries in complex AZ class.
$A$ and $\bar A$ represent symmetry and
 anti-symmetry, respectively.
The subscript
 of $A_{\eta_{\Gamma}}$ specifies the (anti-)commutation relation 
 $\Gamma A = \eta_{\Gamma} A \Gamma$.
Symmetries in the same parenthesis are equivalent.
``$2\Z$'' is a weak topological index from zero-dimension. 
$\textcolor{blue}{\Z_2}$ (blue) and $\Z$ represent an emergent topological phase by the additional NSG symmetry and 
an unchanged topological phase under the additional NSG symmetry, respectively (see Appendix \ref{Sec:Sym_Forget} for details).  }
\begin{tabular}[t]{cccccc}
\hline \hline
$t$ & AZ class & Coexisting symm. & Example of physical realization & $K_{\C}^t(\widetilde S^1)$ \\
\hline
$0$ & A & $A$ & TCI with magnetic order-two NSG & ``$2\Z$''
\\
$1$ & AIII & ($A_+$, $\widebar{A}_+$) & TCI with magnetic order-two NSG
	     preserving sublattice& $\Z$ 
\\
$2$ & A & $\widebar{A}$ & TCI with magnetic order-two NSG antisymmetry & $\textcolor{blue}{\Z_2}$ 
\\
$3$ & AIII&($A_-$,$\widebar{A}_+$) & TCI with magnetic order-two NSG
	     exchanging sublattice& $0$ \\
\hline \hline
\end{tabular}
\label{Symmetry_type_AC}
\end{center}
\end{table*}

We denote the $K$-group for TCIs and TCSCs with the data
$(t;d,d_{\parallel})$ as
$K_{\C}^{(t)}(\widetilde S^1 \times T^{d-d_{\parallel}-1} \times
\widebar T^{d_{\parallel}})$.
Then, there are isomorphisms
\begin{align}
&K_{\C}^{t}(\widetilde S^1 \times T^{d-d_{\parallel}-1} \times \widebar T^{d_{\parallel}}) \cong 
K_{\C}^{t}(\widetilde S^1 \times T^{d-d_{\parallel}-2} \times \widebar T^{d_{\parallel}}) \oplus 
K_{\C}^{t-1}(\widetilde S^1 \times T^{d-d_{\parallel}-2} \times \widebar
 T^{d_{\parallel}}), 
\label{eq:KAcomplex1}\\
&K_{\C}^{t}(\widetilde S^1 \times T^{d-d_{\parallel}-1} \times \widebar T^{d_{\parallel}}) \cong 
K_{\C}^{t}(\widetilde S^1 \times T^{d-d_{\parallel}-1} \times \widebar T^{d_{\parallel}-1}) \oplus 
K_{\C}^{t+1}(\widetilde S^1 \times T^{d-d_{\parallel}-1} \times \widebar
 T^{d_{\parallel}-1}), 
\label{eq:KAcomplex2}
\end{align}
where $t$ in the superscript is defined modulo 4.
Here the first terms of the right hand side give weak indices, and only
the second terms contain strong indices.  
%
Using the isomorphisms, one can iterate the dimensional reduction except for
the $k_1$-direction $\widetilde S^1$. 
The dimensional
reduction gives the following formula
\begin{eqnarray}
K_{\C}^t(\widetilde S^1 \times T^{d-d_{\parallel}-1} \times
\widebar T^{d_{\parallel}})|_{\rm strong}
=K_{\C}^{t-d+2d_{\parallel}+1}(\widetilde{S}^{1})|_{\rm strong},
\label{eq:strongKAcomplex}
\end{eqnarray}
for the strong topological index in $d$-dimension. 
We compute the building block $K$-group $K_{\C}^t(\widetilde
S^1)$ in Appendix \ref{Appendix:E}, and
summarize the results in the Table \ref{Symmetry_type_AC}.
 


\subsection{Nonsymmorphic (magnetic) crystalline insulators and
  superconductors in real AZ classes}
\label{sec:NMCI_Real}

Finally, we consider crystalline insulators and superconductors with
an order-two NSG in the presence of TRS and/or PHS.
The types of TRS and/or PHS are specified by real AZ classes. See Table
\ref{Classifying_space}.
The order-two NSG can be unitary or anti-unitary.
We denote the unitary operator as
\begin{align}
\mbox{in AI, AII} : \left\{ \begin{array}{ll}
U_{\eta_T},& \mbox{for $c_{\sigma} =1$}  \\
\widebar U_{\eta_T},& \mbox{for $c_{\sigma} = -1$} \\
\end{array} \right., && 
\mbox{in D, C} : \left\{ \begin{array}{ll}
U_{\eta_C},& \mbox{for $c_{\sigma} =1$} \\
\widebar U_{\eta_C},& \mbox{for $c_{\sigma} = -1$}\\
\end{array} \right., &&
\mbox{in BDI, DIII, CII, CI} : \left\{ \begin{array}{ll}
U_{\eta_T, \eta_C},& \mbox{for $c_{\sigma} =1$} \\
\widebar U_{\eta_T,\eta_C}, & \mbox{for $c_{\sigma} = -1$} \\
\end{array} \right., 
\end{align}
and the anti-unitary operator as
\begin{align}
\mbox{in AI, AII} : \left\{ \begin{array}{ll}
A_{\eta_T},& \mbox{for $c_{\sigma} =1$}  \\
\widebar A_{\eta_T},& \mbox{for $c_{\sigma} = -1$} \\
\end{array} \right., && 
\mbox{in D, C} : \left\{ \begin{array}{ll}
A_{\eta_C},& \mbox{for $c_{\sigma} =1$} \\
\widebar A_{\eta_C}, & \mbox{for $c_{\sigma} = -1$}\\
\end{array} \right., && 
\mbox{in BDI, DIII, CII,CI} : \left\{ \begin{array}{ll}
A_{\eta_T, \eta_C},& \mbox{for $c_{\sigma} =1$} \\
\widebar A_{\eta_T,\eta_C},& \mbox{for $c_{\sigma} = -1$} \\
\end{array} \right.. 
\end{align}
Here the subscripts in the operators identify
$\eta_T$ and $\eta_C$ in
Eq.(\ref{eq:AZNSG}).
Since $\epsilon_{\sigma}$ in Eqs. (\ref{eq:nonsymmorphic_unitary}) and
(\ref{eq:nonsymmorphic_magnetic}) changes the sign by replacing $k_1$
with $k_1+\pi$, the topological classification does not depend on
$\epsilon_{\sigma}$.  
As shown in Table
\ref{Symmetry_type}, there are two nonequivalent subclasses, 
which we label as $t =0,1$ (mod $2$). 

\begin{table*}[!]
\begin{center}
\caption{(Color online)
Possible types [$t=0,1$ (mod 2)] of order-two nonsymmorphic
 symmetries in real AZ
 class [$s=0,1,\dots,7$ (mod 8)].
$U$ and $\bar U$ represent unitary symmetry and antisymmetry,
 respectively, and $A$ and $\bar A$ represent anti-unitary symmetry and
 antisymmetry, respectively. 
The subscript of $S$ ($S=U,\bar{U},A, \bar{A}$) specifies the
 commutation(+)/anti-commutation(-) relation between $S$ and TRS and/or PHS. 
For BDI, DIII, CII and CI, where both TRS and PHS exist, $S$ has two subscripts, in which the first one specifies the
 (anti-)commutation relation between $S$ and $T$ and the second one specifies that between $S$ and $C$.    
Symmetries in the same parenthesis are  equivalent.
``$2\Z$'' and ``$4\Z$'' mean weak indices from zero-dimension. 
$\textcolor{blue}{\Z_2}$ (blue) and $\textcolor{blue}{\Z_4}$ (blue) represent emergent topological phases by the additional NSG symmetry. 
$\Z, \Z_2$ represent unchanged topological phases under the additional NSG symmetry
(See Appendix \ref{Sec:Sym_Forget} for details). 
}
\begin{tabular}[t]{cccccccccccc}
\hline \hline
$s$ & $t$ & AZ class & NSG & Magnetic NSG 
& Example of physical realization& $K_{\R}^{(s,t)}(\widetilde{S}^1)$
\\
\hline
$0$ & $0$ & AI & ($U_+$, $U_-$) & ($A_+$, $A_-$) 
& Spinless TRS TI with order-two NSG & ``$2\Z$'' 
\\
$0$ & $1$ & AI & ($\widebar U_-$, $\widebar U_+$) 
& ($\widebar A_+$, $\widebar A_-$) 
& Spinless TRS TI with order-two NSG antisymmetry & $\textcolor{blue}{\Z_2}$
\\
\hline
$1$ & $0$ & BDI & ($U_{++}$, $U_{--}$, $\widebar{U}_{++}$,
	     $\widebar{U}_{--}$) & ($A_{++}$, $A_{--}$, $\widebar
		 A_{++}$, $\widebar A_{--}$) 
& Spinless TRS  TCSC with $[U,TC]=0$& $\Z\oplus\textcolor{blue}{\Z_2}$ 
\\
$1$ & $1$ & BDI & ($U_{+-}$, $U_{-+}$, $\widebar U_{-+}$,
	     $\widebar U_{+-}$) & ($A_{+-}$, $A_{-+}$, $\widebar
		 A_{-+}$, $\widebar A_{+-}$) 
& Spinless TRS TCSC with $\{U, TC\}=0$ & $\textcolor{blue}{\Z_2}$
\\
\hline
$2$ & $0$ & D & ($U_+$, $U_-$) & ($\widebar A_+$, $\widebar
		 A_-$) 
& TCSC with order-two NSG & $\Z_2\oplus \textcolor{blue}{\Z_2}$
\\
$2$ & $1$ & D & ($\widebar U_+$, $\widebar U_-$) & ($A_+$,
		 $A_-$)
& TCSC with magnetic order-two NSG & $\textcolor{blue}{\Z_4}$
\\
\hline
$3$ & $0$ & DIII & ($U_{++}$, $U_{--}$, $\widebar U_{++}$,
	     $\widebar U_{--}$) & ($A_{++}$, $A_{--}$, $\widebar
		 A_{++}$, $\widebar A_{--}$) 
& Spinful TRS TCSC with $[U, TC]=0$ & $\Z_2$
\\
$3$ & $1$ & DIII & ($U_{-+}$, $U_{+-}$, $\widebar U_{-+}$,
	     $\widebar U_{+-}$) & ($A_{-+}$, $A_{+-}$, $\widebar
		 A_{-+}$, $\widebar A_{+-}$) 
& Spinful TRS TCSC with $\{U, TC\}=0$ & $\Z_2$
\\
\hline
$4$ & $0$ & AII & ($U_+$, $U_-$) & ($A_+$, $A_-$) 
& Spinful TRS TCI with order-two NSG & ``$4\Z$''
\\
$4$ & $1$ & AII & ($\widebar U_-$, $\widebar U_+$) & ($\widebar
		 A_-$, $\widebar A_+$) 
& Spinful TRS TCI with order-two NSG antisymmetry & $\textcolor{blue}{\Z_2}$
\\
\hline
$5$ & $0$ & CII & ($U_{++}$, $U_{--}$, $\widebar U_{++}$,
	     $\widebar U_{--}$) & ($A_{++}$, $A_{--}$, $\widebar
		 A_{++}$, $\widebar A_{--}$) 
& & $2\Z$
\\
$5$ & $1$ & CII & ($U_{+-}$, $U_{-+}$, $\widebar U_{-+}$,
	     $\widebar U_{+-}$) & ($A_{+-}$, $A_{-+}$, $\widebar
		 A_{-+}$, $\widebar A_{+-}$) 
& & 0
\\
\hline
$6$ & $0$ & C & ($U_+$, $U_-$) &($\widebar A_+$, $\widebar A_-$)
& $SU(2)$ symmetric TCSC with order-two NSG & 0
\\
$6$ & $1$ & C & ($\widebar U_+$, $\widebar U_-$) &($A_+$, $A_-$)  
& $SU(2)$ symmetric TCSC with magnetic order-two NSG& 0
\\
\hline
$7$ & $0$ & CI &($U_{++}$, $U_{--}$, $\widebar U_{++}$, $\widebar
	     U_{--}$)&($A_{++}$, $A_{--}$, $\widebar A_{++}$,
		 $\widebar A_{--}$)
& $SU(2)$ symmetric TRS TCSC with $[U, TC]=0$ & 0
\\
$7$ & $1$ & CI & ($U_{-+}$, $U_{+-}$, $\widebar U_{-+}$, $\widebar
	     U_{+-}$)&($A_{-+}$, $A_{+-}$, $\widebar A_{-+}$,
		 $\widebar A_{+-}$)
& $SU(2)$ symmetric TRS TCSC with $\{U, TC\}=0$ & 0 
\\
\hline \hline
\end{tabular}
\label{Symmetry_type}
\end{center}
\end{table*}

In a manner similar to the previous cases, 
we denote the $K$-group for TCIs and TCSCs with the data
$(s,t;d,d_{\parallel})$ by $K_{\R}^{(s,t)}(\widetilde S^1 \times
T^{d-d_{\parallel}-1} \times \widebar T^{d_{\parallel}})$. 
There are isomorphisms 
\begin{align}
&K_{\R}^{(s,t)}(\widetilde S^1 \times T^{d-d_{\parallel}-1} \times \widebar T^{d_{\parallel}}) \cong 
K_{\R}^{(s,t)}(\widetilde S^1 \times T^{d-d_{\parallel}-2} \times \widebar T^{d_{\parallel}}) \oplus 
K_{\R}^{(s-1,t)}(\widetilde S^1 \times T^{d-d_{\parallel}-2} \times
 \widebar T^{d_{\parallel}}), 
\label{eq:Kreal1}\\
&K_{\R}^{(s,t)}(\widetilde S^1 \times T^{d-d_{\parallel}-1} \times \widebar T^{d_{\parallel}}) \cong 
K_{\R}^{(s,t)}(\widetilde S^1 \times T^{d-d_{\parallel}-1} \times \widebar T^{d_{\parallel}-1}) \oplus 
K_{\R}^{(s-1,t-1)}(\widetilde S^1 \times T^{d-d_{\parallel}-1} \times \widebar T^{d_{\parallel}-1}),
\label{eq:Kreal2}
\end{align}
where the first terms of the right hand side give weak
indices, and only the second ones contain strong indices.
Here $s$ and $t$ in the superscript $(s,t)$ are defined modulo 8 and 2,
respectively. 
Using the isomorphisms, we can perform the dimensional reduction except
for $\widetilde{S}^1$, and thus
$K_{\R}^{(s,t)}(\widetilde S^1 \times
T^{d-d_{\parallel}-1} \times \widebar T^{d_{\parallel}})$ is given
by a set of the $K$-groups $\{K_{\R}^{(s,t)}(\widetilde
S^1)\}_{s=1,\dots,8, t=0,1}$.
For the strong topological index in $d$-dimensions, we have
\begin{eqnarray}
K_{\R}^{(s,t)}(\widetilde S^1 \times
T^{d-d_{\parallel}-1} \times \widebar T^{d_{\parallel}})|_{\rm strong}
=K_{\R}^{(s-d+1,t-d_{\parallel})}(\widetilde S^1)|_{\rm strong}.
\label{eq:strongKreal}
\end{eqnarray}
We compute $K_{\R}^{(s,t)}(\widetilde S^1)$ in Appendix
\ref{Appendix:F}, and summarize the results in
Table \ref{Symmetry_type}. 


\section{Topological periodic table}
\label{sec:topological_table}

\subsection{Periodicity of strong topological index in space and flipped
  dimensions}
Like conventional
topological insulators and superconductors,\cite{Kitaev2009}  
the strong indices in Eqs. (\ref{eq:strongKUcomplex}),
(\ref{eq:strongKAcomplex}) and (\ref{eq:strongKreal}) are periodic
in space dimension $d$, because of the Bott periodicity in the index $s$
of AZ classes.
For TCIs and TCSCs with order-two NSGs, the strong indices are also
periodic in the flipped dimension $d_{\parallel}$:
In a manner similar to
the order-two point group symmetry, the periodicity in the flipped
dimension $d_{\parallel}$ is due to the periodicity in
the subclass index $t$,
but in the NSG case, the period reduces to the half. (This is because
the topological classification is independent of $\epsilon_{\sigma}$ in
Eqs. (\ref{eq:nonsymmorphic_unitary}) and (\ref{eq:nonsymmorphic_magnetic}))
As a result, the presence of order-two NSGs gives two different 
families of topological phases: (i) $d_{\parallel}=0$ family:  
The additional symmetry in this family includes nonsymmorphic ${\bm Z}_2
$ symmetry ($d_{\parallel}=0$), two-fold screw ($d_{\parallel}=2$) and
their magnetic symmetries.
(ii) $d_{\parallel}=1$ family: TCIs and TCSCs with glide or magnetic
glide symmetry belong
to this family.
We show the classification tables of strong indices for the $d_{\parallel}=0$ 
and the $d_{\parallel}=1$ families in Tables
\ref{Tab:Periodic_Table_Unitary_d_para=0} and
\ref{Tab:Periodic_Table_Unitary_d_para=1}, respectively.  

\begin{table*}[!]
\begin{center}
\caption{(Color online) Classification table for TCIs and TCSCs in the presence of
 additional (magnetic) order-two NSG symmetry with $d_{\parallel}=0$ (mod 2).
In the first column, ``$U$'' and ``$A$'' mean unitary and anti-unitary (i.e. magnetic) symmetries, respectively. 
$\bar{U}$ and $\bar{A}$ are their ``antisymmetry'' which anticommute with the Hamiltonian. 
$T,C$ and $\Gamma$ represent the time-reversal, particle-hole and chiral (sublattice) transformations, respectively.
Only the strong topological indices are presented. 
$\textcolor{blue}{\Z_2}$ (blue) and $\textcolor{blue}{\Z_4}$ (blue) are emergent
 topological phases by the additional NSG symmetry. 
$1 \in \textcolor{blue}{\Z_2}$ and $2 \in \textcolor{blue}{\Z_4}$ phases show detached surface states. 
}
\begin{tabular}[t]{cccccccccc}
\hline 
Symmetry & AZ class & $d=1$ & $d=2$ & $d=3$ & $d=4$ & $d=5$ & $d=6$ & $d=7$ & $d=8$ \\
\hline 
$U$ &A& $0$ & $\Z$ & $0$ & $\Z$ & $0$ & $\Z$ & $0$ & $\Z$ \\
$[U,\Gamma]=0$ or $[\widebar{U}, \Gamma]=0$&AIII& $\Z$ & $0$ & $\Z$ & $0$ & $\Z$ & $0$ & $\Z$ & $0$ \\
\hline
$\widebar{U}$ &A& $\textcolor{blue}{\Z_2}$ & $0$ & $\textcolor{blue}{\Z_2}$ & $0$ & $\textcolor{blue}{\Z_2}$ & $0$ & $\textcolor{blue}{\Z_2}$ & $0$ \\
$\{U,\Gamma\}=0$ or $\{\widebar{U},\Gamma\}=0$&AIII& $0$ & $\textcolor{blue}{\Z_2}$ & $0$ & $\textcolor{blue}{\Z_2}$ & $0$ & $\textcolor{blue}{\Z_2}$ & $0$ & $\textcolor{blue}{\Z_2}$ \\
\hline
$A$ &A& $0$ & $0$ &$\textcolor{blue}{\Z_2}$ & $\Z$ & $0$ & $0$ &$\textcolor{blue}{\Z_2}$ & $\Z$ \\
$[A,\Gamma]=0$ with $\Gamma^2=1$ &AIII& $\Z$ & $0$ & $0$ &$\textcolor{blue}{\Z_2}$ & $\Z$ & $0$ & $0$ &$\textcolor{blue}{\Z_2}$ \\
$\widebar{A}$& A & $\textcolor{blue}{\Z_2}$ & $\Z$ & $0$ & $0$ &$\textcolor{blue}{\Z_2}$ & $\Z$ & $0$ & $0$ \\
$\{A,\Gamma\}=0$ with $\Gamma^2=1$ &AIII& $0$ & $\textcolor{blue}{\Z_2}$ & $\Z$ & $0$ & $0$ &$\textcolor{blue}{\Z_2}$ & $\Z$ & $0$ \\
\hline 
&AI& $0$ & $0$ &$0$ & $2\Z$ & $0$ & $\Z_2$ & $\Z_2\oplus\textcolor{blue}{\Z_2}$ & $\Z\oplus\textcolor{blue}{\Z_2}$ \\
$U$ or $A$ (for AI, AII) 
&BDI& $\Z\oplus\textcolor{blue}{\Z_2}$ & $0$ & $0$ & $0$ & $2\Z$ & $0$ & $\Z_2$ & $\Z_2\oplus\textcolor{blue}{\Z_2}$ \\
&D& $\Z_2\oplus\textcolor{blue}{\Z_2}$ & $\Z\oplus\textcolor{blue}{\Z_2}$ & $0$ & $0$ & $0$ & $2\Z$ & $0$ & $\Z_2$ \\
$U$ or $\widebar{A}$ (for D, C) 
&DIII& $\Z_2$ & $\Z_2\oplus\textcolor{blue}{\Z_2}$ & $\Z\oplus\textcolor{blue}{\Z_2}$ & $0$ & $0$ & $0$ & $2\Z$ & $0$ \\
&AII& $0$ & $\Z_2$ & $\Z_2\oplus\textcolor{blue}{\Z_2}$ & $\Z\oplus\textcolor{blue}{\Z_2}$ & $0$ & $0$ & $0$ & $2\Z$ \\
$[U,TC]=0$, $[\widebar{U},TC]=0$, 
&CII& $2\Z$ & $0$ & $\Z_2$ & $\Z_2\oplus\textcolor{blue}{\Z_2}$ & $\Z\oplus\textcolor{blue}{\Z_2}$ & $0$ & $0$ & $0$ \\
$[A,TC]=0$ or $[\widebar{A},TC]=0$ 
&C& $0$ & $2\Z$ & $0$ & $\Z_2$ & $\Z_2\oplus\textcolor{blue}{\Z_2}$ & $\Z\oplus\textcolor{blue}{\Z_2}$ & $0$ & $0$ \\
(for BDI, DIII, CII, CI)
&CI& $0$ & $0$ & $2\Z$ & $0$ & $\Z_2$ & $\Z_2\oplus\textcolor{blue}{\Z_2}$ & $\Z\oplus\textcolor{blue}{\Z_2}$ & $0$ \\
\hline 
&AI& $\textcolor{blue}{\Z_2}$ & $0$ &$0$ & $0$ & $\textcolor{blue}{\Z_2}$ & $\Z_2$ & $\textcolor{blue}{\Z_4}$ & $\textcolor{blue}{\Z_2}$ \\
$\widebar{U}$ or $\widebar{A}$ (for AI, AII) 
&BDI& $\textcolor{blue}{\Z_2}$ & $\textcolor{blue}{\Z_2}$ & $0$ & $0$ & $0$ & $\textcolor{blue}{\Z_2}$ & $\Z_2$ & $\textcolor{blue}{\Z_4}$\\
&D& $\textcolor{blue}{\Z_4}$ & $\textcolor{blue}{\Z_2}$ & $\textcolor{blue}{\Z_2}$ & $0$ & $0$ & $0$ & $\textcolor{blue}{\Z_2}$ & $\Z_2$ \\
$\widebar{U}$ or $A$ (for D, C) 
&DIII& $\Z_2$ & $\textcolor{blue}{\Z_4}$ & $\textcolor{blue}{\Z_2}$ & $\textcolor{blue}{\Z_2}$ & $0$ & $0$ & $0$ & $\textcolor{blue}{\Z_2}$ \\
&AII& $\textcolor{blue}{\Z_2}$ & $\Z_2$ & $\textcolor{blue}{\Z_4}$ & $\textcolor{blue}{\Z_2}$ & $\textcolor{blue}{\Z_2}$ & $0$ & $0$ & $0$ \\
$\{U,TC\}=0$, $\{\widebar{U},TC\}=0$, 
&CII& $0$ & $\textcolor{blue}{\Z_2}$ & $\Z_2$ & $\textcolor{blue}{\Z_4}$ & $\textcolor{blue}{\Z_2}$ & $\textcolor{blue}{\Z_2}$ & $0$ & $0$ \\
$\{A,TC\}=0$ or $\{\widebar{A},TC\}=0$ 
&C& $0$ & $0$ & $\textcolor{blue}{\Z_2}$ & $\Z_2$ & $\textcolor{blue}{\Z_4}$ & $\textcolor{blue}{\Z_2}$ & $\textcolor{blue}{\Z_2}$ & $0$ \\
(for BDI, DIII, CII, CI)
&CI& $0$ & $0$ & $0$ & $\textcolor{blue}{\Z_2}$ & $\Z_2$ & $\textcolor{blue}{\Z_4}$ & $\textcolor{blue}{\Z_2}$ & $\textcolor{blue}{\Z_2}$ \\
\hline 
\end{tabular}
\label{Tab:Periodic_Table_Unitary_d_para=0}
\end{center}
\end{table*}

\begin{table*}[!]
\begin{center}
\caption{(Color online)
Classification table for TCIs and TCSCs in the presence of
 additional (magnetic) order-two NSG symmetry with $d_{\parallel}=1$
 (mod 2). 
In the first column, ``$U$'' and ``$A$'' mean unitary and antiunitary (i.e. magnetic) symmetries, respectively. 
$\bar{U}$ and $\bar{A}$ are their ``antisymmetry'' which anticommute
 with the Hamiltonian.  
$T,C$ and $\Gamma$ represent the time-reversal, particle-hole and chiral (sublattice) transformations, respectively. 
Only the strong topological indices are presented.
$\textcolor{blue}{\Z_2}$ (blue) and $\textcolor{blue}{\Z_4}$ (blue) are emergent
 topological phases by the additional NSG symmetry.  
$1 \in \textcolor{blue}{\Z_2}$ and $2 \in \textcolor{blue}{\Z_4}$ phases show detached surface states. 
}
\begin{tabular}[t]{cccccccccc}
\hline 
Symmetry & AZ class & $d=1$ & $d=2$ & $d=3$ & $d=4$ & $d=5$ & $d=6$ & $d=7$ & $d=8$ \\
\hline 
$U$ &A& $\textcolor{blue}{\Z_2}$ & $0$ & $\textcolor{blue}{\Z_2}$ & $0$ & $\textcolor{blue}{\Z_2}$ & $0$ & $\textcolor{blue}{\Z_2}$ & $0$ \\
$[U,\Gamma]=0$ or $[\widebar{U}, \Gamma]=0$ &AIII& $0$ & $\textcolor{blue}{\Z_2}$ & $0$ & $\textcolor{blue}{\Z_2}$ & $0$ & $\textcolor{blue}{\Z_2}$ & $0$ & $\textcolor{blue}{\Z_2}$ \\
\hline
$\widebar{U}$ &A& $0$ & $\Z$ & $0$ & $\Z$ & $0$ & $\Z$ & $0$ & $\Z$ \\
$\{U,\Gamma\}=0$ or $\{\widebar{U},\Gamma\}=0$&AIII& $\Z$ & $0$ & $\Z$ & $0$ & $\Z$ & $0$ & $\Z$ & $0$ \\
\hline
$A$& A & $\textcolor{blue}{\Z_2}$ & $\Z$ & $0$ & $0$ &$\textcolor{blue}{\Z_2}$ & $\Z$ & $0$ & $0$ \\
$[A,\Gamma]=0$ with $\Gamma^2=1$ &AIII& $0$ & $\textcolor{blue}{\Z_2}$ & $\Z$ & $0$ & $0$ &$\textcolor{blue}{\Z_2}$ & $\Z$ & $0$ \\
$\widebar{A}$ &A& $0$ & $0$ &$\textcolor{blue}{\Z_2}$ & $\Z$ & $0$ & $0$ &$\textcolor{blue}{\Z_2}$ & $\Z$ \\
$\{A,\Gamma\}=0$ with $\Gamma^2=1$ &AIII& $\Z$ & $0$ & $0$ &$\textcolor{blue}{\Z_2}$ & $\Z$ & $0$ & $0$ &$\textcolor{blue}{\Z_2}$ \\
\hline 
&AI& $\textcolor{blue}{\Z_2}$ & $0$ &$0$ & $0$ & $\textcolor{blue}{\Z_2}$ & $\Z_2$ & $\textcolor{blue}{\Z_4}$ & $\textcolor{blue}{\Z_2}$ \\
$U$ or $A$ (for AI, AII) 
&BDI& $\textcolor{blue}{\Z_2}$ & $\textcolor{blue}{\Z_2}$ & $0$ & $0$ & $0$ & $\textcolor{blue}{\Z_2}$ & $\Z_2$ & $\textcolor{blue}{\Z_4}$\\
&D& $\textcolor{blue}{\Z_4}$ & $\textcolor{blue}{\Z_2}$ & $\textcolor{blue}{\Z_2}$ & $0$ & $0$ & $0$ & $\textcolor{blue}{\Z_2}$ & $\Z_2$ \\
$U$ or $\widebar{A}$ (for D, C) 
&DIII& $\Z_2$ & $\textcolor{blue}{\Z_4}$ & $\textcolor{blue}{\Z_2}$ & $\textcolor{blue}{\Z_2}$ & $0$ & $0$ & $0$ & $\textcolor{blue}{\Z_2}$ \\
&AII& $\textcolor{blue}{\Z_2}$ & $\Z_2$ & $\textcolor{blue}{\Z_4}$ & $\textcolor{blue}{\Z_2}$ & $\textcolor{blue}{\Z_2}$ & $0$ & $0$ & $0$ \\
$[U,TC]=0$, $[\widebar{U},TC]=0$,
&CII& $0$ & $\textcolor{blue}{\Z_2}$ & $\Z_2$ & $\textcolor{blue}{\Z_4}$ & $\textcolor{blue}{\Z_2}$ & $\textcolor{blue}{\Z_2}$ & $0$ & $0$ \\
$[A,TC]=0$ or $[\widebar{A},TC]=0$
&C& $0$ & $0$ & $\textcolor{blue}{\Z_2}$ & $\Z_2$ & $\textcolor{blue}{\Z_4}$ & $\textcolor{blue}{\Z_2}$ & $\textcolor{blue}{\Z_2}$ & $0$ \\
(for BDI, DIII, CII, CI)
&CI& $0$ & $0$ & $0$ & $\textcolor{blue}{\Z_2}$ & $\Z_2$ & $\textcolor{blue}{\Z_4}$ & $\textcolor{blue}{\Z_2}$ & $\textcolor{blue}{\Z_2}$ \\
\hline 
&AI& $0$ & $0$ &$0$ & $2\Z$ & $0$ & $\Z_2$ & $\Z_2\oplus\textcolor{blue}{\Z_2}$ & $\Z\oplus\textcolor{blue}{\Z_2}$ \\
$\widebar{U}$ or $\widebar{A}$ (for AI, AII) 
&BDI& $\Z\oplus\textcolor{blue}{\Z_2}$ & $0$ & $0$ & $0$ & $2\Z$ & $0$ & $\Z_2$ & $\Z_2\oplus\textcolor{blue}{\Z_2}$ \\
&D& $\Z_2\oplus\textcolor{blue}{\Z_2}$ & $\Z\oplus\textcolor{blue}{\Z_2}$ & $0$ & $0$ & $0$ & $2\Z$ & $0$ & $\Z_2$ \\
$\widebar{U}$ or $A$ (for D, C) 
&DIII& $\Z_2$ & $\Z_2\oplus\textcolor{blue}{\Z_2}$ & $\Z\oplus\textcolor{blue}{\Z_2}$ & $0$ & $0$ & $0$ & $2\Z$ & $0$ \\
&AII& $0$ & $\Z_2$ & $\Z_2\oplus\textcolor{blue}{\Z_2}$ & $\Z\oplus\textcolor{blue}{\Z_2}$ & $0$ & $0$ & $0$ & $2\Z$ \\
$\{U,TC\}=0$, $\{\widebar{U},TC\}=0$, 
&CII& $2\Z$ & $0$ & $\Z_2$ & $\Z_2\oplus\textcolor{blue}{\Z_2}$ & $\Z\oplus\textcolor{blue}{\Z_2}$ & $0$ & $0$ & $0$ \\
$\{A,TC\}=0$ or $\{\widebar{A},TC\}=0$ 
&C& $0$ & $2\Z$ & $0$ & $\Z_2$ & $\Z_2\oplus\textcolor{blue}{\Z_2}$ & $\Z\oplus\textcolor{blue}{\Z_2}$ & $0$ & $0$ \\
(for BDI, DIII, CII, CI)
&CI& $0$ & $0$ & $2\Z$ & $0$ & $\Z_2$ & $\Z_2\oplus\textcolor{blue}{\Z_2}$ & $\Z\oplus\textcolor{blue}{\Z_2}$ & $0$ \\
\hline 
\end{tabular}
\label{Tab:Periodic_Table_Unitary_d_para=1}
\end{center}
\end{table*}

\subsection{$\Z_2$ and $\Z_4$ nonsymmorphic topological phases}

The topological periodic tables in Tables
\ref{Tab:Periodic_Table_Unitary_d_para=0} 
and \ref{Tab:Periodic_Table_Unitary_d_para=1} exhibit novel topological
phases specific to nonsymmorphic TCIs and TCSCs. 
The first one is $\Z_2$ topological phases protected by unitary
NSGs in complex AZ classes (i.e. class A and AIII):
While any $\Z_2$
phase of conventional topological insulators and superconductors requires an antiunitary symmetry such as time-reversal and/or
particle-hole symmetries, the $\Z_2$ phases protected by NSGs do not need any antiunitary symmetry.
Furthermore, the coexistence of NSGs with time-reversal
and/or particle-hole symmetries provide $\Z_4$ topological phases,
which have not been realized
in ordinary topological insulators and superconductor.
These $\Z_2$ and $\Z_4$ nonsymmorphic topological phases realize unique
surface states with a M\"{o}bius twist structure.
We will present concrete models of these exotic topological phases in
Sec.\ref{sec:topological_surface_state}.

\section{Topological boundary state protected by nonsymmorphic space group}
\label{sec:topological_surface_state}

\subsection{$K$-group for boundary gapless state}
\label{Sec:BulkBoundary}
\begin{figure}[!]
 \begin{center}
  \includegraphics[width=0.5\linewidth, trim=0cm 0cm 0cm 0cm]{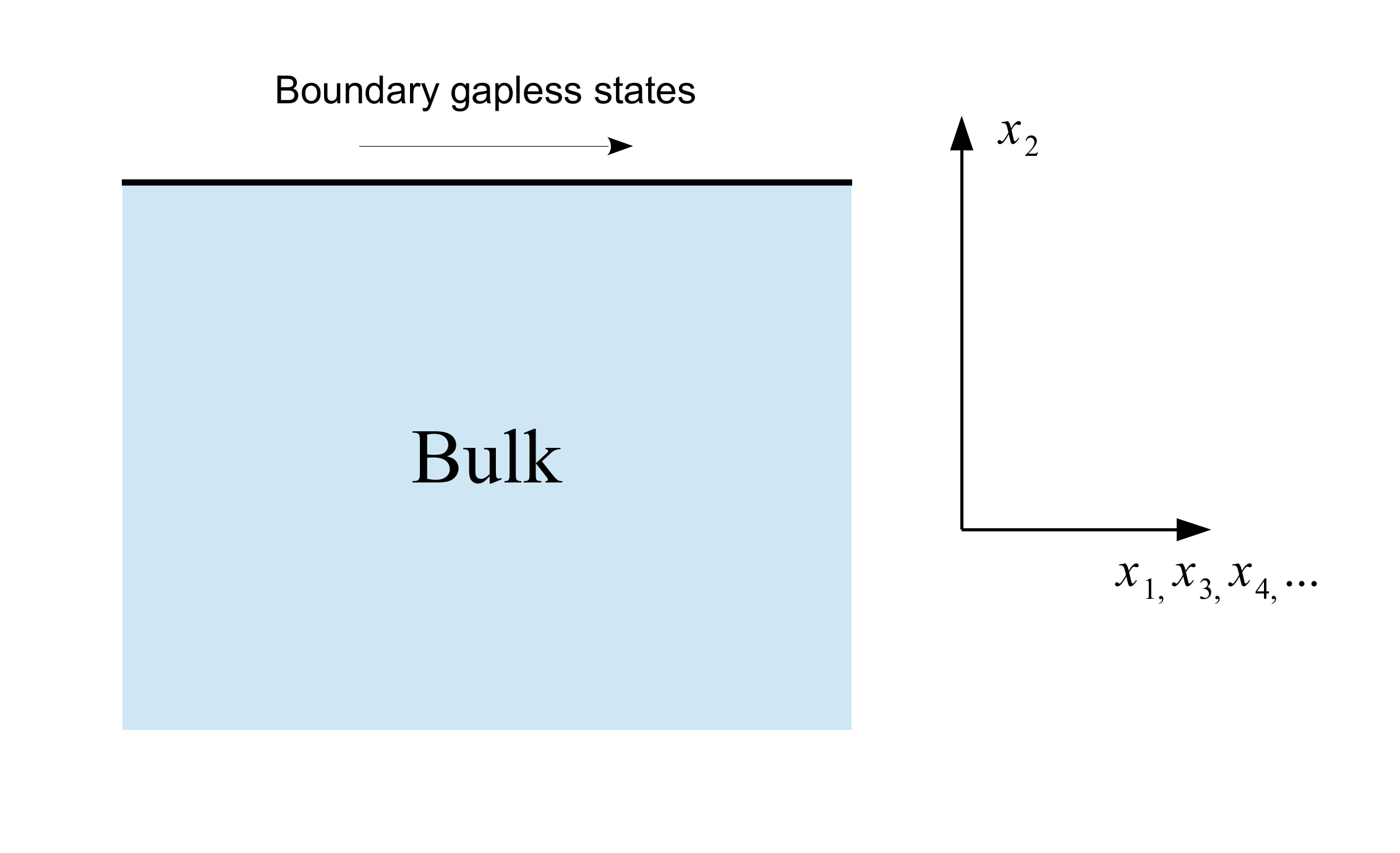}
 \end{center}
 \caption{(Color online) A $(d+1)$-dimensional bulk TCI/TCSC and its $d$-dimensional boundary} 
 \label{Fig:Bulk-Boundary}
\end{figure}

The isomorphisms in Eqs.(\ref{eq:KUcomplex1})-(\ref{eq:KUcomplex2}),
(\ref{eq:KAcomplex1})-(\ref{eq:KAcomplex2}) and
(\ref{eq:Kreal1})-(\ref{eq:Kreal2}), 
and $K(\widetilde{S}^1)$ in Tables \ref{Symmetry_type_UC},
\ref{Symmetry_type_AC} and \ref{Symmetry_type} 
give the complete classification of bulk
TCIs and TCSCs in the presence of an
additional order-two NSG. 
From the bulk-boundary correspondence, these nonsymmorphic TCIs and
TCSCs may support topological gapless boundary states
protected by the symmetry.
In this section, we discuss the topological boundary states.

First, it should be noted that the symmetry protection of the boundary
states requires a symmetry preserving boundary:  
Consider a $d$-dimensional TCI or
TCSC that is invariant under the NSG of
Eq.(\ref{eq:nonsymmorphic_coordinate}). 
To preserve the NSG symmetry, the boundary should
be parallel to both the $x_1$-axis and the hyperplane spanned by
$(x_{d-d_{\parallel}+1},
\dots, x_d)$.
Therefore, the condition $d-d_{\parallel}+1\ge 3$ should be met to have
a boundary preserving the NSG symmetry.
As such a boundary, we consider a surface normal to the $x_2$-axis,
without loss of generality.
See Fig.\ref{Fig:Bulk-Boundary}.
We also note that not all elements of the $K$-groups are relevant to the 
existence of gapless states on the surface.
In general, if a $K$-group is calculated from isomorphisms in the above, 
it contains weak indices as well as strong ones.   
Then, not all the weak indices predict the surface state.
To
see this, consider again the isomorphism of Eqs.(\ref{eq:KUcomplex1}),
(\ref{eq:KAcomplex1}) and (\ref{eq:Kreal1}),
\begin{align}
&K_{\C}^{(s,t)}(\widetilde S^1 \times T^{d-d_{\parallel}-1} \times \widebar T^{d_{\parallel}}) \cong 
K_{\C}^{(s,t)}(\widetilde S^1 \times T^{d-d_{\parallel}-2} \times \widebar T^{d_{\parallel}}) \oplus 
K_{\C}^{(s-1,t)}(\widetilde S^1 \times T^{d-d_{\parallel}-2} \times \widebar T^{d_{\parallel}}), \label{Eq:IsoCU} \\
&K_{\C}^{t}(\widetilde S^1 \times T^{d-d_{\parallel}-1} \times \widebar T^{d_{\parallel}}) \cong 
K_{\C}^{t}(\widetilde S^1 \times T^{d-d_{\parallel}-2} \times \widebar T^{d_{\parallel}}) \oplus 
K_{\C}^{t-1}(\widetilde S^1 \times T^{d-d_{\parallel}-2} \times \widebar T^{d_{\parallel}}), \label{Eq:IsoCAU} \\
&K_{\R}^{(s,t)}(\widetilde S^1 \times T^{d-d_{\parallel}-1} \times \widebar T^{d_{\parallel}}) \cong 
K_{\R}^{(s,t)}(\widetilde S^1 \times T^{d-d_{\parallel}-2} \times \widebar T^{d_{\parallel}}) \oplus 
K_{\R}^{(s-1,t)}(\widetilde S^1 \times T^{d-d_{\parallel}-2} \times \widebar T^{d_{\parallel}}). \label{Eq:IsoR}
\end{align}
As mentioned in the above, the first $K$-groups in the right hand side
provide weak indices that are obtained as topological
indices of stacked $(d-1)$-dimensional systems in the $x_2$-direction. 
These weak indices do not predict any gapless state on the surface normal to the $x_2$-axis. 
%
As a result, only the second $K$-groups are responsible for the
existence of the gapless boundary states.
Taking only the second $K$-groups in the above, we can define $K$-groups
for boundary gapless states in $(d-1)$-dimensions as
\begin{align}
&K_{\C,{\rm BGS}}^{(s,t)}(\widetilde S^1 \times T^{d-d_{\parallel}-2} \times \widebar T^{d_{\parallel}}) \cong 
K_{\C}^{(s-1,t)}(\widetilde S^1 \times T^{d-d_{\parallel}-2} \times \widebar T^{d_{\parallel}}), \label{Eq:BoundaryIndexCU} \\
&K_{\C,{\rm BGS}}^{t}(\widetilde S^1 \times T^{d-d_{\parallel}-2} \times \widebar T^{d_{\parallel}}) \cong 
K_{\C}^{t-1}(\widetilde S^1 \times T^{d-d_{\parallel}-2} \times \widebar T^{d_{\parallel}}), \label{Eq:BoundaryIndexCAU}\\
&K_{\R,{\rm BGS}}^{(s,t)}(\widetilde S^1 \times T^{d-d_{\parallel}-2} \times \widebar T^{d_{\parallel}}) \cong 
K_{\R}^{(s-1,t)}(\widetilde S^1 \times T^{d-d_{\parallel}-2} \times \widebar T^{d_{\parallel}}), \label{Eq:BoundaryIndexR}
\end{align}
where we represent the boundary $K$-groups by the subscript ``BGS''.
It should be noted here that the boundary $K$-groups 
contain weak indices responsible for gapless surface states, as well as
strong indices. 
When the topological numbers of the boundary $K$-groups are nonzero, there
exist corresponding gapless surface states.

In a manner similar to the bulk $K$-groups, using the isomorphisms of Eqs.(\ref{Eq:IsoCU}), (\ref{Eq:IsoCAU}), and
(\ref{Eq:IsoR}), and $K(\widetilde{S}^1)$ in Tables \ref{Symmetry_type_UC},
\ref{Symmetry_type_AC} and \ref{Symmetry_type},  
one can evaluate the boundary $K$-groups. 
In particular, the strong indices of the boundary
$K$-groups coincide with those
of the bulk $K$-groups.  
Therefore, the topological periodic tables in Tables
\ref{Tab:Periodic_Table_Unitary_d_para=0}  and
\ref{Tab:Periodic_Table_Unitary_d_para=1} also give 
the strong indices of the boundary $K$-groups for $(d-1)$-dimensional
gapless states.

In real materials with $d\le 3$, only
$d_{\parallel}=0,1$ can satisfy the condition $d-d_{\parallel}+1\ge 3$
of the symmetry preserving boundary. 
Below, we discuss the surface states protected by these nonsymmorphic
symmetries.

\subsection{Surface state protected by nonsymmorphic global ${\bm Z}_2$ symmetry
  ($d_{\parallel}=0$)}
\label{Sec:half-translation}

\begin{figure}[!]
 \begin{center}
  \includegraphics[width=0.5\linewidth, trim=0cm 0cm 0cm 0cm]{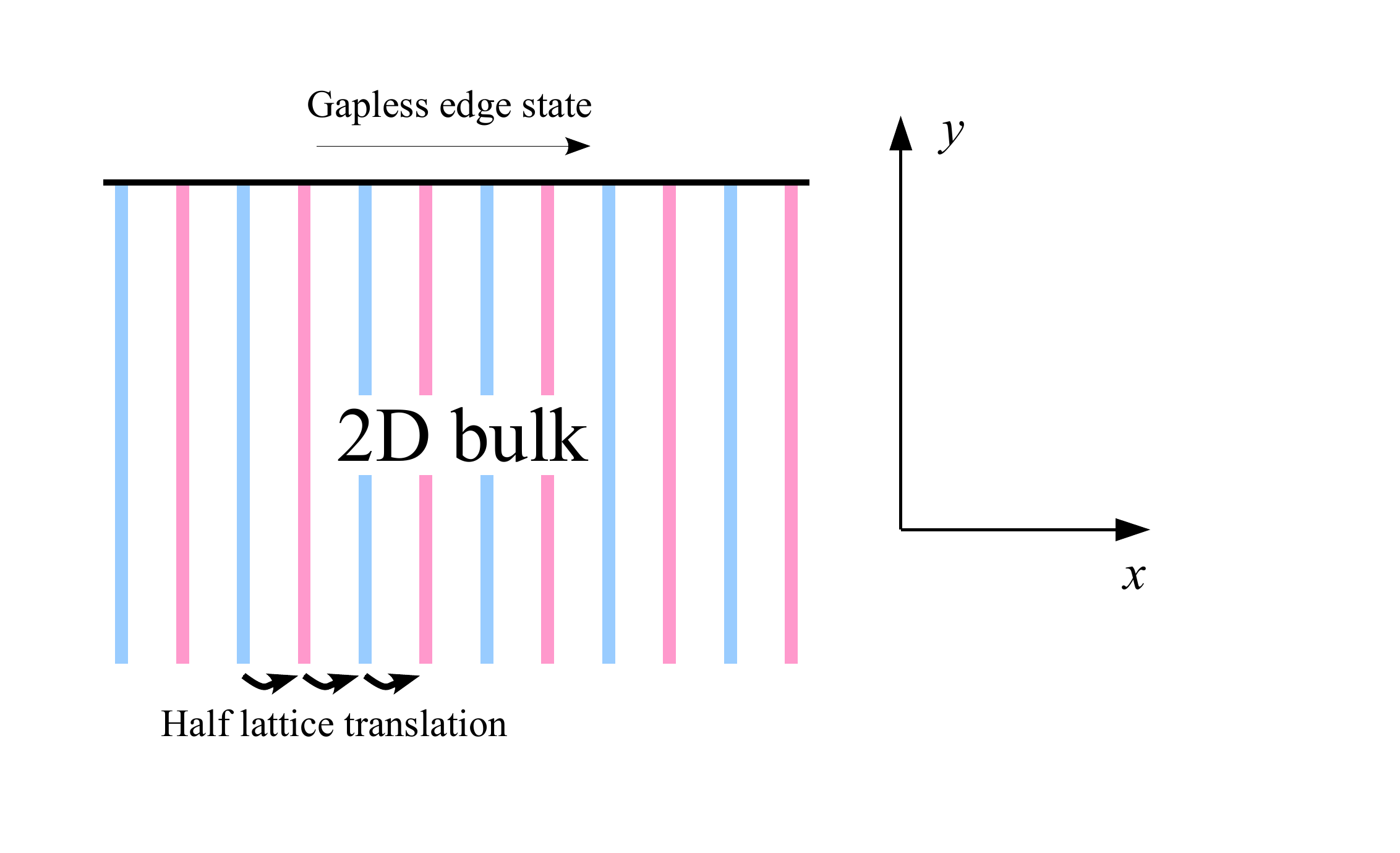}
 \end{center}
 \caption{(Color online) Schematic illustration of order-two nonsymmorphic ${\bm Z}_2$ symmetry in
 two-dimensions. 
The system is invariant under half lattice translation in the
 $x$-direction followed by the exchange of two different (blue and red)
 states. A 
 one-dimensional edge
 parallel to the $x$-direction is compatible with this symmetry. Gapless
 edge states protected by this symmetry appear on this edge when the
 relevant bulk topological number is nonzero.}
 \label{Fig:Bulk-Boundary_2D}
\end{figure}

\begin{figure}[!]
 \begin{center}
  \includegraphics[width=0.5\linewidth, trim=0cm 0cm 0cm 0cm]{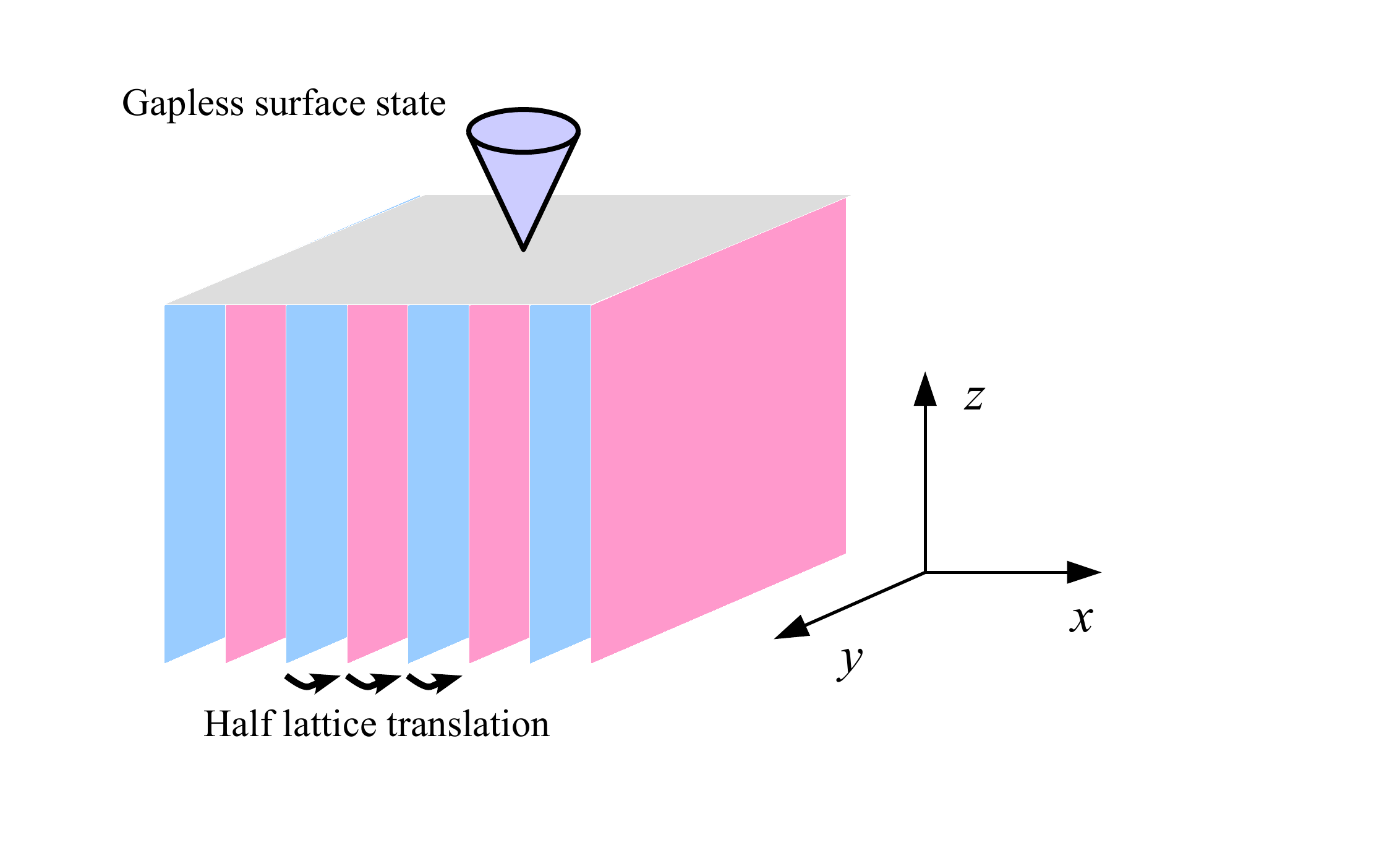}
 \end{center}
 \caption{(Color online) An order-two nonsymmorphic ${\bm Z}_2$ symmetry
 in three-dimensions. The system is invariant under half lattice
 translation in the
 $x$-direction followed by the exchange of blue and red planes. A
 two-dimensional surface parallel to the $x$-direction is compatible
 with this symmetry.
Gapless
 surface states protected by this symmetry appear on this surface when the
 relevant bulk topological number is nonzero.}
\label{Fig:Bulk-Boundary_3D_global}
\end{figure}

In this section, we discuss topological surface states protected by
$d_{\parallel}=0$ NSGs.
For $d\le 3$, two ($d=2$) and three $(d=3)$ dimensional systems have boundaries
compatible with $d_{\parallel}=0$ NSGs. 
Relevant NSGs are
nonsymmorphic ${\bm Z}_2$ symmetry, and its
anti-unitary magnetic version. 
We illustrate nonsymmorphic ${\bm Z}_2$ symmetry with half translation $x \to x+1/2$ 
in two and three dimensions in Figs.\ \ref{Fig:Bulk-Boundary_2D} and \ref{Fig:Bulk-Boundary_3D_global}, respectively. 
A boundary parallel to the $x$-direction is compatible with the half
translation, so gapless states protected by this symmetry appear on
the $(d-1)$-dimensional boundary when the relevant topological number is
nonzero.
The strong topological indices for boundary states are given those in
$d=2$ and $d=3$ of Table 
\ref{Tab:Periodic_Table_Unitary_d_para=0}.

\subsubsection{$\Z_2$ quantum crystalline spin Hall insulator with
   nonsymmorphic ${\bm Z}_2$ symmetry ($d=2$, $U$ in class AII)} 

Consider a two-dimensional time-reversal invariant insulator with nonsymmorphic
${\bm Z}_2$ symmetry,  
\begin{align}
&T H(\bk) T^{-1} = H(-\bk), && T = i s_y K, \\
&G(k_x) H(\bk) G(k_x)^{-1} = H(\bk), && G(k_x) = \begin{pmatrix}
0 & i s_z e^{-i k_x} \\
i s_z & 0 \\
\end{pmatrix}, \\
&G(k_x)^2 = -e^{- i k_x}, && T G(k_x) = G(-k_x) T. 
\end{align}
The additional symmetry is $U$ in class AII of Table
\ref{Tab:Periodic_Table_Unitary_d_para=0}, and thus the topological index
is $\Z_2$.  
The $\Z_2$ topological invariant is nothing but the Kane-Mele's $\Z_2$
invariant\cite{Kane2005}, which   
implies that the nonsymmorphic ${\bm Z}_2$ symmetry is compatible
with the $\Z_2$ classification of quantum spin Hall states in two dimensions.
Here it should be noted that if the ${\bm Z}_2$ symmetry is symmorphic,
we have a different result.
For instance, if we consider the following symmorphic ${\bm Z}_2$ global
symmetry $G'$, instead of $G(k_x)$ 
\begin{align}
G'=
\begin{pmatrix}
0 & is_z\\
is_z & 0
\end{pmatrix}, 
\end{align}
the resultant topological phase becomes a
$\Z$ phase:
Because $G'$ commutes with $H({\bm k})$, 
we can block-diagonalize $H({\bm k})$ into
eigensectors of $G'$ with eigenvalues $\pm i$.   
$T$ interchanges these two sectors, but each sector does not have
its own TRS, so these sectors may have nonzero first Chern numbers (TKNN
integers\cite{Thouless1982, Kohmoto1985}) opposite to
each other.
The TKNN integers define the $\Z$ phase. 
On the other hand, for the nonsymmorphic $G(k_x)$, we cannot define
similar TKNN integers.
In this case, by shifting $k_x$ by $2\pi$, the
eigenvalues $g_{\pm}(k_x)=\pm ie^{-ik_x/2}$ of $G(k_x)$ are exchanged,
so are the eigensectors. 
Therefore, the distinction between the eigensectors is obscured, and
thus the TKNN integers are ill-defined. 



\subsubsection{$\Z\oplus\Z_2$ TCSC with
   nonsymmorphic ${\bm Z}_2$ symmetry ($d=2$, $U$ in class D)}
\label{sec:ZZ_2TCSC}

Next, we consider a two dimensional time-reversal breaking superconductor with
nonsymmorphic ${\bm Z}_2$ symmetry. 
From Table \ref{Tab:Periodic_Table_Unitary_d_para=0}, its topological
phase is $\Z \oplus \Z_2$.  See $U$ in class D. 
Without loss of generality, we assume a spinless superconductor
described by the following BdG Hamiltonian,
\begin{align}
H_{\rm BdG}(\bk) = \begin{pmatrix}
\epsilon(\bk) & \Delta(\bk) \\
\Delta^{\dag}(\bk) & - \epsilon^T(-\bk)
\end{pmatrix}.
\end{align}
For the normal Hamiltonian $\epsilon({\bm k})$, the nonsymmorphic ${\bm Z}_2$
transformation is given by 
\begin{align}
G(k_x) \epsilon(\bk) G(k_x)^{-1}=\epsilon(\bk),
\quad 
G(k_x) = \begin{pmatrix}
0 & e^{-i k_x} \\
1 & 0 \\
\end{pmatrix}.
\end{align}
Here $G(k_x)$ acts on internal degrees of freedom that are exchanged
by the ${\bm Z}_2$ 
transformation.
The system keeps the nonsymmorphic ${\bm Z}_2$ symmetry in the
superconducting state, if the gap function $\Delta({\bm k})$ is even or
odd under $G(k_x)$, 
\begin{align}
G(k_x) \Delta(\bk) G(-k_x)^T = \pm \Delta(\bk). 
\end{align}
%
Actually, defining the nonsymmorphic ${\bm
Z}_2$ symmetry for the BdG Hamiltonian as 
\begin{align}
G^{(\pm)}_{\rm BdG}(k_x) = \begin{pmatrix}
G(k_x) & 0 \\
0 & \pm G(-k_x)^* \\
\end{pmatrix},
\end{align}
we have $G^{(\pm)}_{\rm BdG}(k_x) H_{\rm BdG}({\bm k})G^{(\pm)}_{\rm
BdG}(k_x)^{-1}=H_{\rm BdG}(\bk)$.
The BdG Hamiltonian also has PHS, $CH_{\rm BdG}(\bk)C^{-1}=-H_{\rm
BdG}(-\bk)$ with $C = \tau_x K$ ($\tau_x$ is the Pauli matrix in the
Nambu space.) The PHS $C$ and $G^{(\pm)}_{\rm BdG}(k_x)$ satisfy
$C G^{(\pm)}_{\rm BdG}(k_x) = \pm G^{(\pm)}_{\rm BdG}(-k_x) C$. 


From the above Hamiltonian, we can define three topological invariants. 
The first one is the first Chern number
\begin{align}
ch_1 = \frac{i}{2 \pi} \int_{-\pi}^{\pi}dk_x \int_{-\pi}^{\pi}dk_y 
{\rm tr} {\cal F}(k_x, k_y), 
\end{align}
where ${\cal F}$ is the Berry curvature of
occupied states for the BdG Hamiltonian $H_{\rm
BdG}(\bk)$,\cite{Kohmoto1985} and the
trace is taken for all the occupied states. 
The other two are one-dimensional $\Z_2$ invariants:
To define the $\Z_2$ invariants, we divide $H_{\rm
BdG}(\bk)$ into two eigensectors of $G^{(\pm)}_{\rm BdG}(k_x)$.
On the $k_x=0$ $(k_x=\pi)$ line, the eigenvalues 
of $G^{(+)}(0)$ $(G^{(-)}(\pi))$ 
are $\pm 1$ $(\pm i)$,   
so the relation $CG_{\rm BdG}^{(+)}(0)=G_{\rm BdG}^{(+)}(0)C$ 
($CG_{\rm BdG}^{(-)}(\pi)=-G_{\rm BdG}^{(-)}(\pi)C$) implies that each
eigensector of $G^{(+)}(0)$ $(G^{(-)}(\pi))$ keeps PHS.
%
%
%
%
%
%
Thus, for $G_{\rm BdG}^{(+)}(k_x)$ ($G_{\rm BdG}^{(-)}(k_x)$), 
we can define
$\Z_2$ topological invariants\cite{Qi2008,Sato2010}
\begin{align}
\nu_{\pm} = \frac{i}{\pi} \int_{-\pi}^{\pi} dk_y{\rm tr} {\cal
 A}_{\pm}(k_x^*, k_y) \ \ ({\rm mod\ }2), && k_x^* = 
\left\{ \begin{array}{ll}
0 & {\rm for\ } G^{(+)}_{\rm BdG}(k_x) \\
\pi & {\rm for\ } G^{(-)}_{\rm BdG}(k_x) \\
\end{array}\right. 
\end{align}
where 
${\cal A}_{\pm}$ is the Berry connection for the occupied states in the $G_{\rm
BdG}(0) = \pm 1$ ($G_{\rm BdG}(\pi)=\pm i$) sector.  
A direct calculation shows that the topological invariants $(ch_1,
\nu_+, \nu_-)$ satisfy the constraint $\nu_+ + \nu_- = ch_1$ (mod
$2$), and thus they describe a $\Z\oplus\Z_2$ phase.

\begin{figure}[!]
 \begin{center}
  \includegraphics[width=0.9\linewidth, trim=0cm 0cm 0cm
  0cm]{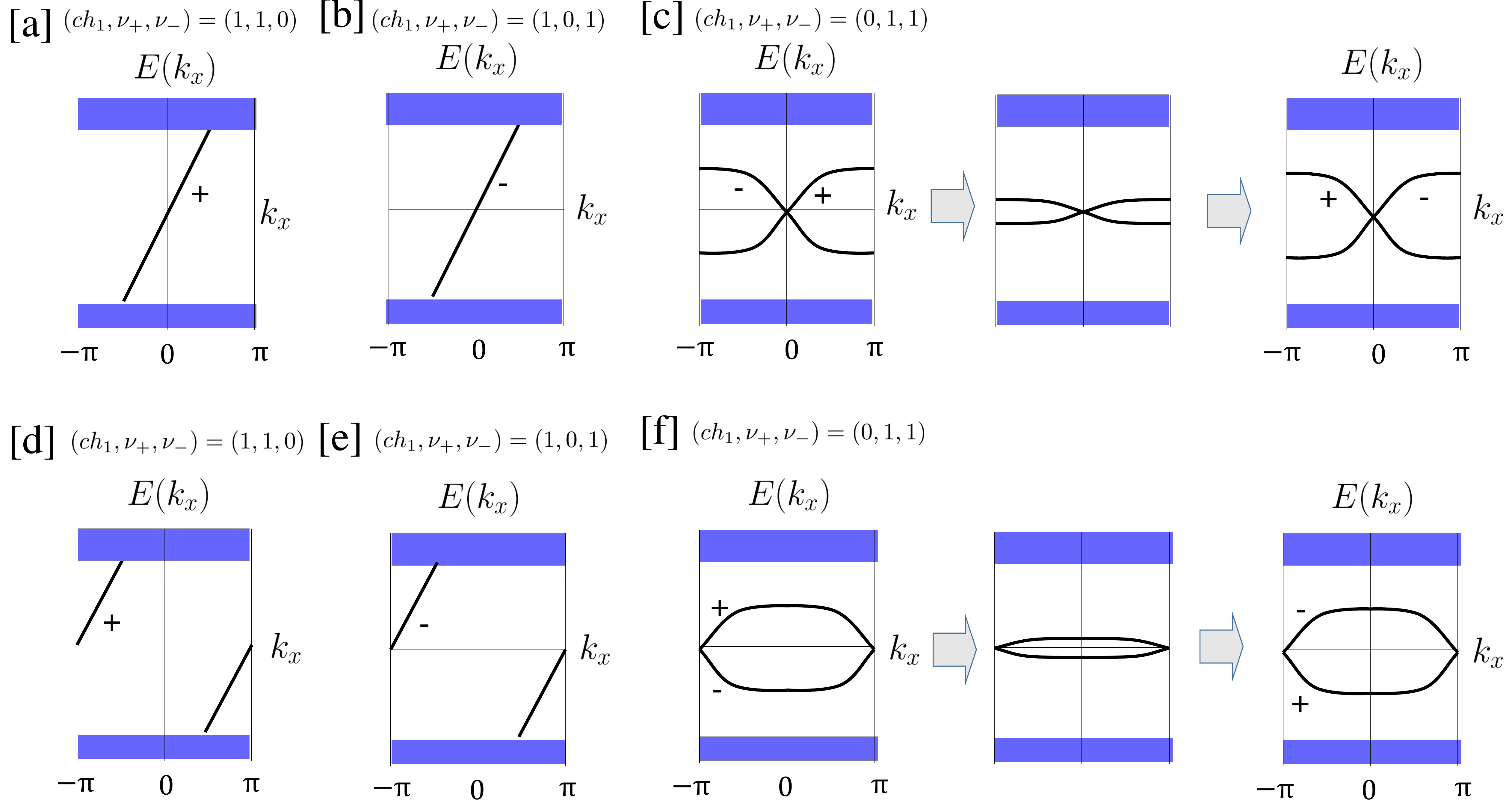} 
 \end{center}
 \caption{(Color online) Gapless edge states in two-dimensional time-reversal
 breaking TCSCs with nonsymmorphic ${\bm Z}_2$ symmetry $G(k_x)$
 ($d_{\parallel}=0$, $d=2$, 
 $U$ in class D).
$G(k_x)^2=e^{-ik_x}$. 
In [a]-[c] ([d]-[f]), the gap function is even (odd) under $G(k_x)$.   
``+" and ``-" in [a]-[c] ([d]-[f]) indicate the eigenvalue $g_{\pm}(k_x)$ of
 $G^{(+)}_{\rm BdG}(k_x)$ ($G^{(-)}_{\rm BdG}(k_x)$) for edge states.  }
 \label{Fig:2D_D_Glide}
\end{figure}



To see more details of the $\Z \oplus \Z_2$ structure, 
we now consider
corresponding edge states.
First, consider a superconductor with an even gap function
under $G(k_x)$, where the nonsymmorphic ${\bm Z}_2$ symmetry is given by
$G_{\rm BdG}^{(+)}(k_x)$. 
When $ch_1=1$, we have a single chiral Majorana edge state, but the
nonsymmorphic ${\bm Z}_2$ symmetry provides a constraint on the spectrum:
From PHS, the spectrum should be $E(k_x)=ck_x$ or
$E(k_x)=c(k_x-\pi)$
with a constant $c$, but the latter is not allowed by $G_{\rm
BdG}^{(+)}(k_x)$.   
Indeed, at $k_x=\pi$, the eigenvalues of $G_{\rm BdG}^{(+)}(k_z)$ are
$\pm i$, so the relation $CG_{\rm BdG}^{(+)}(\pi)=G_{\rm BdG}^{(+)}(\pi)C
$ implies that two eigensectors  of $G_{\rm BdG}^{(+)}(\pi)$ are
exchanged by PHS. In particular, if there is a zero energy state at
$k_x=\pi$, there should be another zero energy state with an opposite
eigenvalue of $G_{\rm BdG}^{(+)}(\pi)$.   
The latter spectrum does not satisfy this constraint, so only the
former is realized. 
Like the bulk modes, the chiral edge state has a
definite eigenvalue of $G_{\rm BdG}^{(+)}(k_x)$, so two different chiral
Majorana modes with different eigenvalues of $G_{\rm BdG}^{(+)}(k_x)$
are possible, as shown in
Figs.\ref{Fig:2D_D_Glide} [a] and [b], respectively.
For the chiral edge state with
the eigenvalue $g_+(k_x)=e^{-ik_x/2}$  ($g_-(k_x)=-e^{-ik_x/2}$), $\nu_+$ ($\nu_-$) should be
nontrivial, i.e. $\nu_+=1$ $(\nu_-=1)$, 
 since the chiral energy
state has a zero
energy state with the eigenvalue +1 (-1) at $k_x=0$.    
Therefore, the topological numbers $(ch_1, \nu_+, \nu_-)$ of 
the edge states $e_+$ and $e_-$ in Figs.~\ref{Fig:2D_D_Glide} [a] and [b]
are given by $e_+ = (1,1,0)$ and $e_- = (1,0,1)$, respectively.
Note that these topological numbers are consistent with the constraint
$\nu_++\nu_-=ch_1$ (mod 2).

The edge states $e_+$ and $e_-$ have the minimal unit of $ch_1$.
They are also independent of each
other. 
Therefore, they are generators of edge states in the $\Z\oplus \Z_2$ phase.    
Any topological gapless state in the present system can be
obtained by combining 
$e_+$ and $e_-$ as $e = n_+ e_+ + n_- e_-$ ($n_{\pm}\in \Z$).  
In particular, 
we can construct a non-chiral mode $e=e_+-e_-$, which only has the
$\Z_2$ numbers $\nu_+=\nu_-=1$, as illustrated in Fig.\ref{Fig:2D_D_Glide}[c].  
Being different from helical edge modes in quantum spin Hall states, this
non-chiral mode does not require TRS, but it is protected by PHS and the
nonsymmorphic ${\bm Z}_2$ symmetry.   
As discussed in Refs.\onlinecite{Shiozaki2015} and \onlinecite{Wang2015}, 
the non-chiral mode can
be detached from the bulk spectrum. See Fig.\ref{Fig:2D_D_Glide}[c].

\begin{figure}[!]
 \begin{center}
  \includegraphics[width=0.7\linewidth, trim=0cm 0cm 0cm 0cm]{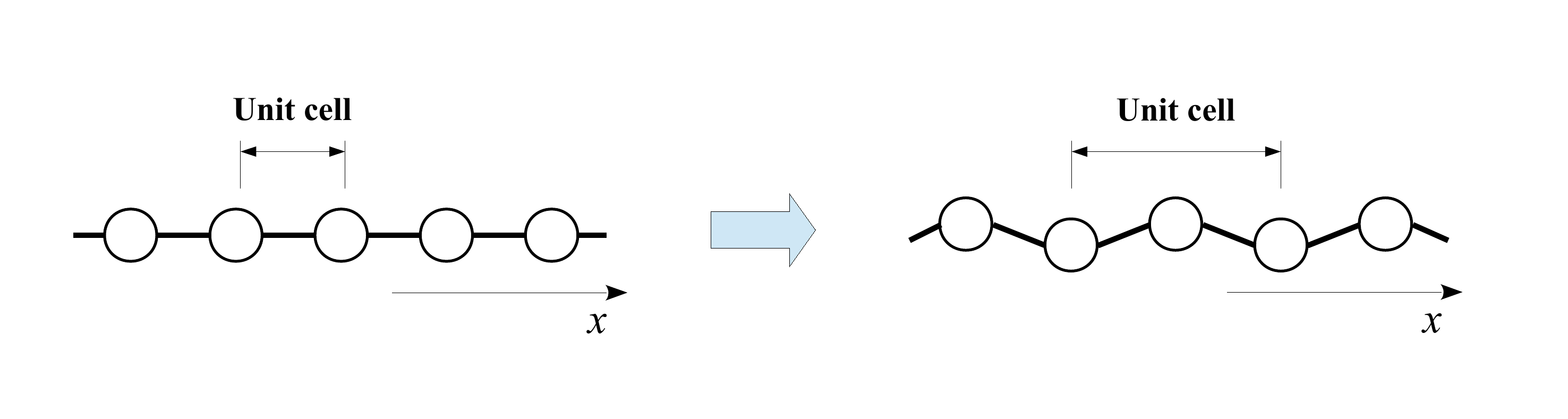}
 \end{center}
 \caption{(Color online) Side view of the square lattice. 
In the presence of the
 staggered modulation in the right figure, the size of the unit cell
 is doubled in the $x$-direction. The right lattice configuration has a
 nonsymmorphic ${\bm Z}_2 $ symmetry: Turn it upside down and  move it
 by half of the unit cell in the $x$-direction, then the system goes back
 to the original configuration.  
 }
 \label{Fig:staggered}
\end{figure}

Bulk model Hamiltonians for $e_\pm$ are
constructed as follows:
First consider spinless chiral $p$-wave
superconductors on the square lattice, 
\begin{align}
H_{p_x\pm ip_y}(\bk) = \left( \mp t \cos (k_x) -t \cos (k_y) - \mu
\right) \tau_z \pm \Delta \sin (k_x) \tau_x - \Delta \sin (k_y)
\tau_y, \quad (t>0, \Delta>0) 
\end{align}
where $\tau_i$ $(i=x,y,z)$ are the Pauli matrices in the Nambu space, 
$t$,$\mu$, and $\Delta$ are the hopping parameter, the chemical potential, 
and the pairing amplitude, respectively.
For $-2t <\mu < 0$,  the both chiral superconductors realize the $ch_1
= 1$ phase that support a chiral edge state near $k_x = 0$.
Then to introduce the nonsymmorphic ${\bm Z}_2$ symmetry, 
we add a staggered modulation of the lattice in the $x$-direction. 
See Fig.\ref{Fig:staggered}. 
The staggered modulation doubles the size of the unit cell in the
$x$-direction,
making two inequivalent sites in each extended unit cell.
Then the resultant systems host ${\bm Z}_2$ nonsymmorphic symmetry as glide
symmetry with respect to the $xy$-plane, giving the model Hamiltonians for
$e_{\pm}$.
By denoting the Pauli matrix in the two inequivalent sites as $\sigma_i$,  
the bulk model Hamiltonians $H_{e_{\pm}}(\bk)$ for $e_{\pm}$ are
given by 
\begin{align}
H_{e_{\pm}}(\bk) 
= \left( \mp t \cos (k_x/2) \sigma_x - t \cos k_y -
 \mu \right) \tau_z\pm \Delta \sin (k_x/2) \sigma_x
 \tau_x -
\Delta \sin k_y \tau_y,
\end{align}
where $k_x$ has been rescaled as $k_x\to k_x/2$ since the lattice
spacing between equivalent sites is doubled in the $x$-direction.
(For simplicity, we have neglected here a staggered potential induced by the
lattice modulation.)
The nonsymmorphic ${\bm Z}_2$ symmetry for $H_{e_\pm}(\bk)$ is given by
$G_{\rm BdG}^{(+)}(k_x)=e^{-ik_x/2}\sigma_x$, which obeys 
$CG_{\rm BdG}^{(+)}(k_x)=G_{\rm BdG}^{(+)}(-k_x)C$.
Whereas the above $H_{e_\pm}(\bk)$ and $G_{\rm BdG}^{(+)}(k_x)$ do not have
$2\pi$-periodicity in the $k_x$-direction,  
using the ${\bm
k}$-dependent unitary transformation in Appendix \ref{sec:Periodic_Bloch}, 
\begin{align}
V({\bm k})=
\left(
\begin{array}{cc}
1 &0 \\
0 & e^{-ik_x/2}
\end{array}
\right)_{\sigma} \tau_0, 
\end{align}
we obtain their periodic version as
\begin{align}
H_{e_{\pm}}(\bk) 
= \left[\mp (t/2)\left(1+\cos k_x\right)\sigma_x-(t/2)\sin k_x
 \sigma_y 
- t \cos k_y -
 \mu \right] \tau_z
\pm(\Delta/2)\left[
\sin k_x \sigma_x+\left(1-\cos k_x\right)
\sigma_y
\right] \tau_x 
-
\Delta \sin k_y \tau_y,
\end{align}
with
\begin{align}
G^{(+)}_{\rm BdG}(k_x)=
\left(
\begin{array}{cc}
0 & e^{-ik_x} \\
1 & 0
\end{array}
\right)_{\sigma} \tau_0.
\end{align}
From the construction of the model, one can easily show that
$H_{e_{\pm}}$ give the correct topological numbers, $e_+=(1,1,0)$,
$e_-=(1,0,1)$.


In general, a model Hamiltonian for the non-chiral edge state
$e_+-e_-$ is constructed as the
direct product $H_{e_+}(\bk)\oplus [-H_{e_-}(\bk)]$.
However, we find the following much simpler Hamiltonian for $e_+-e_-$,
\begin{align}
H_{e_+-e_-}({\bm k}) = (-t \cos k_y - \mu) \tau_z 
+ f(k_x)\left[\cos(k_x/2)\sigma_x
+\sin(k_x/2)\sigma_y \right] 
\tau_x +  \Delta \sin k_y \tau_y. 
\end{align}
with a real function $f(k_x)$. PHS and the $2\pi$-periodicity in
$k_x$ impose the constraints $f(-k_x) =
-f(k_x)$ and $f(k_x)=-f(k_x+2\pi)$, respectively. 
On a boundary parallel to the $x$-direction, the system has a non-chiral
mode with the dispersion $E(k_x)=\pm f(k_x)$. 
If one considers $f(k_x)=c\sin (k_x/2)$ with small $c$, the non-chiral edge state has a spectrum like Fig
\ref{Fig:2D_D_Glide} [c]. 

A similar consideration can be done in the case with an odd gap
function under $G(k_x)$, where the nonsymmorphic ${\bm Z}_2$ symmetry is
given by $G_{\rm BdG}^{(-)}(k_x)$. 
Figures \ref{Fig:2D_D_Glide}[d]-[f] illustrate edge states in
this case.

\subsubsection{$\Z_4$ time-reversal invariant TCSC
   with nonsymmorphic  ${\bm Z}_2$ symmetry ($d=2$, $\{U,
   \Gamma\}=0$ in class DIII)} 
\label{Sec:2DZ4TSC}

Consider a two-dimensional time-reversal invariant superconductor with
nonsymmorphic ${\bm Z}_2$ symmetry. 
The BdG Hamiltonian has both TRS and PHS, 
\begin{align}
H_{\rm BdG}(\bk) = \begin{pmatrix}
\epsilon(\bk) & \Delta(\bk) \\
\Delta^{\dag}(\bk) & - \epsilon^T(-\bk)
\end{pmatrix},
&& T H_{\rm BdG}(\bk) T^{-1}=H_{\rm BdG}(-\bk),
&& C H_{\rm BdG}(\bk) C^{-1}=-C_{\rm BdG}(-\bk),
\end{align}
with $T=is_y K$ and $C=\tau_x K$.
The nonsymmorphic ${\bm Z}_2$ symmetry for the normal Hamiltonian
$\epsilon(\bk)$ is 
\begin{align}
G(k_x) \epsilon(\bk) G(k_x)^{-1} = \epsilon(\bk), && G(k_x) = \begin{pmatrix}
0 & i s_z e^{- ik_x} \\
i s_z & 0 \\
\end{pmatrix}, 
\end{align}
which satisfies $T G(k_x) = G(-k_x) T$.
In a manner similar to time-reversal breaking case in
Sec.\ref{sec:ZZ_2TCSC}, the superconductor retains the
nonsymmorphic ${\bm Z}_2$ symmetry if the gap function $\Delta({\bm k})$
is even or odd under $G(k_x)$. However, in the present case, the
topological property depends on the parity under $G(k_x)$. We first
consider the odd parity case in this subsection. 
The even parity case will
be discussed in the next subsection.   

Assuming that the gap function $\Delta(\bk)$ is odd under $G(k_x)$, i.e. 
$G(k_x) \Delta(\bk) G(-k_x)^{T} = - \Delta(\bk)$, the nonsymmorphic 
${\bm Z}_2$ operator for the BdG Hamiltonian is given by  
\begin{align}
G^{(-)}_{\rm BdG}(k_x) = \begin{pmatrix}
G(k_x) & 0 \\
0 & -G(-k_x)^* \\
\end{pmatrix}, 
\end{align}
which satisfies 
\begin{align}
G^{(-)}_{\rm BdG}(k_x)^2 = - e^{- i k_x}, 
&&
T G^{(-)}_{\rm BdG}(k_x) = G_{\rm BdG}^{(-)}(-k_x) T, &&
C G^{(-)}_{\rm BdG}(k_x) = -G_{\rm BdG}^{(-)}(-k_x) C. 
\label{eq:GTC}
\end{align}
In particular, $G^{(-)}_{\rm BdG}(k_x)$ and $TC$
anticommute, $\{G^{(-)}_{\rm BdG}(k_x), TC \}=0$.
Thus the corresponding symmetry in Table
\ref{Tab:Periodic_Table_Unitary_d_para=0} 
is $\{U,TC \}=0$ of class DIII. 
From Table \ref{Tab:Periodic_Table_Unitary_d_para=0}, its topological
number is $\Z_4$.

\begin{figure}[!]
 \begin{center}
  \includegraphics[width=0.7\linewidth, trim=0cm 0cm 0cm 0cm]{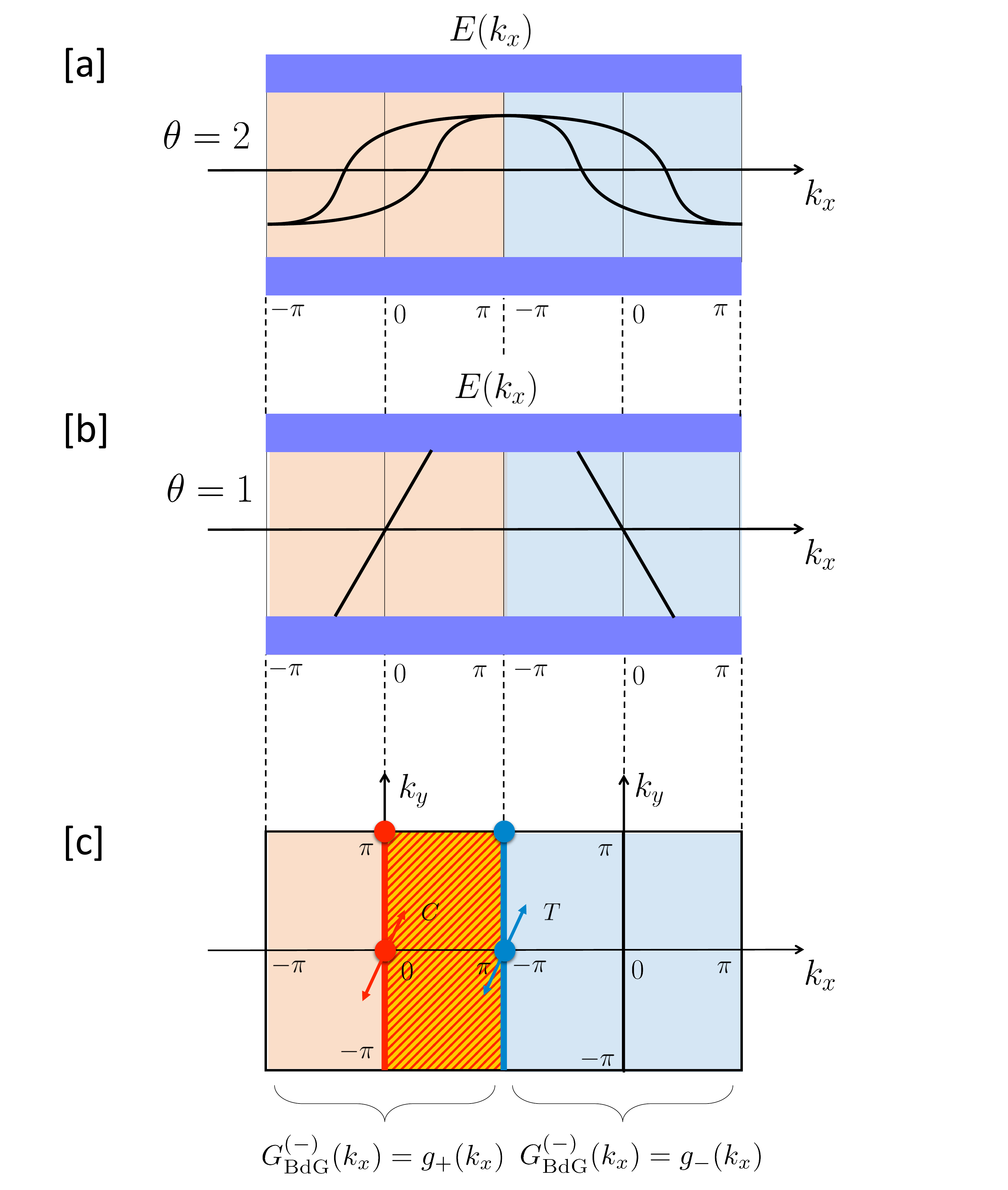}
 \end{center}
 \caption{(Color online) 
[a] ([b]) A typical edge spectrum for $\theta=2$ ($\theta = 1$). 
[c] The integral region for the $\Z_4$ invariant (\ref{Eq:Z4_Inv_2D_Class_DIII}). 
The left hand side and right hand side regions represent the positive and negative glide eigensectors, respectively, which are glued together on their boundary lines $k_x = -\pi, \pi$. 
In figure [c], the red circles at $(k_x,k_y) = (0,0)$ and $(0,\pi)$ are the centers of the particle-hole transformation, 
and the blue circles at $(k_x,k_y) = (\pi,0)$ and $(\pi,\pi)$ are the centers of the time-reversal transformation. 
The integral region for the $\Z_4$ invariant
 (\ref{Eq:Z4_Inv_2D_Class_DIII}) is the shaded portion in [c].
}
 \label{Fig:Z4Invariant_kspace} 
\end{figure}

Now construct the $\Z_4$ topological invariant. 
First, we divide the Hamiltonian $H_{\rm BdG}(\bk)$ into two
eigensectors of $G_{\rm BdG}(k_x)$ with eigenvalues $g_{\pm}(k_x) = \pm
i e^{-i k_x/2}$.   
At $k_x=0$, because of $g_{\pm}(0)=\pm i$, 
Eq.(\ref{eq:GTC}) yields that PHS maps the $g_{\pm}(0)$ sector to itself, but
TRS exchanges the $g_{+}(0)$ and $g_{-}(0)$ sectors.  
In other words, each $g_{\pm}(0)$ subsector at $k_x=0$ 
(the red line in Fig.\ \ref{Fig:Z4Invariant_kspace} [c] for the $g_+(0)$ sector) 
keeps PHS but not TRS.
Therefore it can be regarded as a
one-dimensional time-reversal breaking superconductor belonging to class
D, where we can define the $\Z_2$ invariants 
\begin{align}
\nu_{\pm}(0) = \frac{i}{\pi}\int_{-\pi}^{\pi}dk_y
{\rm tr} {\cal A}_{\pm}(0, k_y) \ \
 ({\rm mod\ } 2). 
\label{Eq:Z4_Inv_2D_Class_DIII_Z2part}
\end{align}
Here ${\cal A}_{\pm}$ is the Berry connection of occupied states in the
$g_{\pm}(0)$ sector and the trace is taken for all occupied states in
each subsector.
From TRS, $\nu_+(0)=\nu_-(0)$ (mod 2), and thus only one of
them, say $\nu_+(0)$ is independent.
As is shown below, the $\Z_2$ invariant $\nu_+(0)$ gives a $\Z_2$
part of the $\Z_4$ invariant.  

To introduce the $\Z_4$ invariant, we also need to consider the $k_x=\pi$
line.
At $k_x=\pi$, because of $g_{\pm}(\pi)=\pm 1$, 
each $g_{\pm}(\pi)$ subsector has its own TRS, but not PHS.
Thus, it can be regarded as a one-dimensional time-reversal invariant
insulator belonging to class AII.
Whereas no one-dimensional topological number exists in class AII, we
can use TRS in the subsector to define the $\Z_4$ invariant:
The key is the Kramers degeneracy.  
From TRS, occupied states in the $g^{(+)}(\pi)$ sector 
(the blue line in Fig.\ \ref{Fig:Z4Invariant_kspace} [c]) 
form  Kramers
pairs, $|{u^{(+) {\rm I}}_n(\pi, k_y)}\rangle$, $|u^{(+){\rm
II}}_n(\pi, k_y)\rangle$ with $|u^{(+)\rm I}_n(\pi, k_y)\rangle = T |u^{(+)\rm
II}_n(\pi, -k_y)\rangle$.  
Using the Berry connection ${\cal A}_+^{\rm I, II}$ of $|u_n^{(+)\rm
I,II}(\pi,k_y)\rangle$, we define the $\Z_4$ invariant by
\begin{align}
\theta := \frac{2i}{\pi}\int_{-\pi}^{\pi}dk_y
{\rm tr}{\cal A}_+^{\rm I}(\pi, k_y)
- \frac{i}{\pi} \int_0^{\pi} d k_x \int_{-\pi}^{\pi} dk_y 
{\rm tr} {\cal F}_+(k_x,k_y) 
\ \ ({\rm mod\ } 4) , 
\label{Eq:Z4_Inv_2D_Class_DIII}
\end{align}
where ${\cal F}_+$ is the Berry curvature for occupied states in the
$g_+(k_x)$ eigensector. 
The integral region of the second term in (\ref{Eq:Z4_Inv_2D_Class_DIII}) 
connects the two Berry phases, says, $\nu_+(0)$ defined in Eq.\ (\ref{Eq:Z4_Inv_2D_Class_DIII_Z2part}) 
and the first term in (\ref{Eq:Z4_Inv_2D_Class_DIII}). 
The modulo-4 ambiguity in Eq.(\ref{Eq:Z4_Inv_2D_Class_DIII}) comes
the $U(1)$ gauge freedom of the Berry connection ${\cal A}_+^{\rm I}$.
In order for $\theta$ in Eq.(\ref{Eq:Z4_Inv_2D_Class_DIII}) to define
the $\Z_4$ invariant actually, $\theta$ must be quantized to be an
integer, which we
shall show below:
From the Stokes' theorem, the second term of $\theta$ is written as
\begin{align}
-\frac{i}{\pi}\int_0^{\pi}dk_x \int_{-\pi}^{\pi} dk_y 
{\rm tr} {\cal F}_+(k_x, k_y)
=&-\frac{i}{\pi}\int_{-\pi}^{\pi}dk_y{\rm tr} {\cal A}_+(\pi, k_y) 
+\frac{i}{\pi}\int_{-\pi}^{\pi}dk_y {\rm tr} {\cal A}_+(0, k_y) 
\ \ (\mbox{mod 2}), 
\nonumber\\
=&-\frac{2i}{\pi}\int_{-\pi}^{\pi} dk_y {\rm tr} {\cal A}^{\rm I}_+(\pi, k_y) 
+\frac{i}{\pi}\int dk_y
{\rm tr} {\cal A}_+(0, k_y) 
\ \ (\mbox{mod 2}), 
\end{align}
where we have used the identities,
$
{\rm tr}{\cal A}_+(\pi, k_y)={\rm tr}{\cal A}^{\rm I}_+(\pi, k_y)+
{\rm tr}{\cal A}^{\rm II}_+(\pi, k_y),
$
and 
$
{\rm tr}{\cal A}^{\rm I}_+(\pi, k_y)={\rm tr}{\cal A}^{\rm II}_+(\pi, -k_y).
$
Thus, $\theta$ is recast into
\begin{align}
\theta=\frac{i}{\pi}
\int_{-\pi}^{\pi}dk_y{\rm tr}{\cal A}_+(0, k_y)=\nu_+(0) 
\ \ ({\rm mod\ } 2 ).
\end{align}
Since $\nu_+(0)$ is an integer, $\theta$ is also quantized to be an integer. 
Taking into account the modulo-4 ambiguity in
Eq.(\ref{Eq:Z4_Inv_2D_Class_DIII}), we have four distinct values of
$\theta$, $\theta=0, 1, 2, 3$, thus $\theta$ defines a $\Z_4$
invariant.  
A typical edge spectrum for $\theta=2$ ($\theta=1$) is shown 
in Fig.\ \ref{Fig:Z4Invariant_kspace} [a] ([b]). 
The above equation also implies that 
${\Z_2}$ invariant $\nu_+(0)$ describes a $\Z_2$ subgroup of $\Z_4$,
as mentioned before.

\begin{figure}[!]
 \begin{center}
\includegraphics[height=0.7\linewidth, trim=0cm 0cm 0cm 0cm]{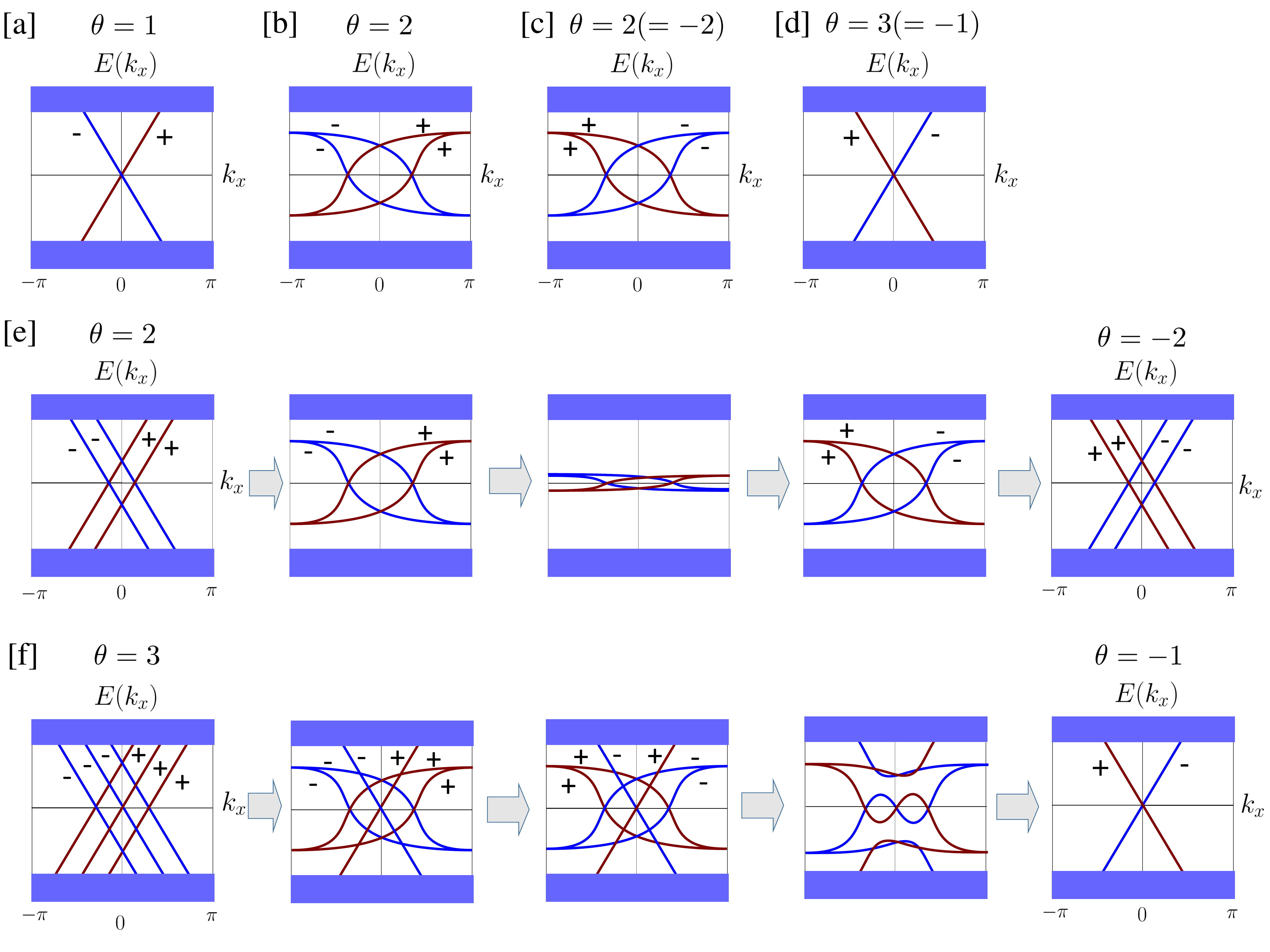}
 \end{center}
 \caption{(Color online) Gapless edge states in two-dimensional
 time-reversal invariant TCSCs with nonsymmorphic ${\bm Z}_2$ symmetry
 $G^{(-)}(k_x)$ ($d_{\parallel}=0$, $d=2$, $\{U, \Gamma\}=0$ in class
 DIII). Here $G^{(-)}(k_x)^2=-e^{-ik_x}$. The gap function
 is odd under the nonsymmorphic ${\bm Z}_2$ symmetry.
Red (+) and blue (-) lines indicate edge
 states in the $g_{\pm}(k_x)$ sector. 
[a] Edge states in the $\theta=1$ phase.
[b]-[c] Edge states in the $\theta=2$ phase.
[d] Edge states in the $\theta=3$ phase.
Figure [e] ([f]) shows adiabatic transformation of edge
 states in the $\theta=2$ ($\theta=3$) phase.
The modulo-4 ambiguity of $\theta$ is evident in Figs.[e] and [f].}
 \label{Fig:2D_DIII_Glide}
\end{figure}

The $\Z_4$ structure can be confirmed in corresponding edge states: 
When $\theta=1$,  we have a helical Majorana fermion shown in Fig.\
\ref{Fig:2D_DIII_Glide}[a]. 
The right (left) moving mode is an  eigenstate of $G_{\rm
BdG}^{(-)}(k_x)$ with the eigenvalue $g_{+}(k_x)$ ($g_{-}(k_x)$). 
Using an argument similar to that for chiral edge modes in
Sec.\ref{sec:ZZ_2TCSC}, one can show that the Dirac point of the
helical mode should be at $k_x=0$. (Note that 
$G(k_x)^2$ in the present case has an opposite sign to that in
Sec.\ref{sec:ZZ_2TCSC}. For this reason, an odd
gap function under $G(k_x)$ provides a Dirac point at $k_x=0$.) 
Being different from Fig.\ref{Fig:2D_D_Glide}[c], this helical mode can
not be detached from the bulk spectrum.
Indeed, 
TRS at $k_x=\pi$ requires
two-fold Kramers degeneracy in each $g_{\pm}(\pi)$ sector, but this
requirement cannot be met by the single helical edge mode.

For $\theta=2$, we have a pair of helical Majorana fermions, 
as shown in Fig.\ \ref{Fig:2D_DIII_Glide}[b].
In this case, the helical Majorana modes can be detached from
the bulk spectrum without contradiction to symmetry.
Since the counter propagating modes have different eigenvalues
of $G_{\rm BdG}^{(-)}(k_x)$,  
the edge states cannot be gapped out.
Furthermore, because CS $\Gamma=iTC$ requires a symmetric
energy spectrum with respect to $E(k_x)\to -E(k_x)$,  
the edge states cannot go up or go down to merge into the bulk spectrum.
As a result, the detached states are topologically stable.
It is remarkable that
the edge modes in Fig.\ \ref{Fig:2D_DIII_Glide}[b]
can be smoothly deformed into those in Fig.\ \ref{Fig:2D_DIII_Glide}[c],
as illustrated in Fig.\ \ref{Fig:2D_DIII_Glide}[e] 

When $\theta=3$, edge states  
are obtained by adding those in
$\theta=1$ (Fig.\ref{Fig:2D_DIII_Glide}[a]) and $\theta=2$
(Fig.\ref{Fig:2D_DIII_Glide}[b] or [c]).
Since one of the helical edge modes in Fig.\ref{Fig:2D_DIII_Glide}[c] can
be pair-annihilated by the helical edge mode in
Fig.\ref{Fig:2D_DIII_Glide}[a], 
the obtained edge state is topologically the same as
a single helical Majorana mode in
Fig.\ref{Fig:2D_DIII_Glide}[d]. See Fig.\ref{Fig:2D_DIII_Glide}[f]. 

Finally, edge states in $\theta=4$ is obtained by adding
helical edge modes in $\theta=1$ (Fig.\ref{Fig:2D_DIII_Glide}[a])
and  $\theta=3$
(Fig.\ref{Fig:2D_DIII_Glide}[d]).
Since these helical edge modes can be pair-annihilated, we confirm the
$\Z_4$ nature of $\theta$, i.e. $\theta=4=0$.



In a manner similar to $H_{e_{\pm }}(\bk)$ in Sec.\ref{sec:ZZ_2TCSC}, a bulk model
Hamiltonian
$H_{1}({\bm k})$ for
$\theta=1$ can be constructed from a helical $(p_x+i p_y)_{\uparrow} +
(p_x-i p_y)_{\downarrow}$-wave superconductor with a staggered
modulation of the lattice in the $k_x$-direction. The obtained model
Hamiltonian in the periodic Bloch basis is
\begin{align}
H_1({\bm k}) = \left[(t/2)(1+\cos k_x) \sigma_x + t\cos k_y - \mu
 \right] \tau_z 
+ (\Delta/2)\left[
\sin k_x \sigma_x+(1-\cos k_x)\sigma_y\right]s_z\tau_x 
+ \Delta\sin k_y \tau_y, 
\end{align}
with
\begin{align}
G_{\rm BdG}^{(-)}(k_x)=
is_z\left(
\begin{array}{cc}
0 & e^{-ik_x}\\
1 & 0
\end{array}
\right)_{\sigma}\tau_0. 
\end{align}

\subsubsection{$\Z_2\oplus \Z_2$ time-reversal invariant TCSC
   with nonsymmorphic  ${\bm Z}_2$ symmetry ($d=2$, $[U,
   \Gamma]=0$ in class DIII)} 
As in Sec.\ref{Sec:2DZ4TSC}, we again consider a two-dimensional
   time-reversal invariant
superconductor with nonsymmorphic ${\bm Z}_2$ symmetry.
In this section, we assume an even gap function under $G(k_x)$,
\begin{align}
G(k_x) \Delta(\bk) G(-k_x)^{T} = \Delta(\bk), && G(k_x) = \begin{pmatrix}
0 & i s_z e^{- ik_x} \\
i s_z & 0 \\
\end{pmatrix}. 
\end{align}
The nonsymmorphic ${\bm Z}_2$ symmetry for the BdG Hamiltonian is given by 
\begin{align}
G^{(+)}_{\rm BdG}(k_x) = \begin{pmatrix}
G(k_x) & 0 \\
0 & G(-k_x)^* \\
\end{pmatrix}, 
\end{align}
which satisfies 
\begin{align}
G^{(+)}_{\rm BdG}(k_x)^2 = - e^{- i k_x}, &&
T G^{(+)}_{\rm BdG}(k_x) = G^{(+)}_{\rm BdG}(-k_x) T, &&
C G^{(+)}_{\rm BdG}(k_x) = G^{(+)}_{\rm BdG}(-k_x) C. 
\label{eq:GTC2}
\end{align}
Since the chiral operator $\Gamma=iTC$ commutes with $G_{\rm
BdG}^{(+)}(k_x)$, the corresponding symmetry in Table
\ref{Tab:Periodic_Table_Unitary_d_para=0} is $[U, \Gamma]=0$ in class DIII.
Thus, the topological phase is $\Z_2 \oplus \Z_2$.

\begin{figure}[!]
 \begin{center}
  \includegraphics[width=1.01\linewidth, trim=0cm 0cm 0cm 0cm]{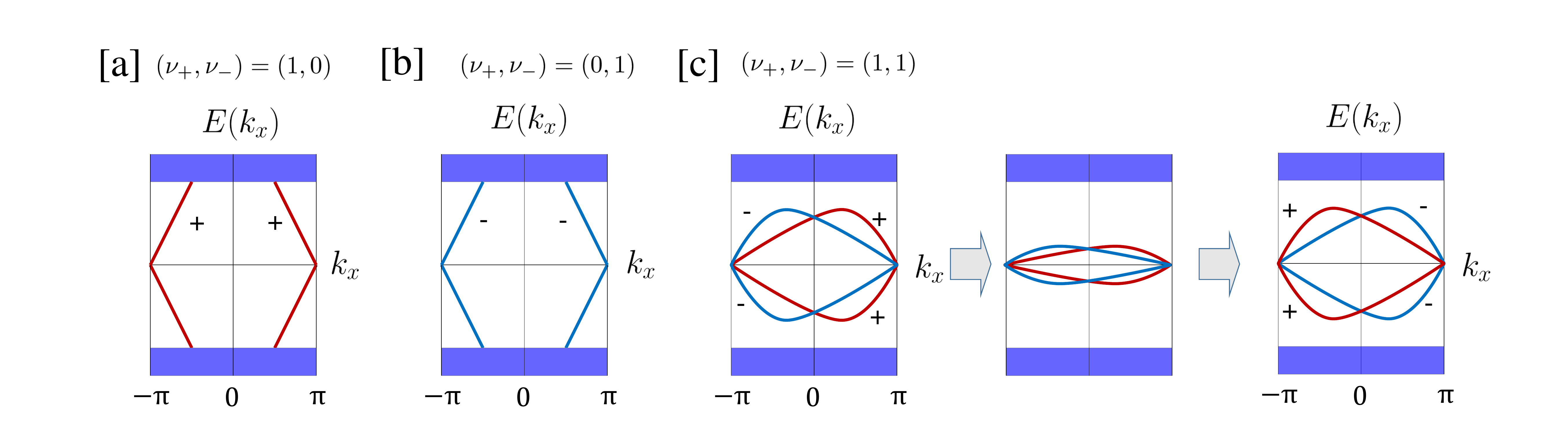}
 \end{center}
 \caption{(Color online) Gapless edge states in two-dimensional
 time-reversal invariant
TCSCs with nonsymmorphic ${\bm Z}_2$ symmetry $G(k_x)$
 ($d_{\parallel}=0$, $d=2$, $[U, \Gamma]=0$ in class DIII). 
$G(k_x)^2=-e^{-ik_x}$. 
The gap function is even under $G(k_x)$.
[a] Edge states in the $(\nu_+, \nu_-)=(1,0)$ phase. 
[b] Edge states in the $(\nu_+, \nu_-)=(0,1)$ phase.
[c] Edge states in the $(\nu_+, \nu_-)=(1,1)$ phase.
Red (+) and blue (-) lines indicate edge states with the
 eigenvalues $g_{\pm}(k_x)=\pm ie^{-ik_x/2}$ of $G_{\rm
 BdG}^{(+)}(k_x)$.
Figure [c] also shows an adiabatic deformation of the edge states.
}
 \label{Fig:2D_DIII_Glide_even}
\end{figure}

The two $\Z_2$ topological invariants are defined at $k_x = \pi$:
We first divide the system into the eigensectors of $G_{\rm
BdG}^{(+)}(k_x)$ with the eigenvalues $g_{\pm}(k_z)=\pm
ie^{-ik_x/2}$. 
From Eq.(\ref{eq:GTC2}),
each eigensector at $k_x=\pi$ has its own TRS and PHS, realizing a
one-dimensional class DIII system.
As a class DIII system, the $g_{\pm}(\pi)$ subsector has
the $\Z_2$ invariant 
\begin{align}
\nu_{\pm} := \frac{i}{\pi} \int_{-\pi}^{\pi} dk_y 
{\rm tr} {\cal A}^{\rm I}_{\pm}(\pi, k_y)
 \ \ ({\rm mod\ } 2), 
\end{align} 
where ${\cal A}^{\rm I}_{\pm}$ is the Berry connection of one of the Kramers
pair with the eigenvalue $g_{\pm}(\pi) = \pm 1$, and the trace is taken
for all the occupied states.
There are four distinct topological phases with the topological numbers 
\begin{align}
(\nu_+, \nu_-) = (0,0), (1,0), (0,1), (1,1). 
\end{align}
%
The $(1,0)$- and $(0,1)$-phases host helical Majorana
fermions in Figs.\ref{Fig:2D_DIII_Glide_even}[a] and [b], respectively. 
The $(1,1)$-phase allows detached edge states in
Fig.\ref{Fig:2D_DIII_Glide_even}[c].  
Like chiral edge modes in Sec.\ref{sec:ZZ_2TCSC}, 
Dirac points of
the edge states should be at $k_x=\pi$ due to the constraint
from the nonsymmorphic ${\bm Z}_2$ symmetry.


\subsubsection{$\Z_2$ class CII-TCSC with nonsymmorphic ${\bm
   Z}_2$ symmetry ($d=2$, $\{U, \Gamma\}=0$ in class CII)} 

We consider here a two-dimensional class CII superconductor. 
As in the previous two cases, the BdG Hamiltonian
$H_{\rm BdG}(\bk)$ has both TRS and PHS, but in this case PHS obeys
$C^2=-1$, not $C^2=1$.
TRS satisfies the same equation $T^2=-1$.
In the presence of nonsymmorphic ${\bm Z}_2$ symmetry, $G_{\rm
BdG}(k_x)H_{\rm BdG}(\bk)G_{\rm BdG}(k_x)^{-1}=H_{\rm BdG}(\bk)$ with
$G_{\rm BdG}(k_x)^2=-e^{-ik_x/2}$, we find
that the system may support a $\Z_2$ TCI phase
if $G_{\rm BdG}(k_x)$ satisfies $\{G_{\rm BdG}(k_x), TC\}=0$. 
See Table \ref{Tab:Periodic_Table_Unitary_d_para=0} with
$d=2$, $\{U,TC\}=0$ in class CII.

The $\Z_2$ invariant is defined as follows.
First, we note that symmetries of the system are the same as those in
Sec.\ref{Sec:2DZ4TSC}, 
except that $C^2$=1 is replaced with $C^2=-1$.
Thus, we can introduce the following topological invariant $\theta'$
analogous to Eq.(\ref{Eq:Z4_Inv_2D_Class_DIII}), 
\begin{align}
\theta' := \frac{2i}{\pi} \int_{-\pi}^{\pi} dk_y {\rm tr}{\cal A}_+^{\rm
 I}(\pi,k_y) 
- \frac{i}{\pi} \int_{0}^{\pi}dk_x \int_{-\pi}^{\pi} dk_y 
{\rm tr}{\cal F}_+(k_x, k_y) \ \ ({\rm mod\ } 4), 
\end{align}
where ${\cal A}_+^{\rm I}$ is the Berry connection of 
$|u_n^{(+){\rm I}}(\bk)\rangle$ for the Kramers pair 
$(|u_n^{(+){\rm I}}(\bk)\rangle, |u_n^{(+){\rm II}}(\bk)\rangle)$ in the
eigensector of $G_{\rm BdG}(k_x)$ with the eigenvalue
$g_+(k_x)=ie^{-ik_x/2}$, ${\cal F}_+$ is the Berry curvature in the
$g_+(k_x)$ sector, and the trace is taken for all the occupied states. 
In a manner similar to $\theta$ in Eq.(\ref{Eq:Z4_Inv_2D_Class_DIII}), 
$\theta'$ obeys
\begin{align}
\theta' = \frac{i}{\pi} \int_{-\pi}^{\pi}dk_y {\rm tr}{\cal
 A}_+(0,k_y) \ \ \ ({\rm mod\ } 2). 
\label{eq:theta'2}
\end{align}
Whereas $\theta'$ is defined in the same manner as $\theta$, the
quantization rule is different.
In the present case, the $g_+(k_x)$ subsector at $k_x=0$
realizes a one-dimensional class C superconductor, instead of class D,
and thus the right hand side of Eq.(\ref{eq:theta'2}) becomes zero (mod
2).\cite{Schnyder2008} Therefore, $\theta'$ takes only two different values
$\theta'=0, 2$ (mod 4).
Hence, we can define the $\Z_2$ invariant $\nu$ by $\nu=\theta'/2$ (mod
2).
A gapless edge state for $\nu = 1$ (mod $2$) has a similar spectrum as
Fig.\ref{Fig:2D_DIII_Glide}[b] and [c].

\subsubsection{$\Z_2$ magnetic insulator with magnetic half
   lattice translation symmetry ($d=3$, $A$ in class A)}

Mong et al. pointed out that a time-reversal
breaking insulator hosts a $\Z_2$ phase in three dimensions if a 
combination of half lattice translation and time-reversal is
preserved.\cite{Mong2010}
In our classification, this symmetry corresponds $A$ in class A of
Table \ref{Tab:Periodic_Table_Unitary_d_para=0}, which reproduces
$\Z_2$ in $d=3$. 
Such a magnetic half lattice translation symmetry is realized in a
three-dimensional material with an antiferromagnetic order.

\subsection{Surface states protected by glide symmetry
  ($d_{\parallel}=1$)}
\label{Sec:3D}

\begin{figure}[!]
 \begin{center}
  \includegraphics[width=0.5\linewidth, trim=0cm 0cm 0cm 0cm]{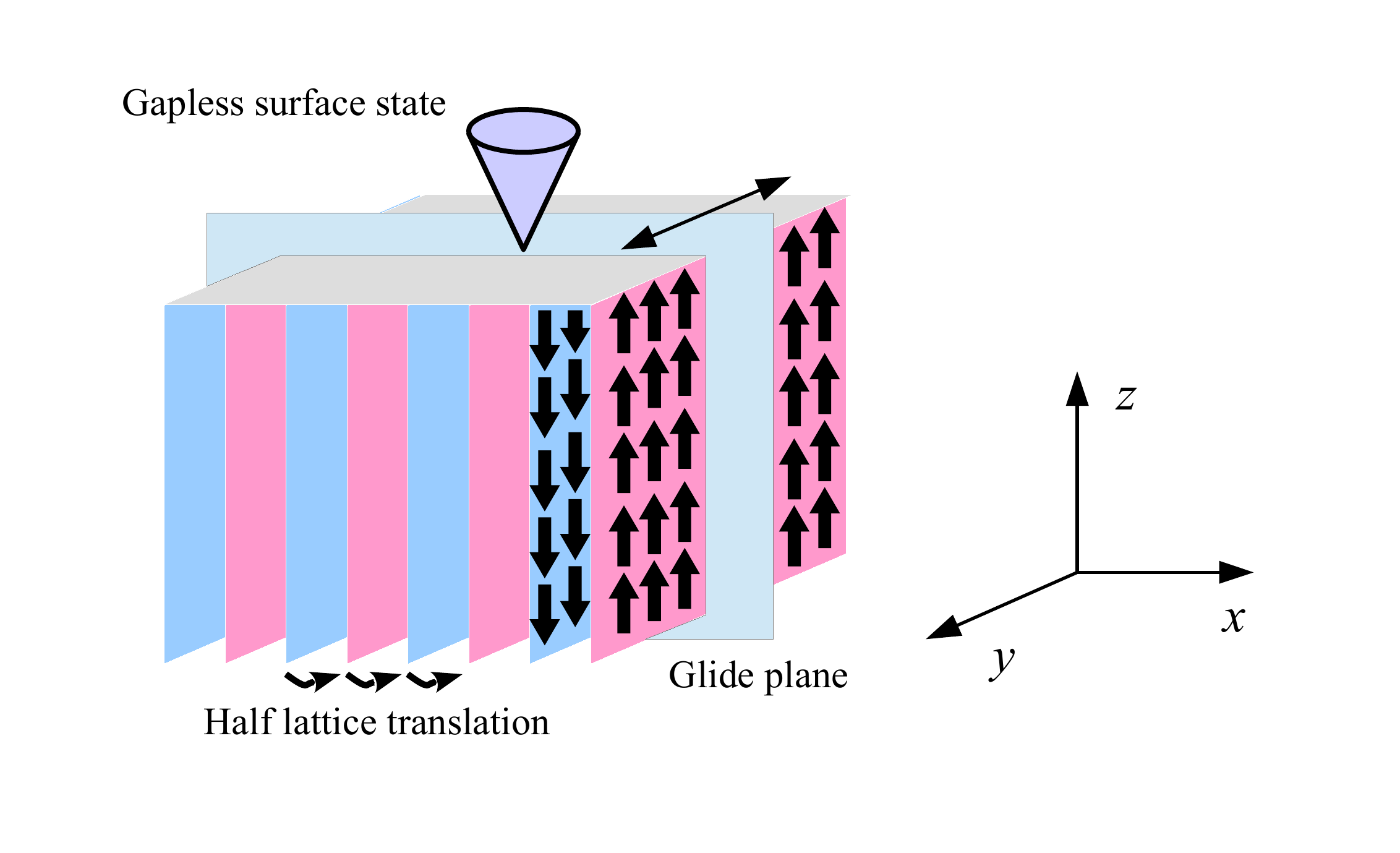}
 \end{center}
 \caption{
(Color online) Schematic illustration of glide symmetry
 in three-dimensions.
The red and blue indicate two different configurations, say spin
 configuration, that are exchanged by mirror reflection with
 respect to the $zx$-plane. 
The system is invariant under glide reflection with respect to the $zx$-plane,
 i.e. combination of mirror reflection and half lattice translation in
 the $x$-direction.
A two-dimensional surface parallel to the $x$-direction is compatible
 with the glide symmetry.
Gapless surface states protected by glide symmetry appear on the glide
 symmetric surface when the 
 relevant bulk topological number is nonzero.}
 \label{Fig:Bulk-Boundary_3D_glide}
\end{figure}

Now we discuss topological surface states in the
nonsymmorphic $d_{\parallel}=1$ family.
In up to three dimensions, only three $(d=3)$ dimensional systems may
have boundaries compatible with the $d_{\parallel}=1$ NSGs, and the relevant
nonsymmorphic symmetry is glide and its magnetic symmetry.
A schematic illustration of glide $(x,y,z) \mapsto
(x+1/2,-y,z)$ is shown in Fig.\ \ref{Fig:Bulk-Boundary_3D_glide}.
A surface normal to the $z$-direction preserves the glide and magnetic
glide symmetries, so gapless states protected by these symmetries appear on
this boundary when the relevant topological number is nonzero. 
The strong topological indices for boundary states are given in
$d=3$ of Table \ref{Tab:Periodic_Table_Unitary_d_para=1}.

\subsubsection{$\Z_2$ TCI with glide symmetry ($d=3$, $U$ in class A)}
\begin{figure}[!]
 \begin{center}
  \includegraphics[width=0.6\linewidth, trim=0cm 0cm 0cm 0cm]{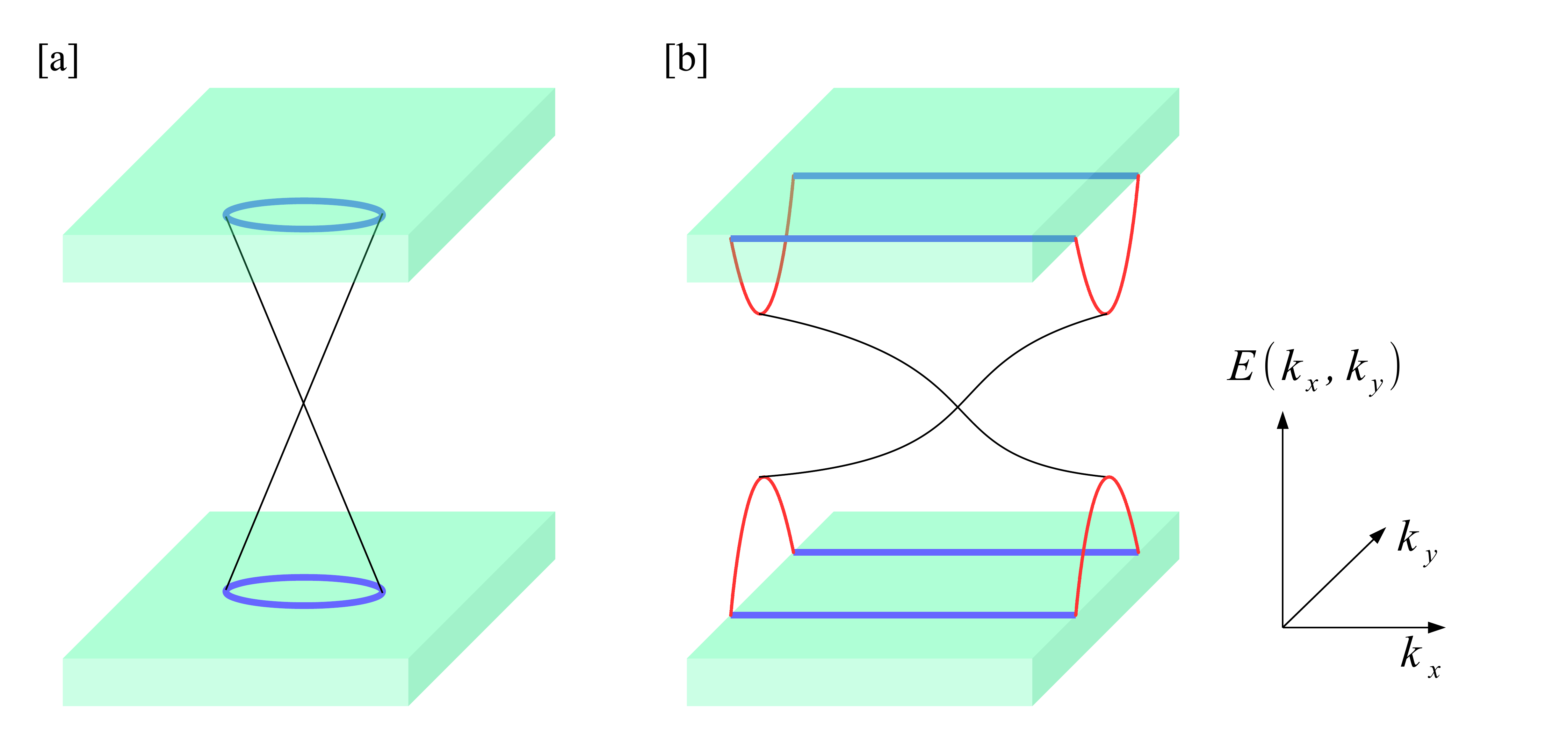}
 \end{center}
 \caption{(Color online) Schematic pictures of $\Z_2$ nontrivial surface
 states with glide symmetry ($d_{\parallel}=1$, $d=3$, $U$ in class A). 
 [a] and [b] show surface states with the same non-trivial $\Z_2$ number. 
 The green volumes represent the bulk spectrum. 
 In [b], the red curves are identified at the zone boundary. }
 \label{Fig:ClassA_Glide_Surface}
\end{figure}

As was shown in Refs.\onlinecite{Fang2015} and
\onlinecite{Shiozaki2015} independently,  
glide symmetry in three dimensions gives  a $\Z_2$ TCI phase, even
in the absence of anti-unitary symmetry such as TRS and PHS.   
In our classification scheme,  the glide symmetry corresponds to $U$ in
class A of Table \ref{Tab:Periodic_Table_Unitary_d_para=1}, which indeed
provides a $\Z_2$ index in three-dimensions. 
The $\Z_2$ topological index is given by\cite{Fang2015, Shiozaki2015}  
\begin{equation}
\begin{split}
\nu
&= \frac{i}{\pi}\int_{-\pi}^{\pi} dk_z
{\rm tr}{\cal A}_+(\pi,0,k_z) 
-\frac{i}{\pi} \int_{-\pi}^{\pi} dk_z 
{\rm tr}{\cal A}_+(\pi,\pi, k_z) \\
&- \frac{i}{2\pi} \int_{-\pi}^{\pi}dk_x\int_{-\pi}^{\pi}dk_z 
{\rm tr}{\cal F}_-(k_x,0,k_z)
+ \frac{i}{2\pi} \int_{-\pi}^{\pi} dk_x \int_{-\pi}^{\pi}dk_z
{\rm tr}{\cal F}_-(k_x, \pi, k_z) \\
&+ \frac{i}{2\pi} \int_{0}^{\pi} dk_y \int_{-\pi}^{\pi} dk_z
{\rm tr}{\cal F}(\pi, k_y, k_z) \ \ ({\rm mod\ } 2), 
\end{split}
\end{equation}
where ${\cal A}$ and ${\cal F}$ are the Berry connection and
curvature of the occupied states, and the the suffix $\pm$ indicates
those in the eigensector of the glide operator $G(k_x)$ with the eigenvalue
$g_{\pm}(k_x)=\pm e^{-ik_x/2}$.  

When $\nu=1$, there exists a gapless state on a surface preserving
glide symmetry.
The low-energy effective Hamiltonian of the $\Z_2$ gapless surface
states is given by\cite{Shiozaki2015}   
\begin{align}
H_{\rm surf}(k_x, k_y) = v k_y \sigma_z + f(k_x) [\cos (k_x/2)\sigma_x+\sin
 (k_x/2)\sigma_y] - \mu,
\label{Eq:Surf_Mode_Glide_A}
\end{align}
with glide symmetry 
\begin{align}
G(k_x) H(k_x,k_y) G(k_x)^{-1} = H(k_x,-k_y), \quad 
G(k_x) =\left(
\begin{array}{cc}
0 & e^{-ik_x} \\
1 & 0
\end{array}
\right).  
\end{align}
Here $f(k_x)$ is an {\it arbitrary} real function satisfying $f(k_x+2 \pi) =
-f(k_x)$. 
For example, we can take $f(k_x) = c\sin (k_x/2)$ with a real
constant $c$.
The surface state has the energy dispersion $E(k_x, k_y)=\pm \sqrt{(vk_y)^2+f(k_x)^2}-\mu$. 
By changing $c$, 
the surface state can be either a usual Dirac fermion in
Fig.\ref{Fig:ClassA_Glide_Surface}[a] 
or a surface state in Fig.\ref{Fig:ClassA_Glide_Surface}[b] which is detached from the bulk spectrum in the
$k_x$-direction. 
The $\Z_2$ nature of the surface state can be explained by the latter detached
property.\cite{Shiozaki2015}

%

\subsubsection{$\Z_4$ time-reversal invariant TCI with glide symmetry
   ($d=3$, $U$ in class AII)}
\label{Sec:Z_4ClassAIIGlide}

When TRS coexists with glide symmetry, a $\Z_4$ topological
phase is obtained in three dimensions.
Let us consider a three-dimensional time-reversal invariant insulator 
\begin{align}
TH(\bk)T^{-1} = H(- \bk), && T^2 = -1,
\end{align}
with glide symmetry,
\begin{align}
G(k_x) H(\bk) G(k_x)^{-1} = H(k_x,-k_y,k_z), && G(k_x) = \begin{pmatrix}
0 & i s_y e^{-i k_x} \\
i s_y & 0 \\
\end{pmatrix}, 
&& TG(k_x)=G(-k_x)T.
\label{eq:glide3D}
\end{align}
The symmetry belongs to $U$ in class AII. 
From Table \ref{Tab:Periodic_Table_Unitary_d_para=1}, the strong
topological index is $\Z_4$.

The $\Z_4$ topological invariant is defined as follows. 
On the $k_y = 0, \pi$ planes in three dimensional Brillouin zone, the
Hamiltonian $H(\bk)$ commutes with the
glide operator $G(k_x)$, so the occupied states are divided into two
sectors according to the eigenvalues $g_{\pm}(k_x)=\pm i e^{-ik_x/2}$ 
of $G(k_x)$.  
We introduce the Berry connection ${\cal A}_+$ and the Berry curvature
${\cal F}_+$ of occupied states in the positive glide sector with $g_+(k_x) = i e^{-i k_x/2}$.  
Furthermore, on the $(k_x,k_y) = (\pi, 0), (\pi,\pi)$ lines,
$g_{+}(k_x)$ is real, and $[G(k_x), T]=0$. Thus TRS maps the positive
glide sector to itself, which implies that the positive glide sector has
its own TRS.
From the Kramers theorem, on the $(k_x,k_y) = (\pi, 0), (\pi,\pi)$
lines, occupied states in the positive glide sector
form Kramers pairs, $(|u_n^{(+)\rm I}({\bm k})\rangle, |u_n^{(+)\rm II}({\bm
k})\rangle)$. 
We introduce the Berry connection ${\cal A}^{\rm I}_+$
of the first occupied states of the Kramers pairs. 
By using these connections and curvature, 
the $\Z_4$ topological invariant $\theta$ is defined as
\begin{align}
\theta 
&:= \frac{2i}{\pi}\int_{-\pi}^{\pi} dk_z
{\rm tr}{\cal A}_+^{\rm I}(\pi,\pi,k_z) -
 \frac{2i}{\pi}\int_{-\pi}^{\pi} dk_z {\rm tr}{\cal A}_+^{\rm I}(\pi,0,k_z) 
\nonumber\\
&+ \frac{i}{\pi} \int_{0}^{\pi} dk_x \int_{-\pi}^{\pi}dk_z 
{\rm tr}{\cal F}_+(k_x, \pi, k_z) 
- \frac{i}{\pi} \int_0^{\pi} dk_x \int _{-\pi}^{\pi} dk_z 
{\rm tr}{\cal
 F}_+(k_x, 0, k_z) 
\nonumber\\
&- \frac{i}{2\pi} \int_0^{\pi} dk_y \int_{-\pi}^{\pi} dk_z {\rm tr}{\cal F}(0, k_y, k_z) \ \ ({\rm mod\ } 4), 
\end{align}
where ${\cal F}$ is the Berry curvature of the occupied states. 
The modulo-4 ambiguity in the above equation originates from the $U(1)$
gauge ambiguity of ${\cal
A}_{+}^{I}$.
Below, we prove that $\theta$ is indeed quantized so as to define a
$\Z_4$ index.
Using the Stokes' theorem, we first rewrite the third term of the right hand
side as
\begin{align}
\frac{i}{\pi} \int_{0}^{\pi} dk_x \int_{-\pi}^{\pi}dk_z 
{\rm tr}{\cal F}_+(k_x, \pi, k_z) 
&=-\frac{i}{\pi}\int_{-\pi}^{\pi} dk_z{\rm tr}{\cal
 A}_+(\pi,\pi,k_z)
+\frac{i}{\pi}\int_{-\pi}^{\pi} dk_z{\rm tr}{\cal A}_+(0,\pi,k_z) \ \ ({\rm
 mod\ } 2).
\label{eq:third}
\end{align}
As mentioned above, on the $(k_x,k_y)=(\pi,\pi)$ line, the positive glide
sector has its own TRS, so the first term of the right hand side in
Eq.(\ref{eq:third}) is recast into
\begin{align}
-\frac{i}{\pi}\int_{-\pi}^{\pi} dk_z{\rm tr}{\cal
 A}_+(\pi,\pi,k_z)= -\frac{2i}{\pi}\int_{-\pi}^{\pi} dk_z{\rm tr}{\cal
 A}_+^{\rm I}(\pi,\pi,k_z),
\ \ ({\rm mod\ } 2).
\end{align}
On the other hand, on the $(k_x, k_y)=(0, \pi)$ line, because
$g_{\pm}(k_x=0)=\pm i$, the glide subsectors are exchanged by TRS.
Thus, an occupied state $|u_n^{(+)}(0,\pi, k_z)\rangle$ in the
positive glide sector forms a Kramers pair with an occupied state
$|u_n^{(-)}(0,\pi, k_z)\rangle$ in the negative glide sector. 
By denoting the Kramers pair as $(|u_n^{(+)}(0,\pi,k_z)\rangle,
|u_n^{(-)}(0,\pi,k_z)\rangle)=(|u_n^{\rm I}(0,\pi,k_z)\rangle, |u_n^{\rm
II}(0,\pi,k_z)\rangle)$, 
the second
term in Eq.(\ref{eq:third}) becomes
\begin{align}
\frac{i}{\pi}\int_{-\pi}^{\pi} dk_z{\rm tr}{\cal
 A}_+(0,\pi,k_z)= \frac{i}{\pi}\int_{-\pi}^{\pi} dk_z{\rm tr}{\cal
 A}^{\rm I}(0,\pi,k_z),
\ \ ({\rm mod\ } 2),
\end{align}
with the Berry connection ${\cal A}^{\rm I}$ of $|u_n^{\rm I}\rangle$.
As a result, we obtain
\begin{align}
\frac{i}{\pi} \int_{0}^{\pi} dk_x \int_{-\pi}^{\pi}dk_z 
{\rm tr}{\cal F}_+(k_x, \pi, k_z) 
=-\frac{2i}{\pi}\int_{-\pi}^{\pi} dk_z{\rm tr}{\cal
 A}_+^{\rm I}(\pi,\pi,k_z)
+\frac{i}{\pi}\int_{-\pi}^{\pi} dk_z{\rm tr}{\cal A}^{\rm I}(0,\pi,k_z)
\ \ ({\rm mod\ } 2).
\end{align}
We can also rewrite the fourth term in $\theta$ in a similar form.
Using these relations, we find that $\theta$ obeys
\begin{align}
\theta =\frac{i}{\pi} \int_{-\pi}^{\pi}dk_z{\rm tr}{\cal A}^{\rm I}(0,\pi,k_z) 
- \frac{i}{\pi}\int_{-\pi}^{\pi} dk_z {\rm tr}{\cal A}^{\rm I}(0,0,k_z) 
- \frac{i}{2\pi} \int_0^{\pi} dk_y \int_{-\pi}^{\pi}dk_z 
{\rm tr}{\cal F}(0,k_y,k_z) \ \ ({\rm mod\ } 2). 
\label{eq:thetaKaneMele}
\end{align}
The right hand side of the above equation is nothing
but the Kane-Mele's $\Z_2$-invariant on the $k_x = 0$ plane,\cite{Fu2006}
so $\theta$ is quantized to be an integer.
By taking into account the modulo-4 ambiguity in the definition of
$\theta$,  $\theta$ takes only four distinct values, $\theta=0, 1, 2,
3$ (mod 4), which implies that $\theta$ defines a $\Z_4$ index.

\begin{figure}[!]
 \begin{center}
  \includegraphics[width=0.7\linewidth, trim=0cm 0cm 0cm 0cm]{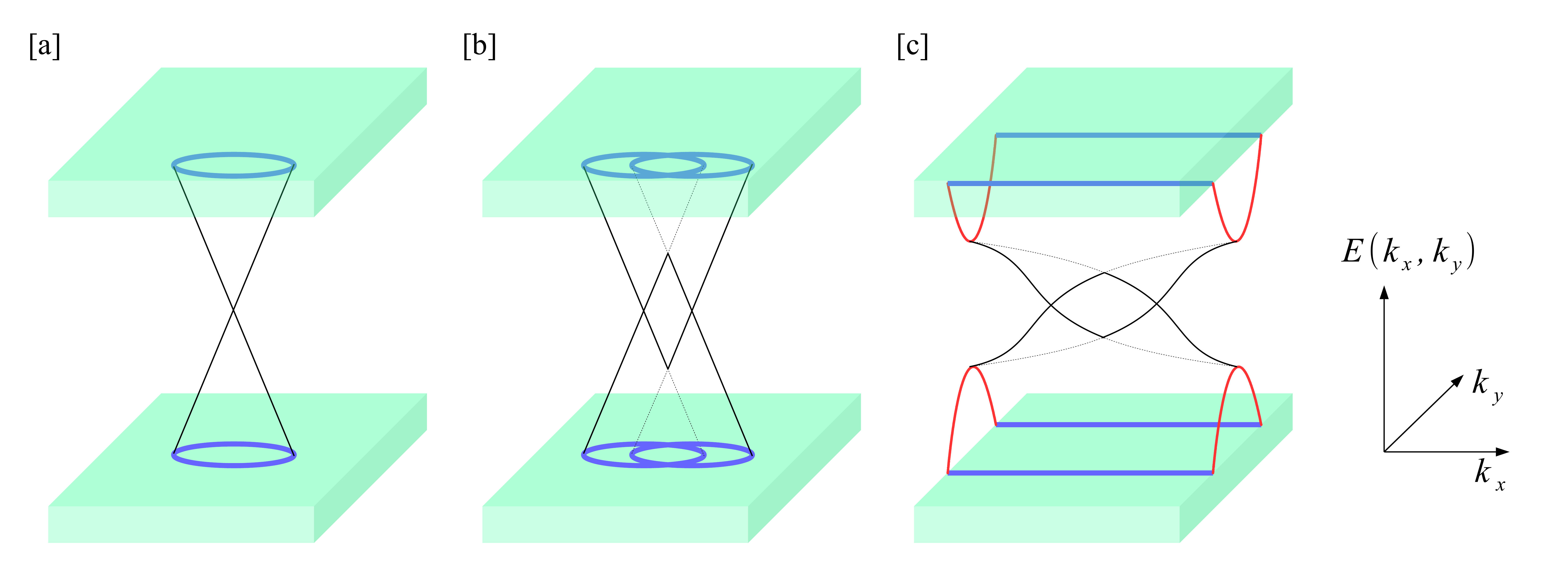}
 \end{center}
 \caption{(Color online) Schematic pictures of $\Z_4$ nontrivial surface
 states with glide symmetry ($d_{\parallel}=1$, $d=3$, $U$ in class AII). 
 The green volumes represent the bulk spectrum. 
 [a] $\Z_4 = 1$ ($\theta=1$) phase with single a surface Dirac
 fermion. 
 [b] and [c] $\Z_4 = 2$ ($\theta=2$) phase. Two Dirac fermions ([b])
 can be adiabatically transformed into a detached surface state ([c]). 
 In [c], the red curves are identified at the zone boundary.}
 \label{Fig:3DAII_Glide}
\end{figure}
\begin{figure}[!]
 \begin{center}
  \includegraphics[width=\linewidth, trim=0cm 0cm 0cm 0cm]{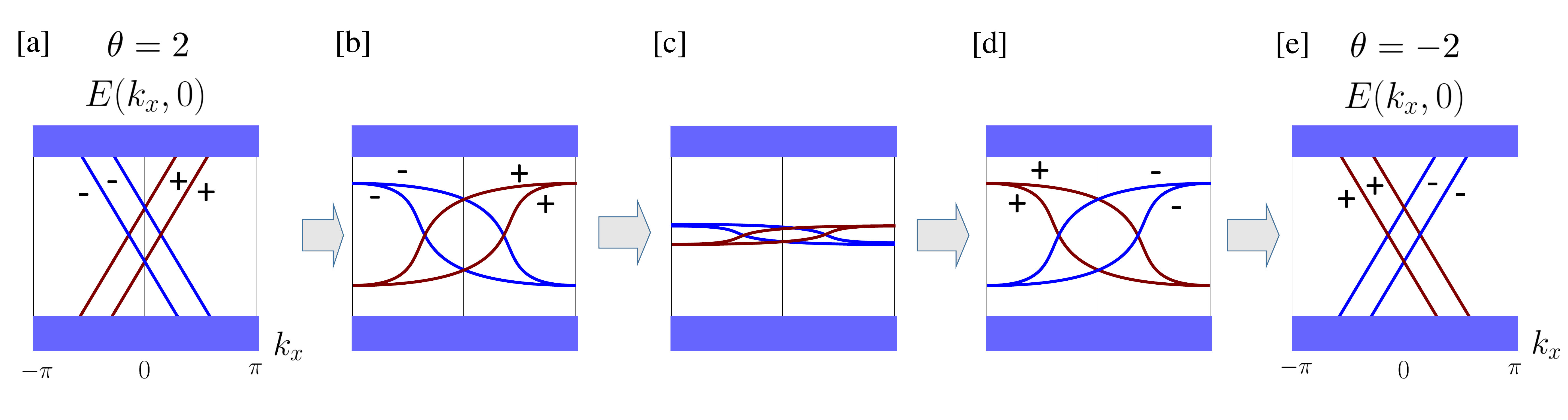}
 \end{center}
 \caption{(Color online) Adiabatic deformation of gapless surface states
 in the $\Z_4=2$ ($\theta=2$) phase. 
The figures show the energy spectra at the glide invariant line $k_y =
 0$. Red (+) and blue (-) lines show surface states in the
 $g_{\pm}(k_x)=\pm ie^{-ik_x/2}$ sectors.}
 \label{Fig:ClassAII_Glide_Surface}
\end{figure}

The $\Z_4$ classification is also confirmed by examining surface gapless
states.
Consider a surface normal to the $z$-direction, which keeps the glide
symmetry in the above. 
When $\theta=1$, we have a two-dimensional surface helical Dirac
fermion.
See Fig.\ref{Fig:3DAII_Glide}[a].
The surface Dirac fermion is very similar to that in an ordinary time-reversal
invariant topological insulator, but there is a constraint from
glide symmetry:
From Eq.(\ref{eq:thetaKaneMele}), the Kane-Mele's $\Z_2$ invariant at $k_x=0$ is
nontrivial when $\theta=1$, so the Dirac point of the surface state is
pinned at either $(k_x,
k_y)=(0,0)$ or $(0, \pi)$.  
Below we suppose the former case without loss of generality. 
For $\theta=2$, the surface Dirac fermion is doubled, as shown in
Fig.\ref{Fig:3DAII_Glide}[b].
In contrast to ordinary topological insulators, the pair of Dirac
fermions are stable due to glide symmetry:
On the $k_y=0$ line (i.e. the $k_x$-axis) in the surface Brillouin zone, 
the glide operator commutes with the Hamiltonian,
so each branch of the surface modes has a definite glide
eigenvalue as illustrated in Fig.\ref{Fig:ClassAII_Glide_Surface}[a].  
Due to TRS, the counter propagating modes has an opposite glide
eigenvalue.
Since band mixing between different glide sectors is prohibited, 
no gap opens for the paired Dirac fermions.
Now consider a deformation process of
the paired Dirac fermions from Figs.\ref{Fig:ClassAII_Glide_Surface}[a] to [e].
By widen the angle of the Dirac points, the surface Dirac fermions in
Fig.\ref{Fig:ClassAII_Glide_Surface}[a]  reach the
zone boundary at $k_x=\pm \pi$ in the $k_x$-axis.
In this situation, 
the positive glide sector of the Dirac fermions can merge with the negative one
at the zone boundary because the eigenvalues of these sectors
coincide, i.e. $g_+(\pi)=g_-(-\pi)$. 
Each merged branch should be doubly degenerate at the zone boundary
due to TRS. 
As illustrated in Fig.\ref{Fig:ClassAII_Glide_Surface}[b], 
the merged Dirac fermions can be detached from the bulk spectrum in
the $k_y=0$ line.
Importantly, the detached surface modes can be deformed from
Fig.\ref{Fig:ClassAII_Glide_Surface}[b] to [d].
Then, finally, the surface modes in Fig.\ref{Fig:ClassAII_Glide_Surface}[d] become
the paired Dirac fermions in Fig.\ref{Fig:ClassAII_Glide_Surface}[e] by 
reversing a process from [a] to [c].
%
%
In comparison with the initial Dirac fermions in
 Fig.\ref{Fig:ClassAII_Glide_Surface}[a],  
the final Dirac fermions in Fig.\ref{Fig:ClassAII_Glide_Surface}[e] have
opposite glide eigenvalues, so they  can be regarded as Dirac fermions in the
 $\theta=-2$ phase. 
As a result, the surface states satisfies the identity $2=-2$ in
 $\theta$, which is nothing but the 
the $\Z_4$ nature of $\theta$.


A bulk model Hamiltonian $H_{1}(\bk)$ for $\theta=1$ is given by 
\begin{align}
H_{1}(\bk) 
&= (1/2)\left[
\sin k_x \sigma_x +(1-\cos k_x)\sigma_y\right]
s_y \mu_x + \sin k_y s_x \mu_x + \sin k_z
 \mu_y 
\nonumber\\
&+ \left[
m+(1/2)(1+\cos k_x )\sigma_x+(1/2)\sin k_x \sigma_y 
+ \cos k_y + \cos k_z\right] \mu_z 
\quad (-3 <
 m < -1),
\end{align}
where $s_\mu$, $\mu_\mu$ and
$\sigma_\mu$ 
are the Pauli matrices
in the spin and two different internal orbital spaces, respectively.
The glide operator is given by
\begin{align}
G(k_x)=i\left(
\begin{array}{cc}
0 & e^{-ik_x}\\
1 & 0
\end{array}
\right)_\sigma s_y.
\end{align}
Among the two orbital spaces, the glide reflection interchanges only the
orbital degrees of freedom in the $\sigma_\mu$ space. The glide
reflection also changes the spin direction in a manner similar to the
mirror reflection with respect to the $zx$-plane.
$H_{1}(\bk)$ supports a surface helical Dirac fermion when $-3<m<-1$.

We also find that a model Hamiltonian $H_{2}(\bk)$ for $\theta=2$
can be obtained by stacking two-dimensional time-reversal invariant
topological insulators: Consider a set of two-dimensional time-reversal
topological insulators, 
\begin{align}
H_{\rm 2d}(k_y, k_z) 
=\left[ \sin k_y s_x \mu_x +\sin k_z \mu_y+ (m+ \cos k_y + \cos k_z)
 \mu_z \right] \otimes \sigma_0 \ \ (-2<m <0 ),
\end{align}
with the glide symmetry,
\begin{align}
G(k_x)H_{\rm 2d}(k_y, k_z)G(k_x)^{-1}=H_{\rm 2d}(-k_y, k_z), 
\quad 
G(k_x)=i\left(
\begin{array}{cc}
0 & e^{-ik_x}\\
1 & 0
\end{array}
\right)_\sigma s_y. 
\end{align}
Just stacking them in the $x$-direction with neglecting the interlayer
coupling, we can obtain $H_{2}(\bk)$ as
$H_2(\bk)=H_{\rm 2d}(k_y, k_z)$. 
Since $H_2(\bk)$ is $k_x$-independent, the corresponding surface state is
completely flat in the $k_x$-direction. 
A surface state in Fig.\ref{Fig:3DAII_Glide}[c] 
is realized
by adding a proper symmetry preserving perturbation to $H_2(\bk)$.

\subsection{M\"{o}bius twist in surface states}

A remarkable feature of nonsymmorphic topological phases is possible detached
spectra of surface states. A detached surface state also has
a unique topological structure, i.e. M\"{o}bius twist
structure,\cite{Shiozaki2015} if it is an eigen state of order-two NSG
operator $G(k_x)$.
This property originates from that the eigenvalue of $G(k_x)$
does not have the $2\pi$-periodicity in $k_x$ in spite that the operator $G(k_x)$
itself does. 

For instance, let us consider the glide operator $G(k_x)$ in
Eq.(\ref{eq:glide3D}).
The glide operator has the correct $2\pi$-periodicity in $k_x$, 
but its eigenvalues $g_{\pm}(k_x)=\pm i e^{-ik_x/2}$ do not.
These eigenvalues are interchanged when $k_x$ is changed by $2\pi$.
Therefore, if a surface state is detached from the bulk spectrum in the
$k_x$ direction, a branch 
of the surface state in the $g_{+}(k_x)=ie^{-ik_x/2}$ sector is
smoothly connected to a branch in the $g_{-}(k_x)=e^{-ik_x/2}$ sector
in the $k_x$-direction.
The above structure results in M\"obius twist in the surface state 
since the $k_x$-space can be regarded as a one-dimensional circle by
identifying $k_x=-\pi$ with $k_x=\pi$: Two different sectors are
interchanged when one goes round the circle in the $k_x$ direction, as
is seen in Figs.\ref{Fig:ClassAII_Glide_Surface}[b] and [c]. 
Similar M\"{o}bius twist structures can be seen in the
surface states in Figs. \ref{Fig:2D_D_Glide}[c],
\ref{Fig:2D_D_Glide}[f], \ref{Fig:2D_DIII_Glide}[b], 
\ref{Fig:2D_DIII_Glide}[c] and \ref{Fig:2D_DIII_Glide_even}[c].

\section{Conclusion}
\label{Sec:Conc}
In this paper, on the basis of the twisted equivariant $K$-theory, we
present the
classification of TCIs and TCSCs that
support order-two NSG symmetries, besides AZ symmetries.
With the classification of TCIs and TCSCs protected by order-two
point group symmetries in Ref.\onlinecite{Shiozaki2014}, 
this work completes the classification of TCIs and TCSCs in the presence of
order-two space groups.
The obtained results for strong topological phases are summarized
in Tables \ref{Tab:Periodic_Table_Unitary_d_para=0} and
\ref{Tab:Periodic_Table_Unitary_d_para=1}.
Various nonsymmorphic TCIs and TCSCs and their surface
states are identified in a
unified framework.
In particular, we discover several new $\Z_2$ and $\Z_4$ phases that support 
unique boundary states specific to nonsymmorphic TCIs and
TCSCs. We also provide analytic expressions of the corresponding $\Z_2$ and
$\Z_4$ topological invariants.

In the present work, we concentrate on the classification of  full
gapped TCIs and TCSCs. 
However, NSGs also may
stabilize bulk topological gapless modes in semimetals and nodal
superconductors.  
Our $K$-theory framework presented in this paper is applicable to such
gapless systems. 
Whereas stability of the bulk gapless modes can be examined by local
topology around the gapless nodes in the momentum space, our $K$-theory
approach can reveal global topological constraints for the node
structure at the same time.


\vspace{5ex}

{\it Note added}. After the submission of the present work, there
appeared an independent and complementary work \cite{Wang2016} proposing
material realization of the M\"{o}bius twisted surface state in
Sec.\ref{Sec:Z_4ClassAIIGlide}. \footnote{A different and  interesting approach to nonsymmorphic topological phases was proposed in a very recent paper \cite{Aris}. This paper also claimed that  $D^4_{6h}$ of the material in Ref.\onlinecite{Wang2016} falls outside our $K$ theoretic classification. However, the discrepancy pointed out is superficial, and it is resolved by taking a proper unit cell which is consistent with the surface considered. Therefore, $D^4_{6h}$ is consistent with our result.}
Another material realization of the M\"{o}bius twisted surface
state was also proposed very recently\cite{Chang2016}. 

\begin{acknowledgments}
K.S. thanks to useful discussions with S. Fujimoto, S. Kobayashi,
 C. -X. Liu, T. Nomoto and Y. Yanase. 
K.S. is supported by a JSPS Postdoctoral Fellowship for Research Abroad. 
This work was supported
by JSPS (Nos. 25287085 and 15K04871) and Topological Quantum Phenomena
(No. 22103005) and Topological Materials Science
(No. 15H05855) KAKENHI on innovation areas from MEXT.
\end{acknowledgments}

\appendix

\section{Periodic Bloch basis in Brillouin zone}
\label{sec:Periodic_Bloch}

In this section, we summarize the relation between the L\"{o}wdin orbitals and 
the ``periodic'' Bloch basis in tight-binding models for crystalline materials. 
Let $\ket{\bm{R}, \alpha, i}$ be a one-particle basis in real space, where 
$\bm{R} \in \Z^d$ is an element of a Bravais lattice, 
$\alpha$ is the label for localized atoms, and $i$ is the label for
internal degrees of
freedoms inside of the atoms $\alpha$. 
The L\"{o}wdin orbital $\ket{\bk,\alpha,i}$ is defined with retaining
the localized positions of
$\alpha$-atoms as 
\begin{align}
\ket{\bk,\alpha,i} := \sum_{\bm{R}} \ket{\bm{R}, \alpha,i} e^{i \bk \cdot (\bm{R}+\bm{x}_{\alpha})}, 
\end{align}
where $\bm{x}_{\alpha}$ is the localized position of $\alpha$-atom measured by the center of unit cell, says, $\bm{R}$. 
An important point in using the L\"{o}wdin orbitals is 
the non-periodicity in the Brillouin zone torus: 
Let $\bm{G}$ is a reciprocal lattice vector. 
Then, the L\"{o}wdin orbital is transformed as 
\begin{align}
\ket{\bk + \bm{G},\alpha,i} = \ket{\bk,\alpha,i} e^{i \bm{G} \cdot \bm{x}_{\alpha}}.
\end{align} 
Usually, we have to employ the L\"owdin orbitals for crystalline
materials in which there are multiple atoms in a unit cells, not to  
miss the localized positions of atoms. 
However, for topological classification, the ``periodic'' Bloch basis
$\ket{\bk,\alpha,i}_{\rm P}$ is useful.
$\ket{\bk,\alpha,i}_{\rm P}$ is defined as 
\begin{align}
\ket{\bk,\alpha,i}_{\rm P} := \sum_{\bm{R}} \ket{\bm{R}, \alpha,i} e^{i
 \bk \cdot \bm{R}}. 
\end{align}
It is obvious that $\ket{\bk,\alpha,i}_{\rm P}$ is periodic under the
shift by the reciprocal lattice vector,  
\begin{align}
\ket{\bk + \bm{G},\alpha,i}_{\rm P} = \ket{\bk,\alpha,i}_{\rm P}. 
\end{align}
Using the L\"owdin orbital and the periodic Bloch basis does not changes
the topological classifications,  
since the both are related by the $\bk$-dependent unitary transformation, 
\begin{align} 
\ket{\bk,\alpha,i}_{\rm P} = \sum_{\beta, j} \ket{\bk,\beta,j} V({\bm
 k})_{\beta j, \alpha i},
\quad V(\bk)_{\alpha i, \beta j}
=e^{-i \bk \cdot \bm{x}_{\alpha}}\delta_{\alpha\beta}\delta_{i,j},  
\end{align}
which means the shift of localized positions of atoms into the center of
unit cell $\bm{R}$.  
From the above unitary transformation,
the Hamiltonian for the L\"owdin orbitals,
$
H(\bk)_{\alpha i, \beta j}=\langle \bk, \alpha, i|H|\bk, \beta, j\rangle, 
$
is related to that in the periodic Bloch basis, 
$
H_{\rm P}(\bk)_{\alpha i, \beta j}={}_{\rm P}\langle \bk, \alpha,
 i|H|\bk, \beta, j \rangle_{\rm P} 
$
as 
\begin{align} 
H_{\rm P}(\bk)_{\alpha i, \beta j}
=\left[
V({\bm k})^{\dagger}H(\bk)
V({\bm k})
\right]_{\alpha i, \beta j}
=e^{i{\bm k}\cdot{\bm x}_{\alpha}}
H(\bk)_{\alpha i, \beta j}e^{-i{\bm k}\cdot{\bm x}_{\beta}}.
\end{align}
For crystalline materials, 
an element of space group $\sigma=\{p|{\bm a}\}$ inducing the
map ${\bm x}\rightarrow p{\bm x}+{\bm a}$ acts on the L\"owdin orbitals as
\begin{align}
U_\sigma(\bk)_{\alpha i,\beta j}=e^{-ip{\bm k}\cdot{\bm a}}U(p)_{\alpha
 i, \beta j}, 
\end{align}
where $U(p)$ is a unitary representation of $p$ on the internal space
$(\alpha, i)$.
In the periodic Bloch basis, the group action is given as
\begin{align}
U^{\rm P}_{\sigma}(\bk)_{\alpha i,\beta j}
=\left[
V(p{\bm k})^{\dagger}U_{\sigma}(\bk)V(\bk)
\right]_{\alpha i,\beta j}
=
e^{-ip{\bm k}\cdot{\bm \Delta}_{\alpha\beta}(p)}
U(p)_{\alpha
 i, \beta j}, 
\end{align}
with ${\bm \Delta}_{\alpha\beta}(p)=p{\bm x}_\beta-{\bm x}_\alpha+{\bm a}$.
In order for $\{p|{\bm a}\}$ to be an element of space group, ${\bm
\Delta}_{\alpha\beta}(p)$ must be an element of the Bravais lattice.
Thus, $U_{\sigma}^{\rm P}(\bk)$ is periodic in the Brillouin zone. 
An advantage in using the periodic Bloch basis is the periodicity of
Hamiltonian $H_{\rm P}(\bk)$ and group action $U^{\rm P}_{\sigma}(\bk)$
in the Brillouin zone.
Our twisted $K$-theory classification in Secs.\
\ref{Appendix:TEK-theory}-\ref{Appendix:F} is based on the periodic
Bloch basis.

\section{Dimensional reduction}
\label{sec:Dimensional reduction}
Using the technique used in Ref.\onlinecite{Shiozaki2014}, we can shift
the dimension of the base space for the $K$-group, except in the
$k_1(\equiv k_x)$-direction where a twist exists due to order-two NSGs. 
%
(Mathematically, the same result is obtained by the Gysin exact
sequence in the twisted equivariant $K$-theory.) 
For instance, consider the $K$-group $K_{\R}^{(s,t)}(\widetilde
S_x^1\times S_y^1\times \widebar S_z^1)$ for three-dimensional
nonsymmorphic TCIs/TCSCs in
real AZ classes. (For the definition of $K_{\R}^{(s,t)}(\widetilde
S_x^1\times S_y^1\times \widebar S_z^1)$, 
see Sec.\ref{sec:NMCI_Real}.)   
When we collapse the $k_y$-direction on which the NSG
acts trivially, we have 
\begin{align}
K_{\R}^{(s,t)}(\widetilde S_x^1 \times S^1_y\times \widebar S_z^1)
\cong K_{\R}^{(s,t)}(\widetilde S^1_x \times \widebar S^1_z) \oplus
 K_{\R}^{(s-1,t)}(\widetilde S^1_x \times \widebar S^1_z).
\label{Gysin_1}
\end{align}
The former part in the right hand side comes from a weak index
independent of $k_y$, so no shift of the symmetry indices
$(s,t)$ happens.  
On the other hand, the latter part gives a strong index, in which the
dimensional reduction shifts the AZ
symmetry as $s\rightarrow s-1$.\cite{Kitaev2009, Teo2010} 
We can also collapse the $k_z$-direction on which the NSG acts nontrivially.
In this case, we have
\begin{align}
K_{\R}^{(s,t)}(\widetilde S_x^1 \times S^1_y\times \widebar S_z^1)
\cong K_{\R}^{(s,t)}(\widetilde S^1_x \times S^1_y) \oplus
 K_{\R}^{(s-1,t-1)}(\widetilde S^1_x \times  S^1_y).
\label{Gysin_2}
\end{align}
Here the former part of the right hand side is a weak index
independent of $k_z$, so there is no symmetry shift again.
However, in the latter part, both $s$ and $t$ are shifted since both AZ
and NSG symmetries act on $k_z$ nontrivially.\cite{Shiozaki2014}

Using Eqs.(\ref{Gysin_1}) and (\ref{Gysin_2}) iteratively, we obtain
\begin{align}
K_{\R}^{(s,t)}(\widetilde S_x^1 \times S^1_y\times \widebar S_z^1)
\cong K_{\R}^{(s,t)}(\widetilde S^1_x) \oplus
 K_{\R}^{(s-1,t-1)}(\widetilde S^1_x) \oplus
 K_{\R}^{(s-1,t)}(\widetilde S^1_x) \oplus
 K_{\R}^{(s-2,t-1)}(\widetilde S^1_x).
\label{Gysin_3}
\end{align}
Here only the last part gives a strong index in three-dimensions, and 
the others are weak indices in three-dimensions.
From (\ref{Gysin_3}), the evaluation of the $K$-group in three-dimensions
reduces to that in one-dimension.

%
%

\section{Dimension-raising map}
\label{Sec:Dim_Raise_Map}

The isomorphisms in Eqs.(\ref{eq:KUcomplex1})-(\ref{eq:KUcomplex2}),
(\ref{eq:KAcomplex1})-(\ref{eq:KAcomplex2}) and
(\ref{eq:Kreal1})-(\ref{eq:Kreal2})
imply that all representative Hamiltonians with order-two NSGs in
$d$-dimensions can be 
obtained from one-dimensional Hamiltonians $H(k_x)$ over $\widetilde S^1$.
In this section, we explain how to construct a higher dimensional
system from a lower dimensional one.


\begin{figure}[!]
 \begin{center}
  \includegraphics[width=\linewidth, trim=0cm 0cm 0cm 0cm]{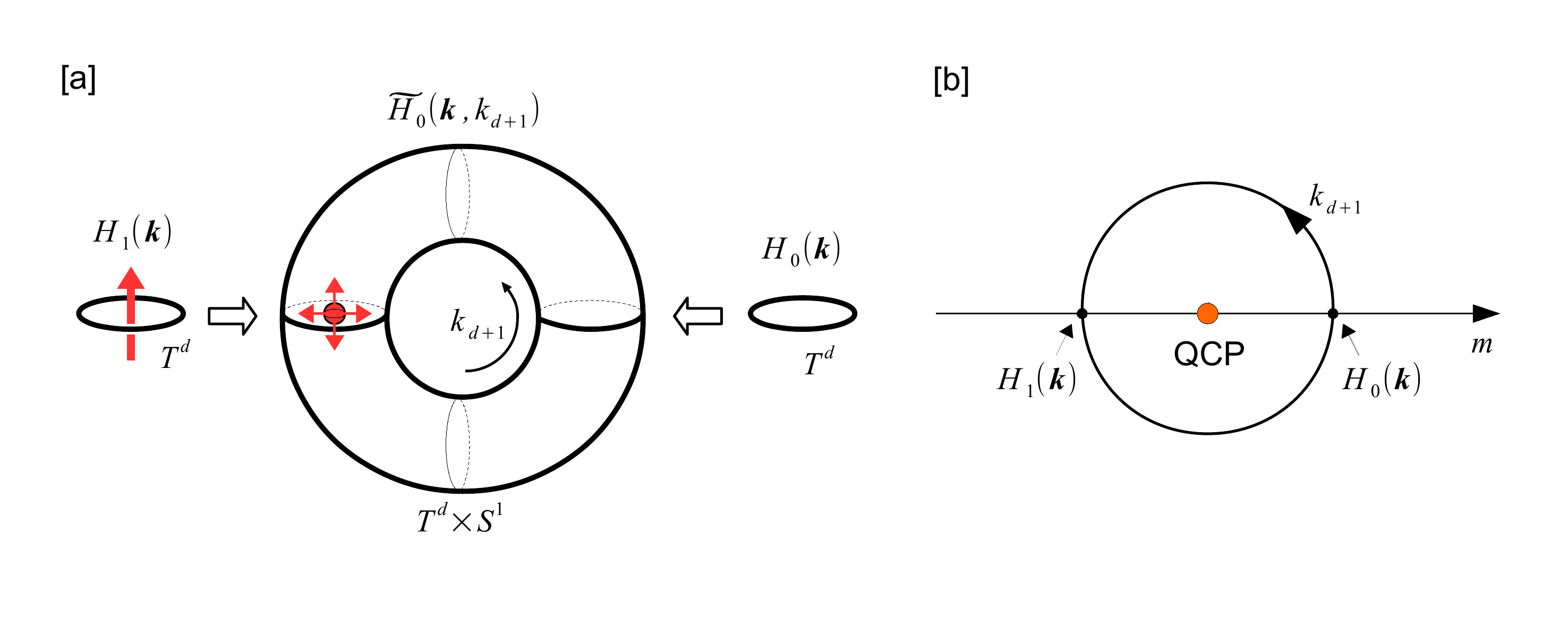}
 \end{center}
 \caption{(Color online) Hamiltonian raising map from $T^d$ to $T^d \times S^1$.}
 \label{Fig:HamiltonianMap}
\end{figure}

For concreteness, we consider TCIs with unitary order-two NSGs in complex
AZ classes.
The isomorphisms in Eqs.(\ref{eq:KUcomplex1}) and (\ref{eq:KUcomplex2})
in this case 
take the following forms 
\begin{align}
&K_{\C}^{(s+1,t)}(T^d \times S^1) \cong 
\underbrace{K_{\C}^{(s+1,t)}(T^d)}_{\rm weak} \oplus 
\underbrace{K_{\C}^{(s,t)}(T^d)}_{\rm strong}, 
\label{eq:DRM1}
\\
&K_{\C}^{(s+1,t+1)}(T^d \times \widebar S^1) \cong 
\underbrace{K_{\C}^{(s+1,t+1)}(T^d)}_{\rm weak} \oplus 
\underbrace{K_{\C}^{(s,t)}(T^d)}_{\rm strong}, 
\label{eq:DRM2}
\end{align}
where $S^1\ni k_{d+1}$ ($\widebar S^1\ni k_{d+1}$) represents a
one-dimensional circle on which
the NSG acts trivially as $k_{d+1}\mapsto k_{d+1}$ (nontrivially as
$k_{d+1} \mapsto -k_{d+1}$).  
These relations suggest that there are two different ways to construct a
$(d+1)$-dimensional system from a $d$-dimensional one.  
The first one is rather trivial:
Corresponding to the first terms of the isomorphisms, 
one can obtain a $(d+1)$-dimensional weak topological system by
just considering a $d$-dimensional Hamiltonian as a
$(d+1)$-dimensional but $k_{d+1}$-independent one.
The obtained Hamiltonian has the same symmetry as the original
$d$-dimensional one.

A $(d+1)$-dimensional strong topological system can be constructed in the
second way. 
Let $(H_0(\bk), H_1(\bk))$ be a set of representative Hamiltonians for the $K$-group $K_{\C}^{(s,t)}(T^d)$ on the $d$-dimensional torus $T^d$. 
They satisfy
\begin{align}
&U(\bk) H_i(\bk) U(\bk)^{-1} = (-1)^t H_i(\sigma \bk), 
\quad &&\mbox{(for $s=0$)}, \\
& U(\bk) H_i(\bk) U(\bk)^{-1} = H_i(\sigma \bk), \quad
\{\Gamma, H_i(\bk) \}=0, 
\quad
U(\bk) \Gamma = (-1)^t \Gamma U(\bk),
\quad 
&&\mbox{(for $s=1$)},
\end{align}
with $U(\sigma \bk) U(\bk) = e^{-i k_x}$.
Without lost of generality, we can take $H_0({\bm k})$ as a
topologically trivial reference Hamiltonian.
A $(d+1)$-dimensional system is constructed from the $d$-dimensional
Hamiltonian $H_1(\bk)$ in the following manner.\cite{Shiozaki2015} 

First, let $H(\bk,m)$ be a $d$-dimensional Hamiltonian that interpolates
$H_0(\bk)$ and $H_1({\bk})$ as
$H(\bk,1) = H_0(\bk)$ and $H(\bk,-1) = H_1(\bk)$. 
We also assume that $H(\bk, m)$ keeps the same
symmetry as $H_0({\bm k})$ and $H_1({\bm k})$.
In general, $H_1({\bm k})$ is topologically non-trivial, so
there is a gapless point in $H(\bk, m)$ between $m=1$ and $m=-1$, as
shown in Fig.\ \ref{Fig:HamiltonianMap}[b].
From $H(\bk, m)$, a $(d+1)$-dimensional Hamiltonian $\widetilde
H(\bk,k_{d+1})$ is constructed as 
\begin{align}
&\widetilde H(\bk,k_{d+1}) := H(\bk,\cos (k_{d+1})) \otimes \tau_z +
 \sin (k_{d+1}) \tau_x, \ \   && 
\mbox{(for $s=0$)}, 
\label{Eq:Dim_Raise_H_Class_A} \\
&\widetilde H(\bk,k_{d+1}) := H(\bk,\cos (k_{d+1})) + \sin (k_{d+1}) \Gamma, 
&& \mbox{(for $s=1$)}, 
\label{Eq:Dim_Raise_H_Class_AIII} 
\end{align}
which is fully gapped: $\widetilde H(\bk,k_{d+1})^2 =
H(\bk,\cos (k_{d+1}))^2 + \sin^2 (k_{d+1}) > 0$.   
When $k_{d+1}=0$ and $k_{d+1}=\pi$, the obtained Hamiltonian is
essentially the same as $H_0(\bk)$ and $H_1(\bk)$, respectively.
Then, keeping gaps of the systems, 
$H_0(\bk)$ and $H_1(\bk)$ are extended in the $k_{d+1}$-direction 
and they are glued together at $k_{d+1}=\pm \pi/2$.
See Fig.\ref{Fig:HamiltonianMap}[a]. 
If $H_1(\bk)$ is topologically nontrivial, there is a source of
the topological charge (``vortex'') inside the $d$-dimensional torus $T^d$.
In the above construction, the vortex becomes a ``monopole'' inside the
$(d+1)$-dimensional torus $T^d\times S^1$. 
Therefore, $\widetilde H(\bk, k_{d+1})$ has the same topological number
as $H_1(\bk)$.

Note that the $\widetilde H(\bk, k_{d+1})$ has a different symmetry than
$H_1(\bm k)$ has:
For $s=0$, $\widetilde H(\bk, k_{d+1})$ has CS with
$\Gamma=\tau_y$, so $s$ becomes $1$. On the other hand, for $s=1$, $\widetilde H(\bk,
k_{d+1})$ loses CS, so $s$ becomes zero. 
Therefore, the dimension-raising maps in Eqs.(\ref{Eq:Dim_Raise_H_Class_A})
and (\ref{Eq:Dim_Raise_H_Class_AIII}) shift $s$ as $s \mapsto s+1$ (mod 2). 
For each $(s,t)$, $\widetilde H(\bk, k_{d+1})$ has two different
order-two NSGs. 
The first one does not flip $k_{d+1}$, so the order-two NSG acts as
$(\bk,k_{d+1}) \mapsto (\sigma \bk, k_{d+1})$.  
With this action for the momentum space, 
$\widetilde H(\bk, k_{d+1})$ has the $(s+1, t)$ order-two NSG symmetry
with unitary operator,
\begin{align}
\widetilde U_0(\bk,k_{d+1}) = 
\left\{ \begin{array}{ll}
U(\bk) \otimes \tau_0, & \mbox{for $(s,t)=(0,0)$} \\
U(\bk) \otimes \tau_x, & \mbox{for $(s,t)=(0,1)$} \\
U(\bk), & \mbox{for $(s,t)=(1,0)$} \\
\Gamma U(\bk), & \mbox{for $(s,t)=(1,1)$} \\
\end{array}\right., 
\end{align}
The second one flips $k_{d+1}$, so it acts as $(\bk,k_{d+1}) \mapsto
(\sigma \bk, -k_{d+1})$. With this action for the momentum space,   
$\widetilde H(\bk, k_{d+1})$ has the $(s+1, t+1)$ order-two NSG symmetry
with unitary operator,
\begin{align}
\widetilde U_1(\bk,k_{d+1}) = 
\left\{ \begin{array}{ll}
U(\bk) \otimes \tau_z, & \mbox{for $(s,t)=(0,0)$} \\
U(\bk) \otimes \tau_y, & \mbox{for $(s,t)=(0,1)$} \\
\Gamma U(\bk), & \mbox{for $(s,t)=(1,0)$} \\
U(\bk), & \mbox{for $(s,t)=(1,1)$} \\
\end{array}\right. , 
\end{align}
The first NSG defines the dimension-raising map from
$K_{\C}^{(s,t)}(T^d)$ to $K_{\C}^{(s+1,t)}(T^d\times S^1)$ in
Eq.(\ref{eq:DRM1}), and the second one defines that from
$K_{\C}^{(s,t)}(T^d)$ to $K_{\C}^{(s+1,t+1)}(T^d\times \widebar S^1)$ in
Eq.(\ref{eq:DRM2}).
%

Here we only describe the case with the unitary order-two NSGs in complex
AZ classes, 
however, the extension to the other cases is straightforward: 
The dimension-raising map is the same as
Eq.(\ref{Eq:Dim_Raise_H_Class_A}) for non-chiral cases and
Eq. (\ref{Eq:Dim_Raise_H_Class_AIII}) for chiral cases, respectively. 
The only difference is the existence of antiunitary symmetry in the
other cases.

\subsection{Example: dimension-raising map for emergent detached
topological phases}

In Appendix \ref{Sec:Rep_Model_1-dim}, we will present representative
Hamiltonians for one-dimensional $\Z_2$ phases that is emergent due to NSGs.
As an application of the dimension-raising map, we construct
model Hamiltonians in higher dimensions from the emergent $\Z_2$
topological phases in one-dimension.

For example, consider the one-dimensional emergent $\Z_2$ phase of Table
\ref{Tab:Rep_Model_1-dim} with NSG $\bar U$ in class D. 
The relevant isomorphism is as follows.  
\begin{align}
\Z_4 \cong 
\underbrace{K^{(s=2,t=1)}_{\R}(\widetilde S^1)}_{{\rm Class\ D,\ }\widebar U}
\cong 
\underbrace{K^{(s=3,t=1)}_{\R}(\widetilde S^1 \times S^1)|_{\rm
 strong}}_{{\rm Class\ DIII,\ } \{U,TC\}=0}  
\cong 
\underbrace{K^{(s=4,t=0)}_{\R}(\widetilde S^1 \times S^1 \times
 \overline S^1)|_{\rm strong}}_{{\rm Class\ AII,\ } U}.  
\end{align}
A nontrivial element of the $\Z_2$ subgroup of $\Z_4$, which is characterized
by $\theta=2$ (mod 4) with $\theta$ in
Eq.(\ref{Eq:Z4_inv_1D}), is the emergent $\Z_2$ phase.  
%
From Table \ref{Tab:Rep_Model_1-dim}, 
a set of representative Hamiltonians for $\theta =2$ and the
symmetry operators are given by   
\begin{align}
(H^{1d}_0(k_x), H^{1d}_1(k_x)) = (\sigma_z \tau_z, - \sigma_z \tau_z), && 
C = \sigma_0 \tau_x K, && 
U^{1d}(k_x) = \begin{pmatrix}
0 & e^{-i k_x} \\
1 & 0 \\
\end{pmatrix}_{\sigma} \tau_0. 
\end{align}
$H^{1d}(k_x,m_1) = m_1 \sigma_z \tau_z$ interpolates 
$H^{1d}_0(k_x)$ and $H^{1d}_1(k_x)$. Thus from
Eq.(\ref{Eq:Dim_Raise_H_Class_A}), we can construct   
a two-dimensional model Hamiltonian $H^{2d}(k_x,k_y)$ as
\begin{align}
H^{2d}(k_x,k_y) = \cos k_y \sigma_z \tau_z s_z + \sin k_y \sigma_0
 \tau_0 s_x,
\end{align}
which has PHS $C = \sigma_0 \tau_x s_0 K$, TRS $T = i \sigma_0 \tau_x
s_y K$, and nonsymmorphic ${\bm Z}_2$ symmetry
\begin{align}
U^{2d}(k_x)H^{2d}(k_x, k_y)U^{2d}(k_x)^{-1}=H^{2d}(k_x, k_y),
&&
U^{2d}(k_x) = \begin{pmatrix}
0 & e^{-i k_x} \\
1 & 0 \\
\end{pmatrix}_{\sigma} \tau_0 s_x. 
\end{align}
Here $U^{2d}(k_x)$ satisfies $\{U^{2d}(k_x), TC\}=0$.
In a similar manner, we can construct a three-dimensional Hamiltonian
from $H_1(k_x, k_y)\equiv H^{2d}(k_x,k_y)$.
A reference Hamiltonian $H_0(k_x, k_y)$ can be taken as
\begin{align}
H_0(k_x,k_y)= (-2+ \cos k_y) \sigma_z \tau_z s_z + \sin k_y \sigma_0 \tau_0 s_x,
\end{align}
so the interpolating Hamiltonian is given by
\begin{align}
H^{2d}(k_x,k_y,m_2) = (m_2 -1 + \cos k_y) \sigma_z \tau_z s_z + \sin
 k_y \sigma_0 \tau_0 s_x.
\end{align}
Then, from Eq.(\ref{Eq:Dim_Raise_H_Class_A}), we have  
\begin{align}
H^{3d}(k_x,k_y,k_z) = (\cos k_z-1+\cos k_y ) \sigma_z \tau_z s_z + \sin k_y \sigma_0 \tau_0 s_x + \sin k_z \sigma_0 \tau_0 s_y, 
\end{align}
which has TRS $T = i \sigma_0 \tau_x s_y K$ and glide symmetry
\begin{align}
U^{3d}(k_x)H^{3d}(k_x, k_y, k_z)U^{3d}(k_x)^{-1}=
H^{3d}(k_x, k_y , -k_z),
&&
U^{3d}(k_x) = 
\begin{pmatrix}
0 & e^{-i k_x} \\
1 & 0 \\
\end{pmatrix}_{\sigma} \tau_0 s_x. 
\end{align}

\section{Symmetry forgetting functor: emergent and disappeared phases by additional symmetry}
\label{Sec:Sym_Forget}
Here we introduce a useful tool to identify emergent and disappeared
phases by an additional symmetry. 
Let $K^s_{\R}(S^1)$ ($K^s_{\C}(S^1)$) be the $K$-group in real (complex)
AZ class $s$ in one-dimension. 
In general, 
$K^s_{\R}(S^1)$ and $K^s_{\C}(S^1)$ have zero-dimensional weak indices. 
With imposing an additional oder-two NSG, these $K$-groups are changed as 
$K^{(s,t)}_{\R}(\widetilde S^1)$, $K^{(s,t)}_{\C}(\widetilde S^1)$ or $K^t_{\C}(\widetilde S^1)$. 
``Forgetting the additional symmetry'' defines functors between the $K$-groups:
\begin{align}
&f^{(s,t)}_{\C} : K^{(s,t)}_{\C}(\widetilde S^1) \to K^s_{\C}(S^1), \\
&f^t_{\C} : K^{t}_{\C}(\widetilde S^1) \to K^s_{\C}(S^1), \ \ (s = t\ {\rm mod\ }2), \\
&f^{(s,t)}_{\R} : K^{(s,t)}_{\R}(\widetilde S^1) \to K^s_{\R}(S^1). 
\end{align}
The kernel of the symmetry forgetting functor, say ${\rm
Ker}(f^{(s,t)}_{\R})$, identifies all emergent topological phases in
$K^{(s,t)}_{\R}(\widetilde S^1)$ by the additional symmetry,  
i.e. the topological phases that can not be stabilized in the absence of the additional symmetry. 
On the other hand, the cokernel, ${\rm Coker}(f^{(s,t)}_{\R}) :=
K^s_{\R}(S^1)/{\rm Im}(f^{(s,t)}_{\R})$, identifies disappeared
topological phases in $K^s_{\R}(S^1)$ by adding the symmetry. 
Relevant Abelian group structures are summarized in the following exact
sequence,
\begin{align}
\begin{CD}
0 @>>> 
\underbrace{{\rm Ker}(f^{(s,t)}_{\R})}_{\rm New\ phases} @>>> 
K^{(s,t)}_{\R}(\widetilde S^1) @>f^{(s,t)}_{\R}>>
K^s_{\R}(S^1) @>>> 
\underbrace{{\rm Coker}(f^{(s,t)}_{\R})}_{\rm Disappeared\ phase} @>>> 
0. 
\end{CD}
\end{align}
In Table \ref{Table:SFF}, we summarize all symmetry forgetting
functors for additional order-two NSGs. 
In the tables, the double-quotation ``$A$'' ($A = \Z, 2\Z, 4\Z, \Z_2,
\Z/2\Z, 2\Z/4\Z$) implies the weak index from zero-dimension.  

\begin{table}
\begin{center}
\caption{Symmetry forgetting functors for (a) $K_{\C}^{(s,t)}(\widetilde
 S^1)$, (b) $K_{\C}^{(t)}(\widetilde S^1)$, and
 (c) $K_{\R}^{(s,t)}(\widetilde S^1)$. 
} 
\begin{tabular}[c]{ccccccccc}
(a)\\
\hline 
$s$ & $t$ & AZ class & Additional NSG & ${\rm Ker}(f^{(s,t)}_{\C})$ & $K^{(s,t)}_{\C}(\widetilde S^1)$ & $K^s_{\C}(S^1)$ & ${\rm Coker}(f^{(s,t)}_{\C})$ & $f^{(s,t)}_{\C}$ \\
\hline 
\hline 
$0$ & $0$ & A & $U$ & $0$ & ``$2\Z$'' & ``$\Z$'' & ``$\Z/2\Z$'' & $2p \mapsto 2p$ \\  
$1$ & $0$ & AIII & $[U,\Gamma]=0$ & $0$ & $\Z$ & $\Z$ & $0$ & $p \mapsto p$ \\  
$0$ & $1$ & A & $\widebar{U}$ & $\Z_2$ & $\Z_2$ & ``$\Z$'' & ``$\Z$'' & $p \mapsto 0$ \\  
$1$ & $1$ & AIII & $\{U,\Gamma\}=0$ & $0$ & $0$ & $\Z$ & $\Z$ & $0$ \\  
\hline 
\\
(b)\\
\hline 
$s$ & $t$ & AZ class & Additional NSG & ${\rm Ker}(f^{t}_{\C})$ & $K^{t}_{\C}(\widetilde S^1)$ & $K^s_{\C}(S^1)$ & ${\rm Coker}(f^{t}_{\C})$ & $f^{t}_{\C}$ \\
\hline 
\hline 
$0$ & $0$ & A & $A$ & $0$ & ``$2\Z$'' & ``$\Z$'' & ``$\Z/2\Z$'' & $2p \mapsto 2p$ \\  
$1$ & $1$ & AIII & $[A,\Gamma]=0, \Gamma^2=1$ & $0$ & $\Z$ & $\Z$ & $0$ & $p \mapsto p$ \\  
$0$ & $2$ & A & $\widebar{A}$ & $\Z_2$ & $\Z_2$ & ``$\Z$'' & ``$\Z$'' & $p \mapsto 0$ \\  
$1$ & $3$ & AIII & $\{A,\Gamma\}=0, \Gamma^2=1$ & $0$ & $0$ & $\Z$ & $\Z$ & $0$ \\  
\hline 
\\
(c)\\
\hline 
$s$ & $t$ & AZ class & Additional NSG & ${\rm Ker}(f^{(s,t)}_{\R})$ & $K^{(s,t)}_{\R}(\widetilde S^1)$ & $K^s_{\R}(S^1)$ & ${\rm Coker}(f^{(s,t)}_{\R})$ & $f^{(s,t)}_{\R}$ \\
\hline 
\hline 
$0$ & $0$ & AI & $U,A$ & $0$ & ``$2\Z$'' & ``$\Z$'' & ``$\Z/2\Z$'' & $2p \mapsto 2p$ \\  
$1$ & $0$ & BDI & $[U,TC]=0, [A,TC]=0$ & $\Z_2$ & $\Z \oplus \Z_2$ & $\Z \oplus$``$\Z_2$'' & ``$\Z_2$'' & $(p,\nu) \mapsto (p,0)$ \\  
$2$ & $0$ & D & $U, \widebar{A}$ & $\Z_2$ & $\Z_2\oplus\Z_2$ & $\Z_2 \oplus$``$\Z_2$'' & ``$\Z_2$'' & $(\nu_1,\nu_2)\mapsto (\nu_1+\nu_2,0)$ \\  
$3$ & $0$ & DIII & $[U,TC]=0, [A,TC]=0$ & $0$ & $\Z_2$ & $\Z_2$ & $0$ & $\nu \mapsto \nu$ \\  
$4$ & $0$ & AII & $U,A$ & $0$ & ``$4\Z$'' & ``$2\Z$'' & ``$2\Z/4\Z$'' & $4p \mapsto 4p$ \\  
$5$ & $0$ & CII & $[U,TC]=0, [A,TC]=0$ & $0$ & $2\Z$ & $2\Z$ & $0$ & $2p \mapsto 2p$ \\  
$6$ & $0$ & C & $U, \widebar{A}$ & $0$ & $0$ & $0$ & $0$ & $0$ \\  
$7$ & $0$ & CI & $[U,TC]=0, [A,TC]=0$ & $0$ & $0$ & $0$ & $0$ & $0$ \\  
\hline 
$0$ & $1$ & AI & $\widebar{U},\widebar{A}$ & $\Z_2$ & $\Z_2$ & ``$\Z$'' & ``$\Z$'' & $\nu \mapsto 0$ \\  
$1$ & $1$ & BDI & $\{U,TC\}=0, \{A,TC\}=0$ & $\Z_2$ & $\Z_2$ & $\Z \oplus$``$\Z_2$'' & $\Z \oplus$``$\Z_2$'' & $\nu \mapsto (0,0)$ \\  
$2$ & $1$ & D & $\widebar{U},A$ & $\Z_2$ & $\Z_4$ & $\Z_2 \oplus$``$\Z_2$'' & ``$\Z_2$'' & $\theta \mapsto (\theta,0)$ \\  
$3$ & $1$ & DIII & $\{U,TC\}=0, \{A,TC\}=0$ & $0$ & $\Z_2$ & $\Z_2$ & $0$ & $\nu \mapsto \nu$ \\  
$4$ & $1$ & AII & $\widebar{U},\widebar{A}$ & $\Z_2$ & $\Z_2$ & ``$2\Z$'' & ``$2\Z$'' & $\nu \mapsto 0$ \\  
$5$ & $1$ & CII & $\{U,TC\}=0, \{A,TC\}=0$ & $0$ & $0$ & $2\Z$ & $2 \Z$ & $0$ \\  
$6$ & $1$ & C & $\widebar{U},A$ & $0$ & $0$ & $0$ & $0$ & $0$ \\  
$7$ & $1$ & CI & $\{U,TC\}=0, \{A,TC\}=0$ & $0$ & $0$ & $0$ & $0$ & $0$ \\  
\hline 
\label{Table:SFF}
\end{tabular}
\end{center}
\end{table}

There are eight emergent topological phases with either ${\rm
Ker}(f^{(s,t)}_{\R (\C)})=\Z_2$ or ${\rm Ker}(f^{t}_{\C}) = \Z_2$.  
In Appendix \ref{Sec:Rep_Model_1-dim}, we show that representative
Hamiltonians of these emergent phases can be $k_x$-independent. 
From this property, the emergent phases exhibit boundary states
detached from the bulk spectrum in the $k_x$-direction if
the space dimension is increased by dimension-raising maps in Appendix
\ref{Sec:Dim_Raise_Map}.

\section{Representative Hamiltonians of emergent topological phases in
one-dimension}
\label{Sec:Rep_Model_1-dim}

We summarize in Table \ref{Tab:Rep_Model_1-dim} all representative
Hamiltonians of emergent one-dimensional $\Z_2$ topological phases due to
order-two (magnetic) NSGs.
The representative Hamiltonians can be justified by calculating the
topological invariants in the tables of Appendices \ref{Appendix:D},
\ref{Appendix:E} and \ref{Appendix:F}.
In Table \ref{Tab:Rep_Model_1-dim},  
we present a set of representative
Hamiltonians $(H_0(k_x),H_1(k_x))$. 
In the case of the emergent topological phases,
the representative Hamiltonian $H_1(k_x)$ is momentum-independent, so a
reference
Hamiltonian $H_0(k_x)$ is needed to define 
the topological invariant.
$H_1(k_x)$ is topologically non-trivial in the basis where 
$H_0(k_x)$ is not.
%
For instance, consider an order-two nonsymmorphic antiunitary
antisymmetry $\widebar A$ in class A, 
\begin{align}
\widebar A(k_x) H(k_x) \widebar A(k_x)^{-1} = - H(k_x), && \widebar A(k_x) = \begin{pmatrix}
0 & e^{-i k_x} \\
1 & 0 \\
\end{pmatrix} K. 
\end{align}
Table \ref{Table:SFF}(b) indicates that ${\rm Ker}(f_{\C}^{t=2}) =
K_{\C}^{t=2}(\widetilde S^1) = \Z_2$, so the $\Z_2$ phase of this system
is an emergent topological phase due to the nonsymmorphic antiunitary
antisymmetry $\widebar A$. 
As shown in the table of Appendix \ref{Appendix:E} (see $n=0$), the
topological number is given by the $\Z_2$ invariant at $k_x=0$. Indeed,
because $\widebar A(k_x)$ at $k_x=0$ reduces to PHS $C$ in class D with the
identification $\widebar A(0)=\sigma_x K=C$, one can introduce the
$\Z_2$ invariant $\nu$ (mod 2) as $(-1)^{\nu}={\rm sgn}\left[{\rm
Pf}(H(0)\sigma_x)\right]$ like a zero-dimensional class D
system.\cite{Schnyder2008}
Then, it is easy to find $\nu=1$ ($\nu=0$) for $H_1(k_x)=-\sigma_z$
($H_0(k_z)=\sigma_z$).
Mathematically, the sets of Hamiltonians give the Karoubi's grading
description of the $K$-theories,\cite{Karoubi2006, Kitaev2009}
where the pair $(H_0(\bk),H_1(\bk))$ expresses the topological
``difference'' between $H_0(\bk)$ and $H_1(\bk)$.
In a similar way, we can specify sets of the representative Hamiltonians
$(H_0(k_x), H_1(k_x))$ for all other emergent topological phases in
one-dimension. 
The sets of Hamiltonians can be used to construct the dimension-raising
maps in Appendix \ref{Sec:Dim_Raise_Map}.


\begin{table*}[!]
\begin{center}
\caption{
Emergent topological phases in one-dimension with order-two (magnetic)
 NSGs.  
$\sigma_i, \tau_i, s_i (i=x,y,z)$ represent the Pauli matrices and $K$
 is complex conjugation. $(H_0(k_x), H_1(k_x))$ is a set of
 representative Hamiltonians for the element of the $K$-group in the table.
}
\label{Tab:Rep_Model_1-dim}
\begin{tabular}[c]{ccccccl}
\hline 
$s$ & $t$ & AZ class & Additional NSG & element of $K$-group & $(H_0(k_x), H_1(k_x))$ & \multicolumn{1}{c}{Representation of symmetries}\\
\hline 
\hline 
$0$ & $1$ & A & $\widebar{U}$ & $1 \in \Z_2$ &  
$(\sigma_z,-\sigma_z)$ & 
$\widebar{U}(k_x) = \begin{pmatrix}
0 & e^{-i k_x} \\
1 & 0 \\
\end{pmatrix}_{\sigma}$ \\  
\hline 
$0$ & $2$ & A & $\widebar{A}$ & $1 \in \Z_2$ & 
$(\sigma_z,-\sigma_z)$ & 
$\widebar{A}(k_x) = \begin{pmatrix}
0 & e^{-i k_x} \\
1 & 0 \\
\end{pmatrix}_{\sigma} K$ \\  
\hline 
$1$ & $0$ & BDI & $[U,TC]=0, [A,TC]=0$ & $(0,1) \in \Z \oplus \Z_2$ & 
$(\sigma_0 \tau_z,-\sigma_0 \tau_z)$ & 
$T=K, \ \ C=\tau_x K, \ \ 
U(k_x) = \begin{pmatrix}
0 & e^{-i k_x} \\
1 & 0 \\
\end{pmatrix}_{\sigma} \tau_0 $ \\  
$2$ & $0$ & D & $U, \widebar{A}$ & $(1,1) \in \Z_2 \oplus \Z_2$ & 
$(\sigma_0 \tau_z,-\sigma_0 \tau_z)$ & 
$C=\tau_x K, \ \ 
U(k_x) = \begin{pmatrix}
0 & e^{-i k_x} \\
1 & 0 \\
\end{pmatrix}_{\sigma} \tau_0 $ \\  
$0$ & $1$ & AI & $\widebar{U},\widebar{A}$ & $1 \in \Z_2$ & 
$(\sigma_z,-\sigma_z)$ & 
$T=K, \ \ 
\widebar{U}(k_x) = \begin{pmatrix}
0 & e^{-i k_x} \\
1 & 0 \\
\end{pmatrix}_{\sigma} $ \\  
$1$ & $1$ & BDI & $\{U,TC\}=0, \{A,TC\}=0$ & $1 \in \Z_2$ & 
$(\sigma_0 \tau_z,-\sigma_0 \tau_z)$ & 
$T=K, \ \ C=\tau_x K, \ \ 
U(k_x) = \begin{pmatrix}
0 & e^{-i k_x} \\
1 & 0 \\
\end{pmatrix}_{\sigma} \tau_z $ \\  
$2$ & $1$ & D & $\widebar{U},A$ & $2 \in \Z_4$ & 
$(\sigma_z \tau_z,-\sigma_z \tau_z)$ & 
$C=\tau_x K, \ \ 
\widebar{U}(k_x) = \begin{pmatrix}
0 & e^{-i k_x} \\
1 & 0 \\
\end{pmatrix}_{\sigma} \tau_0 $ \\  
$4$ & $1$ & AII & $\widebar{U},\widebar{A}$ & $1 \in \Z_2$ & 
$(\sigma_z s_0,-\sigma_z s_0)$ & 
$T=i s_y K, \ \ 
\widebar{U}(k_x) = \begin{pmatrix}
0 & e^{-i k_x} \\
1 & 0 \\
\end{pmatrix}_{\sigma} s_0 $ \\  
\hline 
\end{tabular}
\end{center}
\end{table*}

\section{Twisted and equivariant $K$-theory}
\label{Appendix:TEK-theory}

In this section, we summarize the notation of the twisted equivariant
$K$-theory,\cite{Freed2013} which we use in Appendices
\ref{sec:MV}-\ref{Appendix:F}
The $K$-group over $\widetilde S^1$ with a twist $\tau$ is denoted by
$K_G^{\tau+n}(\widetilde S^1)$. 
For order-two nonsymmorphic TCIs/TCSCs with real AZ symmetries, 
the superscript $n = 0, \dots, 7$ (mod $8$) specifies the real AZ
class $s= -n$ 
(mod 8) in Table \ref{Classifying_space}, 
and 
the symmetry group $G$ is
$G = \Z_2 \times \Z_2$, 
where $\Z_2 \times 1 = \{1, \sigma_1\}$ represents an order-two NSG
symmetry and 
$1 \times \Z_2 = \{1,\sigma_2\}$ represents  TRS or PHS in the AZ class. 
The group $G$ acts on $k_1\equiv k_x\in \widetilde S^1$ as 
\begin{align}
\sigma_1 : k_x \mapsto k_x, && 
\sigma_2 : k_x \mapsto -k_x. 
\end{align}
For $n=0$ (class AI), 
the twist $\tau$ is determined by the following three data $\tau =
(c,\phi,\zeta)$:  
\begin{itemize}
\item A homomorphism $c : G \to \Z_2 = \{1,-1\}$, which specifies the symmetry or ``antisymmetry''
\begin{align}
U_{g}(k_x) H(k_x) = c(g) H(g k_x) U_{g}(k_x), && g \in G. 
\end{align}
\item A homomorphism $\phi : G \to \Z_2 = \{1,-1\}$, which specifies
      unitary symmetry for $\phi(g) = 1$ or antiunitary symmetry for
      $\phi(g) = -1$ ($g\in G$). 
\item $\phi$-twisted 2-cocycle $\zeta_{g,g'}(k_x) \in Z^2_{\rm group}(G ;
      C(S^1,U(1)))$ [Here $C(S^1,U(1))$ is a set of $U(1)$-valued functions
      on $S^1$.] 
The $\phi$-twisted 2-cocycle specifies the twisting of the group action 
\begin{align}
U_{g}(g' k_x) U_{g'}(k_x) = \zeta_{g,g'}(g g' k_x) U_{gg'}(k_x), && g,g' \in G. 
\end{align}
(``$\phi$-twisted'' means that the action of $g \in G$ on $f \in C(S^1,U(1))$ is given by $(g \cdot f)(k_x) = \overline{f(g^{-1} k_x)}$ for $\phi(g) = -1$.)
\end{itemize}
From Table \ref{Symmetry_type}, there are two inequivalent order-two
NSG symmetries $t=0,1$ (mod 2) in class AI. 
They have the following twist $\tau = (c,\phi,\zeta)$:
$c$ and $\phi$ are given by 
\begin{align}
&c(\sigma_1) = \left\{ \begin{array}{ll}
1 & (t=0) \\
-1 & (t=1) \\
\end{array} \right. &&
c(\sigma_2) = 1, \\
&\phi(\sigma_1) = 1, && 
\phi(\sigma_2) = -1. 
\end{align}
$\zeta$ is determined by 
\begin{align}
&U_{\sigma_1}(k_x)^2 = e^{-i k_x}, && 
U_{\sigma_2}(-k_x) U_{\sigma_2}(k_x) = 1, \\
&U_{\sigma_1}(-k_x) U_{\sigma_2}(k_x) = 
\left\{ \begin{array}{ll}
U_{\sigma_2}(k_x) U_{\sigma_1}(k_x) & (t= 0) \\
-U_{\sigma_2}(k_x) U_{\sigma_1}(k_x) & (t=1) \\
\end{array} \right. .
\end{align}
The other real AZ classes with $n > 0$ are obtained by adding the chiral
symmetries $\Gamma_1, \dots, \Gamma_n$ satisfying
\begin{align}
&\Gamma_i \Gamma_j + \Gamma_j \Gamma_i = -2 \delta_{ij}, \\
&\Gamma_i U_{g}(k_x) = c(g) U_g(k_x) \Gamma_i, && i = 1,\dots n, \ \ g \in G. 
\end{align}
The order-two NSG symmetry $t = 0,1$ (mod 2) of the system is unchanged
by this process.

For order-two (magnetic) nonsymmorphic TCIs in complex AZ classes, $G$ in
$K_G^{\tau+n}(\widetilde{S}^1)$ is given by
$G=\Z_2=\{1, \sigma_1\}$ with an order-two (magnetic) NSG symmetry $\sigma_1$. 
For unitary order-two NSGs, $n$ specifies the complex AZ class
$s=-n$ (mod 2) in Table \ref{Classifying_space}, and 
the additional order-two NSG symmetries $t=0,1$ (mod 2) are given in
Table \ref{Symmetry_type_UC}.
For magnetic order-two NSGs, $n=0,\dots, 7$ (mod 8) specifies the
additional anti unitary symmetry in the table in Sec.\ref{Appendix:E}. (
Whereas the number of inequivalent oder-two magnetic NSG symmetries is
four, as summarized in Table \ref{Symmetry_type_AC}, it is convenient to
refine the classification by using $\epsilon_{\sigma}$ of
Eq.(\ref{eq:nonsymmorphic_magnetic}), in the actual evaluation of the
$K$-groups. See Sec.\ref{Appendix:E}
) 
The twist $\tau$ in these cases are given in a similar manner as that
in TCIs/TCSCs in real AZ classes.

\section{Mayer-Vietoris sequence}
\label{sec:MV}

To evaluate $K$-groups on $\widetilde S^1$, we use the Mayer-Vietoris exact sequence.\cite{Bott-Tu, Freed2011}
We decompose the one-dimensional base space $\widetilde S^1$ into
$\widetilde S^1 = U \cup V$ with 
\begin{align}
U = \{e^{i k_x} | -\frac{\pi}{2} \leq k_x \leq \frac{\pi}{2} \}, && 
V = \{e^{i k_x} | \frac{\pi}{2} \leq k_x \leq \frac{3 \pi}{2} \}. 
\end{align}
Then, the sequence of the inclusion 
\begin{equation}
\begin{CD}
\xy*\cir<1cm>{}\endxy 
@<<<
\xygraph{
!{<0cm,0cm>;<1cm,0cm>:<0cm,1cm>::}
!{(-0.3,0)}*{\cir<1cm>{l^r}}
!{(0.3,0)}*{\cir<1cm>{r^l}}
}@<<<
\xygraph{
!{<0cm,0cm>;<1cm,0cm>:<0cm,1cm>::}
!{(0,0.7)}*{\bullet}
!{(0.5,0.7)}*{\{i \}}
!{(0,-0.7)}*{\bullet}
!{(0.6,-0.7)}*{\{-i \}}
}\\
S^1 = U \cup V @<(i_U, i_V)<< U \sqcup V @<(j_U,j_V)<< U \cap V \\
\end{CD}
\end{equation}
induces the long exact sequence of the twisted equivariant $K$-theory 
\begin{align}
\begin{CD}
@>>> 
K^{\tau+n}_{G}(\widetilde S^1) @>>> 
K^{\tau|_U+n}_{G}(U) \oplus K^{\tau|_V+n}_{G}(V) @>\Delta_n>>
K^{\tau|_{U\cap V}+n}_{G}(U\cap V) @>d_n>> 
K^{\tau+n+1}_{G}(\widetilde S^1) @>>>
\end{CD}
\label{eq:MV}
\end{align}
where $\Delta_n = j_U^* - j_V^*$ is induced by the inclusions $j_U$ and $j_V$,
$d_n$ is the connecting homomorphism, and $\tau|_U$, $\tau|_V$ and
$\tau|_{U\cap V}$ are twists induced by $\tau$ on $U$, $V$, and $U\cap
V$, respectively.
Here $K_{G}^{\tau|_U+n}(U)$, $K_{G}^{\tau|_V+n}(V)$ and
$K_{G}^{\tau|_{U\cap V}+n}(U\cap V)$ reduces to $K$-groups in
zero-dimension, so they can be evaluated directly.
Then, using the above sequence, we can compute the $K$-group in one-dimension
$K_G^{\tau+n}(\widetilde S^1)$.

\section{One-dimensional order-two nonsymmorphic
 insulators in complex AZ classes}
\label{Appendix:D}

\subsection{$t=0$ order-two NSG}
The data for the Mayer-Vietoris sequence 
and the topological invariant
for $K_{\Z_2}^{\tau+n}(\widetilde S^1)$ 
are summarized as follows: 
$$
\begin{tabular}[t]{|c|c|c|c|c|c|}
\hline 
$n$ & AZ class & $K^{\tau+n}_{\Z_2}(\widetilde S^1)$ &
	     $K^{\tau|_U+n}_{\Z_2}(U) \oplus K^{\tau|_V+n}_{\Z_2}(V)$ &
		 $K^{\tau|_{U\cap V}+n}_{\Z_2}(U\cap V)$ & Topological
		     invariant for $K^{\tau+n}_{\Z_2}(\widetilde S^1)$ \\ 
\hline 
$1$ & AIII &$\Z$ & $0$ & $0$ &1D winding number \\
$0$ & A & $2 \Z$ & $\Z^2 \oplus \Z^2$ & $\Z^4$ & Weak $\Z$ invariant \\
\hline 
\end{tabular}
$$
The map $\Delta_0$ in the Mayer-Vietoris sequence in Eq.(\ref{eq:MV}) is
given by
\begin{align}
\Delta_0 : ((n_1,n_2),(m_1,m_2))
\mapsto (n_1-m_1, n_2-m_2, n_1-m_2, n_2-m_1)
\end{align}
with $(n_1, n_2)\in \Z^2=K_{\Z_2}^{\tau|_U+0}(U)$ and
$(m_1, m_2)\in \Z^2=K_{\Z_2}^{\tau|_V+0}(V)$.

\subsection{$t=1$ order-two NSG}
The data for the Mayer-Vietoris sequence and the topological invariant
for $K_{\Z_2}^{\tau+n}(\widetilde S^1)$ 
are summarized in the following table,
$$
\begin{tabular}[t]{|c|c|c|c|c|c|}
\hline 
$n$ & AZ class & $K^{\tau+n}_{\Z_2}(\widetilde S^1)$ &
	     $K^{\tau|_U+n}_{\Z_2}(U) \oplus K^{\tau|_V+n}_{\Z_2}(V)$ &
		 $K^{\tau|_{U\cap V}+n}_{\Z_2}(U\cap V)$ & Topological
		     invariant for $K^{\tau+n}_{\Z_2}(\widetilde S^1)$ \\ 
\hline 
$1$ & AIII & $0$ & $\Z \oplus \Z$ & $\Z^2$ & \\
$0$ & A &  $\Z_2$ & $0$ & $0$ & $\Z_2$ invariant in Eq.(\ref{Eq:Z2_Inv_1D_A_1}) \\
\hline 
\end{tabular}
$$
The map $\Delta_1$ in Eq.(\ref{eq:MV}) is given by
\begin{align}
\Delta_1 : (n, m) \mapsto (n-m, n+m),
\end{align}
where $n\in \Z=K^{\tau_U+1}_{\Z_2}(U)$ and $n\in \Z=K^{\tau_V+1}_{\Z_2}(V)$.
Here the $\Z_2$ invariant for $K^{\tau+0}_{\Z_2}(S^1) = \Z_2$ is defined as follows. 
Consider a one-dimensional Hamiltonian with $t=1$ order-two NSG symmetry
in class
A $(n=0)$, 
\begin{align}
U(k_x) H(k_x) U(k_x) = - H(k_x), && 
U(k_x) = \begin{pmatrix}
0 & e^{- ik_x} \\
1 & 0
\end{pmatrix}. 
\label{Eq:Sym_1D_A_1}
\end{align}
The Hamiltonian with the symmetry (\ref{Eq:Sym_1D_A_1}) can be written as 
\begin{align}
H(k_x) = \begin{pmatrix}
X(k_x) & i Y(k_x) e^{-i k_x/2} \\
-i Y(k_x) e^{i k_x/2} & - X(k_x)
\end{pmatrix}, && k_x \in[-\pi,\pi], 
\end{align}
where $X(k_x), Y(k_x)$ are hermitian matrices which obey the boundary conditions 
\begin{align}
X(\pi) = X(-\pi), && Y(\pi) = - Y(-\pi). 
\end{align}
We introduce the following unitary matrix 
\begin{align}
Z(k_x) := X(k_x) + i Y(k_x), && Z(\pi) = Z^{\dag}(-\pi). 
\end{align}
The $\Z_2$ invariant $\nu$ is defined as 
\begin{align}
\nu := \frac{i}{\pi}\ln \det Z(\pi) 
- \frac{i}{2\pi} \int_{-\pi}^{\pi} dk_x \partial_{k_x} \ln \det Z(k_x) \ \ ({\rm mod\ }2 ). 
\label{Eq:Z2_Inv_1D_A_1}
\end{align}
Because of the relation $(i/\pi)\ln \det Z(\pi) = - (i/\pi)\ln \det Z(-\pi)$ (mod 2), 
we have $2 \nu = 0$ (mod 2).

\section{One-dimensional order-two nonsymmorphic magnetic insulator
  in complex AZ classes}
\label{Appendix:E} 

The data for the Mayer-Vietoris sequence and the topological numbers are
summarized as
$$
\begin{tabular}[t]{|c|c|c|c|c|c|c|}
\hline 
$n$ & AZ class & $A^{\epsilon_A}_{\eta_{\Gamma}}$ or $\widebar
	 A^{\epsilon_A}_{\eta_{\Gamma}}$ & $K^{\tau+n}_{\Z_2}(\widetilde
	     S^1)$ & $K^{\tau|_U+n}_{\Z_2}(U) \oplus
		 K^{\tau|_V+n}_{\Z_2}(V)$ & $K^{\tau|_{U\cap
		     V}+n}_{\Z_2}(U\cap V)$ & Topological invariant for
			 $K^{\tau+n}_{\Z_2}(\widetilde S^1)$ \\ 
\hline 
$7$ & AIII & $(A^+_+, \widebar A^+_+)$ & $\Z$ & $\Z_2 \oplus 0$ & $0$ & Winding number \\
$6$ & A & $\widebar A^+$ & $\Z_2$ & $\Z_2 \oplus 0$ & $\Z$ & $\Z_2$ invariant at $k_x = 0$ \\
$5$ & AIII & $(A^-_-, \widebar A^+_-)$ & $0$ & $0$ & $0$ & \\
$4$ & A & $A^-$ & $2 \Z$ & $2\Z \oplus \Z $ & $\Z$ & Weak $\Z$ invariant \\
$3$ & AIII & $(A^-_+, \widebar A^-_+)$ & $\Z$ & $0 \oplus \Z_2$ & $0$ & Winding number \\
$2$ & A & $\widebar A^-$ & $\Z_2$ & $0 \oplus \Z_2$ & $\Z$ & $\Z_2$ invariant at $k_x = \pi$ \\
$1$ & AIII & $(A^+_-, \widebar A^-_-)$ & $0$ & $0$ & $0$ & \\
$0$ & A & $A^+$ & $2 \Z$ & $\Z \oplus 2\Z$ & $\Z$ & Weak $\Z$ invariant \\
\hline 
\end{tabular}
$$
Here the superscript $\epsilon_{\sigma}=\pm$ of $A^{\epsilon_{\sigma}}$ and
$\widebar{A}^{\epsilon_{\sigma}}$ specifies the sign of $A^2$ and
$\widebar{A}^2$, i.e. $A^2=\epsilon_{\sigma}$ and
$\widebar{A}^2=\epsilon_{\sigma}$. 
The maps $\Delta_0$ and $\Delta_4$ are given by 
\begin{align}
&\Delta_0 : (p,2q) \mapsto p-2q, \\
&\Delta_4 : (2k,l) \mapsto 2k-l, 
\end{align} 
where $p\in \Z=K_{\Z_2}^{\tau|_U+0}(U)$, 
$2q\in 2\Z=K_{\Z_2}^{\tau|_V+0}(V)$,
$2k\in 2\Z=K_{\Z_2}^{\tau|_U+4}(U)$,
and
$l\in \Z=K_{\Z_2}^{\tau|_V+0}(V)$. 
The other $\Delta_n (n\neq 0,4)$ are zero maps. 
The $K$-groups $K_{\Z_2}^{\tau+n}(\widetilde S^1)$ for $n = 3, 7$ are
not determined uniquely from the
Mayer-Vietoris sequence in this case:  
There remain two possibilities, $\Z$ or $\Z \oplus \Z_2$, for $n=3, 7$.  
We have determined them by comparing them with the Mayer-Vietoris sequence without the 
magnetic translation symmetry. 
(We also have compared the results with solutions of the homotopy problem for
the relevant spaces.)

\section{One-dimensional order-two nonsymmorphic TCIs and TCSCs in real AZ classes}
\label{Appendix:F}

\subsection{$t=0$ order-two NSG}
The data for the Mayer-Vietoris sequence and the topological numbers are
summarized below:
$$
\begin{tabular}[t]{|c|c|c|c|c|c|}
\hline 
$n$ & AZ class & $K^{\tau+n}_{\Z_2 \times \Z_2}(\widetilde{S}^1)$ &
	     $K^{\tau|_U+n}_{\Z_2 \times \Z_2}(U) \oplus
	     K^{\tau|_V+n}_{\Z_2 \times \Z_2}(V)$ & $K^{\tau|_{U\cap
		 V}+n}_{\Z_2 \times \Z_2}(U\cap V)$ & Topological
		     invariant for $K^{\tau+n}_{\Z_2 \times
		     \Z_2}(\widetilde S^1)$ \\ 
\hline 
$7$ & BDI & $\Z \oplus \Z_2$ & $(\Z_2 \oplus \Z_2) \oplus 0$ & $0$ & $\Z$ winding number and $\Z_2$ invariant at $k_x = 0$ \\
$6$ & D & $\Z_2 \oplus \Z_2$ & $(\Z_2 \oplus \Z_2) \oplus \Z$ & $\Z
		 \oplus \Z$ & $\Z_2$ invariant at $k_x = 0$ for each
		     eigensector of $U(0)$ \\
$5$ & DIII & $\Z_2$ & $0$ & $0$ & $\Z_2$ invariant for 1D class DIII \\
$4$ & AII & $4 \Z$ & $(2\Z \oplus 2\Z) \oplus \Z$ & $\Z \oplus \Z$ & $\Z$ weak invariant for class AII \\
$3$ & CII & $2\Z$ & $0$ & $0$ & 1D winding number \\
$2$ & C & $0$ & $0 \oplus \Z$ & $\Z \oplus \Z$ & \\
$1$ & CI & $0$ & $0$ & $0$ & \\
$0$ & AI & $2 \Z$ & $(\Z \oplus \Z) \oplus \Z$ & $\Z \oplus \Z$ & $\Z$
		     weak invariant for class AI \\
\hline 
\end{tabular}
$$
Here we assume that the order-two NSG operator $U(k_x)$ satisfies
$U(k_x)^2=e^{-ik_x}$. 
(For order-two NSGs with $U(k_x)^2=-e^{-ik_x}$, $K_{\Z_2\times
\Z_2}^{\tau|_U+n}(U)$ and $K_{\Z_2\times \Z_2}^{\tau|_V+n}(V)$ in the
above table are exchanged, and the $\Z_2$ invariant in class D is
defined at $k_x=\pi$ for each eigensector of $U(\pi)$. )
To obtain the above results, we have trivialized the
twist $\tau$ of $U(k_x)$ on $U$, $V$ and $U \cap V$ by multiplying $U(k_x)$
by the following factors,
\begin{align}
\beta_U = e^{i k_x/2}, && 
\beta_V = e^{i k_x/2}, && 
\beta_{U \cap V}( \pm \frac{\pi}{2}) = e^{\pm \pi i/4}. 
\label{Trivialization_Twist}
\end{align}
Note that 
\begin{align}
&K^{\tau|_U+n}_{\Z_2 \times \Z_2}(U) \cong K_{\R}^{(s=-n,t=0)}(k_x = 0) = \pi_0({\cal R}_{-n} \times {\cal R}_{-n}), \\
&K^{\tau|_V+n}_{\Z_2 \times \Z_2}(V) \cong K_{\R}^{(s=-n,t=2)}(k_x = \pi) = \pi_0({\cal C}_{-n}), \\
&K^{\tau|_{U \cap V}+n}_{\Z_2 \times \Z_2}(U \cap V) \cong K_{\C}^{(s=-n,t=0)}(k_x =\pi/2) = \pi_0({\cal C}_{-n} \times {\cal C}_{-n}).
\end{align}
Here $K_{\R}^{(s,t)}(pt)$ $(K_{\C}^{(s,t)}(pt))$ represents the
$K$-group over a point in the presence of real (complex) AZ symmetry $s$ and
additional order-two point group $t$ in Ref.\onlinecite{Shiozaki2014}.
${\cal R}_{s} ({\cal C}_s)$ is the classifying space in real (complex) AZ
class $s$ in Table \ref{Classifying_space}. 
The map $\Delta_n$ $(n=0,\dots, 7$ (mod 8)) in Eq.(\ref{eq:MV}) is given by  
\begin{align}
&\Delta_n = 0, \ \ \mbox{(for odd $n$)}, \\
&\Delta_0 : ((p_1,p_2),q) \mapsto (p_1-q,p_2-q), \\
&\Delta_2 : r \mapsto (r,-r), \\
&\Delta_4 : ((2k_1,2k_2),l) \mapsto (2k_1-l,2k_2-l), \\
&\Delta_6 : ((\nu_1,\nu_2),m) \mapsto (m,-m),
\end{align}
where $(p_1, p_2)\in \Z\oplus \Z =K_{\Z_2\times \Z_2}^{\tau_U+0}(U)$, 
$q\in \Z =K_{\Z_2\times \Z_2}^{\tau_V+0}(V)$, 
$r\in \Z =K_{\Z_2\times \Z_2}^{\tau_V+2}(V)$, 
$(2k_1, 2k_2)\in \Z =K_{\Z_2\times \Z_2}^{\tau_U+4}(U)$, 
$l\in \Z =K_{\Z_2\times \Z_2}^{\tau_V+4}(V)$, 
$(\nu_1,\nu_2)\in \Z_2\oplus \Z_2 =K_{\Z_2\times \Z_2}^{\tau_U+6}(U)$,
and  
$m\in\Z=K_{\Z_2\times \Z_2}^{\tau_U+6}(U)$.
From the above map $\Delta_n$ and the $K$-groups on $U$, $V$ and $U\cap
V$ in the Mayer-Vietoris sequence,  we can determine  the $K$-groups on
$\widetilde S^1$, except for $n=7$, as 
shown in the above table. 
The Mayer-Vietoris sequence can not determine the $K$-group
$K^{\tau+7}_{\Z_2 \times \Z_2}(\widetilde S^1)$ uniquely. 
There are two possibilities for $K^{\tau+7}_{\Z_2 \times
\Z_2}(\widetilde S^1)$,
$\Z$ or $\Z \oplus \Z_2$.  
We determine $K^{\tau+7}_{\Z_2 \times \Z_2}(\widetilde S^1)$ in the following subsection.

\subsubsection{$K^{\tau+7}_{\Z_2 \times \Z_2}(\widetilde S^1)$}

In the presence of an order-two NSG symmetry, 
a $K$-group has a $R(\Z_2)$-module structure, where $R(\Z_2)$
is the representation ring of the $\Z_2$ group $U(0)$ for the order-two NSG.
The $\Z_2$ group has two different representations, corresponding
to two eigensectors with eigenvalues $\pm 1$ of $U(0)$, which we denote
as $1, t$,
respectively. Then, $R(\Z_2)$ is the quotient of the polynomial ring
$R(\Z_2)=\Z[t]/(1-t^2)$ .
We use the $R(\Z_2)$-module structure to determine $K^{\tau+7}_{\Z_2
\times \Z_2}(\widetilde S^1)$.

In terms of the $R(\Z_2)$-module structure, the relevant part of the
Mayer-Vietoris sequence is given by
\begin{align}
\begin{CD}
K_{\Z_2\times \Z_2}^{\tau|_U+6}(U)\oplus K_{\Z_2\times \Z_2}^{\tau|_V+6}(V)
@>\Delta_6>> 
K_{\Z_2\times \Z_2}^{\tau|_{U\cap V}+6}(U\cap V)
@>>> K^{\tau+7}_{\Z_2 \times \Z_2}(\widetilde S^1) @>>>
K_{\Z_2\times \Z_2}^{\tau|_U+7}(U)\oplus K_{\Z_2\times \Z_2}^{\tau|_V+7}(V)
@>>> 0 \\ 
@| @| @. @| \\
\Z_2[t]/(1-t^2)\oplus (1-t) && R(\Z_2) && &&  \Z_2[t]/(1-t^2) 
\end{CD}
\end{align}
where $\Delta_6(\nu_1+\nu_2t, m(1-t)) = m-mt$. 
Since ${\rm Coker}(\Delta_6) = (1+t)$, we have the short exact sequence 
\begin{align}
\begin{CD}
0 @>>> (1+t) @>>> K^{\tau+7}_{\Z_2 \times \Z_2}(\widetilde S^1) @>>> \Z_2[t]/(1-t^2) @>>> 0. 
\end{CD}
\label{Eq:SES_t=0}\end{align}
Thus, $K^{\tau+7}_{\Z_2 \times \Z_2}(\widetilde S^1)$ is an extension of
$R(\Z_2)$-module $\Z_2[t]/(1-t^2)$ by $(1+t)$.  
As mentioned in the above, there are two possible extensions, ${\rm
Ext}^1_{R(\Z_2)}(\Z_2[t]/(1-t^2), (1+t)) = \Z_2$, i.e. $(1+t)\oplus
\Z_2[t]/(1-t^2)$ and $R(\Z_2)/(2-2t)$
 which have $\Z \oplus \Z_2 \oplus \Z_2$ and $\Z \oplus \Z_2$ Abelian
 group structures, respectively.
Below, we show that the former has a contradiction, so the latter
holds. 

To show the contradiction,
we
compare the Mayer-Vietoris sequence for $K^{\tau+n}_{\Z_2 \times
\Z_2}(\widetilde S^1)$ 
with that for $K^{\tau+n}_{1 \times \Z_2}(\widetilde S^1)$. The latter $K$-group is
obtained by forgetting the additional order-two NSG symmetry in the former.
We have the commutative diagram as an Abelian group,
\begin{align}
\begin{CD}
\Z_2[t]/(1-t^2) \oplus (1-t) 
@>\Delta_6>> R(\Z_2) @>>> K^{\tau+7}_{\Z_2 \times \Z_2}(\widetilde S^1) @>>> \Z_2[t]/(1-t^2) @>>> 0 \\
@VVf^6_{U \sqcup V}V @VVf^6_{U \cap V}V @VVf^6_{\widetilde S^1}V  @VVf^7_{U \sqcup V}V @VVV \\
K^{\tau+6}_{1 \times \Z_2}(U \sqcup V) @>\Delta_6|_{1 \times \Z_2}>>
 K^{\tau+6}_{1 \times \Z_2}(U \cap V) @>>> K^{\tau+7}_{1 \times
 \Z_2}(\widetilde S^1) @>>> K^{\tau+7}_{1 \times \Z_2}(U \sqcup V) @>>> K^{\tau+7}_{1 \times \Z_2}(U \cap V)\\
@|  @|  @| @| @| \\
\Z_2 \oplus \Z_2 && \Z  && \Z \oplus \Z_2 && \Z_2 \oplus \Z_2 && 0 \\
\end{CD}
\label{Eq:CommSeq_t=0}
\end{align}
where $f^6_{U\cup V}$,  $f^6_{U\cap V}$,  $f^6_{\widetilde S^1}$ and
$f^6_{U\sqcup V}$ are maps neglecting the $\Z_2$ action of the NSG. 
In particular, $f^6_{U \cap V}$ and $f^7_{U \sqcup V}$ are given by
\begin{align}
f^6_{U \cap V} : f(t) \mapsto f(1), && f^7_{U \sqcup V} : \nu(t) \mapsto
 (0, \nu(1)),  
\end{align}
with $f(t)\in R(\Z_2)=\Z[t]/(1-t^2)$ and $\nu(t)\in \Z_2[t]/(1-t^2)$.
In the above diagram,
$K^{\tau+6}_{1 \times \Z_2}(U \cap V)\to
K^{\tau+7}_{1 \times \Z_2}(\widetilde S^1)$ and 
$K^{\tau+7}_{1 \times \Z_2}(\widetilde S^1)\to K^{\tau+7}_{1 \times
\Z_2}(U \sqcup V)$ 
are given by   
\begin{align}
p\in K^{\tau+6}_{1 \times \Z_2}(U \cap V) 
\to (2p, 0)\in 
K^{\tau+7}_{1 \times \Z_2}(\widetilde S^1), 
&&(q,\nu)\in K^{\tau+7}_{1 \times \Z_2}(\widetilde S^1) 
\to 
(q+\nu, q)\in 
K^{\tau+7}_{1 \times \Z_2}(U \sqcup V), 
\end{align}
with $p \in \Z$, $q\in \Z$, and $\nu\in \Z_2$, which can be obtained by  
the $K$-theory for conventional topological insulators and superconductors.
Because $\Delta_6|_{1 \times \Z_2} = 0$, we obtain 
\begin{align}
\begin{CD}
\Z @>>> {\rm Ker}(f^6_{S^1}) @>>> \Z_2 @>>> 0 @>>> {\rm Coker}(f^6_{S^1}) @>>> \Z_2.  \\
\end{CD}
\label{Eq:SeqSnake_t=0}
\end{align}
from the snake lemma.\cite{Weibel} 

Now we assume that the short exact sequence (\ref{Eq:SES_t=0}) is split,
i.e. $K^{\tau+7}_{\Z_2 \times \Z_2}(S^1) \cong (1+t)\oplus
\Z_2[t]/(1-t^2)$.  
Then we have
\begin{align}
&R(\Z_2) \to (1+t)\oplus \Z_2[t]/(1-t^2) : && f(t) \mapsto ( f(1)(1+t) , 0 ), \\
&(1+t)\oplus \Z_2[t]/(1-t^2) \overset{f^6_{S^1}}{\to} \Z \oplus \Z_2: && (p(1+t), \nu(t)) \mapsto (2p, x) 
\label{Eq:f^6_{S^1}_split}
\end{align}
with $f(t)\in R(\Z_2)=\Z[t]/(1-t^2)$, $p\in \Z$ and $\nu(t)\in
\Z_2[t]/(1-t^2)$, 
because of the commutativity of the sequence (\ref{Eq:CommSeq_t=0}). 
(Here $x$ is an unknown $\Z_2$ number. )
Equation (\ref{Eq:f^6_{S^1}_split}) implies that ${\rm Ker}(f^6_{S^1})$
has the $\Z_2 \oplus \Z_2$ subgroup, 
however, this contradicts the snake lemma (\ref{Eq:SeqSnake_t=0}). 
Thus, we get 
\begin{align}
K^{\tau+7}_{\Z_2 \times \Z_2}(\widetilde S^1) \cong R(\Z_2)/(2-2t) = \Z \oplus \Z_2. 
\end{align}

\subsection{$t = 1$ order-two NSG}
To distinguish the twist $\tau$ for $t=1$ from that for $t=0$, 
we denote the twist for $t=1$ by $\tau'$.  
The data for the Mayer-Vietoris sequence and the topological numbers are
summarized below:
$$
\begin{tabular}[t]{|c|c|c|c|c|c|}
\hline 
$n$ & AZ class & $K^{\tau'+n}_{\Z_2 \times \Z_2}(\widetilde S^1)$ &
	     $K^{\tau'|_U+n}_{\Z_2 \times \Z_2}(U) \oplus
	     K^{\tau'|_V+n}_{\Z_2 \times \Z_2}(V)$ & $K^{\tau'|_{U\cap
		 V}+n}_{\Z_2 \times \Z_2}(U\cap V)$ & Topological
		     invariant for $K^{\tau'+n}_{\Z_2 \times
		     \Z_2}(\widetilde S^1)$ \\ 
\hline 
$7$ & BDI & $\Z_2$ & $\Z \oplus \Z_2$ & $\Z$ & $\Z_2$ invariant at $k_x = \pi$ \\
$6$ & D & $\Z_4$ & $\Z_2 \oplus 0$ & $0$ & $\Z_4$ invariant in Eq.(\ref{Eq:Z4_inv_1D}) \\
$5$ & DIII & $\Z_2$ & $\Z_2 \oplus 2\Z$ & $\Z$ & $\Z_2$ invariant at $k_x = 0$ \\
$4$ & AII & $\Z_2$ & $0$ & $0$ & $\Z_2$ invariant in
		     Eq.(\ref{Eq:Z4_inv_1D})\footnote{By using $\theta$ in
		     Eq.(\ref{Eq:Z4_inv_1D}), the $\Z_2$
		     invariant $\nu$ is defined as $\nu=\theta/2$. Here
		     $\tau_x$ in $\theta$ is replaced by $is_y$ of
		     $T=is_y K$.} 
\\
$3$ & CII & $0$ & $2 \Z \oplus 0$ & $\Z$ & \\
$2$ & C & $0$ & $0$ & $0$ & \\
$1$ & CI & $0$ & $0\oplus \Z$ & $\Z$ & \\
$0$ & AI & $\Z_2$ & $0 \oplus \Z_2$ & $0$ & $\Z_2$ invariant at $k_x = \pi$ \\
\hline 
\end{tabular}
$$
To obtain the above results, we have used the trivialization of $\tau'$
with Eq.(\ref{Trivialization_Twist}).
Note that
\begin{align}
&K^{\tau'|_U+n}_{\Z_2 \times \Z_2}(U) \cong K_{\R}^{(s=-n,t=1)}(k_x = 0) = \pi_0({\cal R}_{-n-1}), \\
&K^{\tau'|_V+n}_{\Z_2 \times \Z_2}(V) \cong K_{\R}^{(s=-n,t=3)}(k_x = \pi) = \pi_0({\cal R}_{-n+1}), \\
&K^{\tau'|_{U \cap V}+n}_{\Z_2 \times \Z_2}(U \cap V) \cong K_{\C}^{(s=-n,t=1)}(k_x = \pi/2) = \pi_0({\cal C}_{-n+1}),
\end{align}
where $K_{\R}^{(s,t)}(pt)$ $(K_{\C}^{(s,t)}(pt))$ represents the
$K$-group over a point in the presence of real (complex) AZ symmetry $s$ and
additional order-two point group $t$ in Ref.\onlinecite{Shiozaki2014}.
The map $\Delta_n$ in Eq.(\ref{eq:MV}) is given by 
\begin{align}
&\Delta_n = 0, \ \ \mbox{(for even $n$)}, \\
&\Delta_1 : m \mapsto m, \\
&\Delta_3 : 2r \mapsto 2r, \\
&\Delta_5 : (p,2q) \mapsto 2q, \\
&\Delta_7 : (k,l) \mapsto k, 
\end{align}
where $m\in \Z=K_{\Z_2\times \Z_2}^{\tau'|_{V}+1}(V)$, 
$2r\in 2\Z=K_{\Z_2\times \Z_2}^{\tau'|_{U}+3}(U)$, 
$p\in \Z=K_{\Z_2\times \Z_2}^{\tau'|_{U}+5}(U)$,
$2q\in 2\Z=K_{\Z_2\times \Z_2}^{\tau'|_{V}+5}(V)$, 
$k\in \Z=K_{\Z_2\times \Z_2}^{\tau'|_{U}+7}(U)$, 
and $l\in \Z=K_{\Z_2\times \Z_2}^{\tau'|_{V}+7}(V)$.  
From the expression of $\Delta_n$ and the $K$-groups on $U$,$V$ and
$U\cap V$ in the Mayer-Vietoris sequence, we determine the $K$-group
$K^{\tau'|_U+n}_{\Z_2 \times \Z_2}(\widetilde S^1)$ except for $n= 6$,
as shown in the above table. 

The expression of $\Delta_5$ leads to the following short exact sequence 
\begin{align}
\begin{CD}
0  @>>> {\rm Coker}(\Delta_5) @>>> K^{\tau'+6}_{\Z_2 \times \Z_2}(S^1) @>>> \Z_2 @>>> 0 \\
&& @|  \\
&& \Z_2 \\
\end{CD}
\end{align}
We find two possible $K^{\tau'+6}_{\Z_2 \times \Z_2}(\widetilde S^1)$
consistent with the sequence, i.e. 
$K^{\tau'+6}_{\Z_2 \times \Z_2}(\widetilde S^1) 
\cong \Z_2 \oplus \Z_2$ or $\Z_4$. 
We determine $K^{\tau'+6}_{\Z_2 \times \Z_2}(\widetilde S^1)$ in the
following subsection.

\subsubsection{$K^{\tau'+6}_{\Z_2 \times \Z_2}(\widetilde S^1)$}

To evaluate $K^{\tau'+6}_{\Z_2 \times \Z_2}(S^1)$, consider a
one-dimensional 
Hamiltonian with $t=1$ order-two NSG in $n=6$ real AZ class (class D), 
\begin{align}
\left\{ \begin{array}{ll}
C H(k_x) C^{-1} = - H(-k_x), \quad C^2 = 1, \\
U(k_x) H(k_x) U(k_x)^{-1} = - H(k_x), \quad U(k_x)^2 = e^{- ik_x}, 
\quad CU(k_x) = U(-k_x) C. 
\end{array}\right.
\label{Eq:Sym_1D_D_t=1}
\end{align}
Without loss of generality, we assume 
\begin{align}
C = \sigma_0 \otimes \tau_x K, && U(k_x) = \begin{pmatrix}
0 & e^{-i k_x} \\
1 & 0
\end{pmatrix}_{\sigma} \otimes \tau_0 .
\end{align}
From the $U(k_x)$ symmetry, the Hamiltonian 
is written as
\begin{align}
H(k_x) = \begin{pmatrix}
X(k_x) & i Y(k_x) e^{-i k_x/2} \\
-i Y(k_x) e^{i k_x/2} & -X(k_x) 
\end{pmatrix}_{\sigma}, 
\quad k_x \in [-\pi, \pi],
\end{align}
where $X(k_x)$ and $Y(k_x)$ are hermitian matrices with the boundary conditions 
\begin{align}
X(\pi) = X(-\pi), \quad Y(\pi) = -Y(-\pi). 
\end{align}
We now introduce the unitary matrix $Z(k_x) := X(k_x) + i Y(k_x)$, which
satisfies $Z(\pi) = Z(-\pi)^{\dag}$.  
The particle-hole symmetry acts on $Z(k_x)$ as 
\begin{align}
\tau_x Z(k_x)^* \tau_x = - Z(-k_x), \quad k_x \in [-\pi, \pi]. 
\end{align}
Thus, the individual region of $Z(k_x)$ is $k_x \in [0,\pi]$. 
The evaluation of $K^{\tau'+6}_{\Z_2 \times \Z_2}(S^1)$ reduces to 
the homotopy problem of $Z(k_x)$ ($k_x \in [0,\pi]$)
with the boundary conditions: 
\begin{align}
\tau_x Z(0)^* \tau_x = - Z(0), && \tau_x Z(\pi)^* \tau_x = - Z^{\dag}(\pi). 
\end{align}
To solve the homotopy problem, we define a topological number:
At $k_x=\pi$, $\tau_xZ(\pi)$ is antisymmetric, so we can define the Pfaffian, 
$
{\rm Pf}[\tau_x Z(\pi)].
$
We also find that
\begin{align}
{\rm sgn}({\rm det}[\tau_x Z(0)]) = \pm 1 
\end{align}
at $k_x = 0$. 
Let us introduce 
\begin{align}
\theta := \frac{2i}{\pi}\ln {\rm Pf}[\tau_x Z(\pi)] 
- \frac{i}{\pi}\int_0^{\pi} dk_x \partial_{k_x}
\left[ {\rm ln} {\rm det}[\tau_x Z(k_x)] \right]\ \ ({\rm mod\ } 4), 
\label{Eq:Z4_inv_1D}
\end{align}
where the modulo-4 ambiguity of $\theta$ comes from the ambiguity of $\ln {\rm Pf}[\tau_x Z(\pi)]$.  
Noting that 
\begin{align}
2 \theta 
= \frac{2i}{\pi}\ln {\rm det}[\tau_x Z(\pi)] 
- \frac{2i}{\pi} \int_0^{\pi} dk_x \partial_{k_x}[
{\rm ln} {\rm det}[\tau_x Z(k_x)]]
= \frac{2i}{\pi}\ln {\rm det}[\tau_x Z(0)] 
= 0 \ \ {\rm or}\ \ 2 \ \ ({\rm mod\ } 4), 
\end{align}
we find that $\theta=0,1,2,3$ (mod 4).
Thus, $\theta$ defines a $\Z_4$ topological invariant. 
Combining this with the result of the Mayer-Vietoris sequence in the
previous section 
suggests that $K^{\tau'+6}_{\Z_2 \times \Z_2}(S^1)=\Z_4$. 
Therefore, to complete the classification, we only need to show the
existence of a model
with $\theta = 1$.
We find that the following $Z_1(k_x)$ supports $\theta=1$, 
\begin{align}
Z_1(k_x) 
= i \tau_x e^{i \tau_y k_x/2}, \quad  k_x \in [-\pi, \pi]. 
\end{align}
Note that the direct product $Z_1(k_x)\oplus Z_1(k_x)$ gives a model
with $\theta=2$, which is also consistent with the Abelian structure of
the $K$-group.
Thus, $K^{\tau'+6}_{\Z_2 \times \Z_2}(S^1)=\Z_4$.

\bibliography{GlideTI.bib}

\end{document}